\documentclass[mnsc,nonblindrev]{informs3_hide}

\OneAndAHalfSpacedXI 

\usepackage{algorithmic}
\usepackage{float}
\floatstyle{ruled}
\newfloat{algorithm}{htbp}{lop}
\floatname{algorithm}{Algorithm}

\newcommand{\bI}{\mathbbm{1}}
\newcommand{\bN}{\mathbb{N}}
\newcommand{\bR}{\mathbb{R}}
\newcommand{\bE}{\mathbb{E}}
\newcommand{\mathbbm}[1]{\text{\usefont{U}{bbm}{m}{n}#1}} 

\newcommand{\hS}{\hat{S}}
\newcommand{\pH}{\mathsf{H}}
\newcommand{\pL}{\mathsf{L}}
\newcommand{\pC}{\mathsf{C}}
\newcommand{\OPT}{\mathsf{OPT}_{\mathrm{LP}}}
\newcommand{\Rev}{\mathsf{Rev}}

\newcommand{\Bin}{\mathrm{Bin}}
\newcommand{\Ber}{\mathrm{Ber}}

\newcommand{\cS}{\mathcal{S}}
\newcommand{\ssr}{r^*}
\newcommand{\sss}{s^*}
\newcommand{\ssx}{x^*}
\newcommand{\bS}{\breve{S}}

\newcommand{\bmin}{\underline{b}}

\usepackage{natbib,multirow,multicol,xspace,enumitem,subcaption,caption}
\bibpunct[, ]{(}{)}{,}{a}{}{,}%
\def\bibfont{\small}%
\usepackage{bm}
\usepackage[normalem]{ulem}
\usepackage[dvipsnames]{xcolor}

\TheoremsNumberedThrough     
\ECRepeatTheorems

\EquationsNumberedThrough    

\MANUSCRIPTNO{MS-BDA-18-02864.R2}

\begin{document}


\RUNAUTHOR{Ma, Simchi-Levi, and Zhao}

\RUNTITLE{Dynamic Pricing (and Assortment) under a Static Calendar}

\TITLE{Dynamic Pricing (and Assortment) under a Static Calendar}

\ARTICLEAUTHORS{
\AUTHOR{Will Ma}
\AFF{Graduate School of Business, Columbia University, New York NY 10027, \EMAIL{wm2428@gsb.columbia.edu}}
\AUTHOR{David Simchi-Levi}
\AFF{Institute for Data, Systems, and Society, Department of Civil and Environmental Engineering, and Operations Research Center, Massachusetts Institute of Technology, Cambridge, MA 02139, \EMAIL{dslevi@mit.edu}}
\AUTHOR{Jinglong Zhao}
\AFF{Institute for Data, Systems, and Society, Massachusetts Institute of Technology, Cambridge, MA 02139, \EMAIL{jinglong@mit.edu}}
}

\ABSTRACT{
This work is motivated by our collaboration with a large consumer packaged goods (CPG) company. We have found that while the company appreciates the advantages of dynamic pricing, they deem it operationally much easier to plan out a static price calendar in advance.

We investigate the efficacy of static control policies for revenue management problems whose optimal solution is inherently dynamic. In these problems, a firm has limited inventory to sell over a finite time horizon, over which heterogeneous customers stochastically arrive. We consider both pricing and assortment controls, and derive simple static policies in the form of a price calendar or a planned sequence of assortments, respectively. In the assortment planning problem, we also differentiate between the static vs. dynamic substitution models of customer demand. We show that our policies are within 1-1/e (approximately 0.63) of the optimum under stationary (IID) demand, and 1/2 of the optimum under non-stationary demand, with both guarantees approaching 1 if the starting inventories are large.

We adapt the technique of prophet inequalities from optimal stopping theory to pricing and assortment problems, and our results are relative to the linear programming relaxation. Under the special case of IID single-item pricing, our results improve the understanding of irregular and discrete demand curves, by showing that a static calendar can be $(1-1/e)$-approximate if the prices are sorted high-to-low.

Finally, we demonstrate on both data from the CPG company and synthetic data from the literature that our simple price and assortment calendars are effective.
}



\maketitle

\vspace{-0.65cm}

\section{Introduction}
We consider the following general revenue management problem.
A firm has finite inventory of multiple items to sell over a finite time horizon.
The starting inventory is unreplenishable and exogenously given, having been determined by supply chain constraints or a higher-level managerial decision.
The firm can control its sales through sequential decisions in the form of accepting/rejecting customer requests, pricing, or adjusting the assortment of items offered.
Its objective is to maximize the cumulative revenue earned before the time horizon or inventory runs out.

We consider the setting in which customer demand is \textit{distributionally-known} and \textit{independent} over the time horizon; this can be estimated from, e.g., the historical sales data of our partner consumer packaged goods (CPG) company.
The literature has also considered other settings, where an unknown IID demand distribution \citep{besbes2009dynamic, besbes2012blind, agrawal2017mnl} or an evolving demand process correlated across time \citep{araman2009dynamic, ciocan2012model, ahn2019certainty} must be dynamically learned, or where demand is adversarial \citep{ball2009toward,eren2010monopoly}.
In our setting, the firm's decision at one point in time has no impact on its estimate of the demand at another point in time,
which is supported by our data (see Section~\ref{sec::intro:computational} and Section~\ref{sec::computational} for further discussion).
Instead, the time periods are linked by the inventory constraints, and the firm must trade off between \textit{revenue-centric} decisions that maximize expected revenue irrespective of inventory consumption and \textit{inventory-centric} decisions that maximize the yield from the remaining inventory.

Revenue-centric decisions tend to be myopic and maximize the sales volumes of the most popular items, while inventory-centric decisions tend to be conservative and charge higher prices or prioritize selling highly stocked items.
Intuitively, the optimal control policy would make revenue-centric decisions when the overall remaining inventory is plentiful for the remaining time horizon, and inventory-centric decisions when the overall remaining inventory is scarce relative to the remaining time horizon.

However, not all companies have the infrastructure to query the state of the inventory in real-time or adjust their decisions instantaneously.
In fact, in the case of our partner CPG company, prices must be negotiated with the brick-and-mortar retailers that sell their products.
As a result, a price calendar for the year is planned in advance.
On the one hand, this allows the CPG company's management to estimate its promotional budget, make production plans, and coordinate logistics;
on the other hand, this allows the retailer to make advertisements, estimate marketing budgets, and lay out shelf space and price labels accordingly.


Motivated by this problem, we analyze the performance of \textit{static} policies, which must plan out all of the firm's decisions (in this case, the price for each week) at the start of the time horizon (in this case, one year),
in revenue management problems that are intrinsically \textit{dynamic}, where the optimal control would adapt based on the inventory that remains for the time horizon.
If an item's inventory runs out before the end of the time horizon, then its shelf/catalog price is still marked according to the calendar, but no sales of that item can be realized,
since its shelf at the brick-and-mortar retailer would be empty.
We show that our static policies are effective on data provided by the CPG company.
They are also structurally very simple and have performance guarantees comparable to their dynamic counterparts.

\subsection{Models Considered}
We consider the time horizon to consist of a discrete number of time periods.
This does not lose generality, since a continuous time horizon can be modeled by the limiting case in which the time periods are arbitrarily granular.
Similarly, we model each item as having discrete ``price points'' at which it could be sold.
This allows us to both approximate a continuous price range and capture situations where fixed price points have been predetermined by market standards.
As is common for many retailers, our CPG company typically chooses prices that
end in \$.99 (e.g. \$15.99, \$16.99, \$17.99, \$19.99).
Due to menu costs \citep{mankiw1985small, stamatopoulos2017effects}, such a price ladder is rarely changed.

We will separately consider the following two demand models because the design of effective policies differs significantly between them.
\begin{enumerate}
\item \textbf{Stationary} (Section~\ref{sec::stationary}): the demand distribution for a specific decision, e.g.\ the purchase probability $q_t(p)$ associated with price $p$, is identical for all time periods $t$.
\item \textbf{Non-stationary} (Sections~\ref{sec::nonstationary} and \ref{sec::alternativePolicies}): the demand distribution for any decision can vary arbitrarily over time (but still independent across $t$).
\end{enumerate}
We will also consider two types of decisions made by the firm.
\begin{enumerate}
\item \textbf{Pricing (for a Single Item)}: There is a single item with a discrete starting inventory.
We are given, for each time period $t$ and each feasible price $p$, the probability $q_t(p)$ of earning a sale if price $p$ is offered during period $t$.
The goal is to plan the price to offer during each period $t$, with no sales occurring if inventory has stocked out.
\item \textbf{Assortment (and Pricing)}: There are multiple items each with a discrete starting inventory.
We are given, for each time period $t$ and each assortment $S$ of items that could be offered (as well as corresponding prices), the probability of selling each item in $S$ during period $t$.
The goal is to plan out the assortment of items (and prices) to offer during each period $t$,
with no sales occurring if the customer chooses an item that has stocked out.
\end{enumerate}
If the assortment problem includes pricing, then it captures the pricing problem with a single item.

Our results also generalize to the \textbf{fractional-demand} setting,
where the demand distribution given for each period $t$ and price $p$ is over the continuous interval [0,1] (after normalizing), and the sales in the period equal the minimum of the realized demand and remaining inventory.
The generalization to $[0,1]$-demand gives us considerable modeling power.
The dynamic pricing literature (e.g.\ see \citet{gallego1994optimal,talluri2006theory,den2015dynamic,bitran2003overview,elmaghraby2003dynamic}) has focused on the case of Bernoulli demand because the firm can control the price with arbitrary granularity and hence ensure that at most one sale occurs during any ``time period''.
However, in the case of our CPG company, they can only control prices at the week level, during which the demand distribution can range anywhere from a few hundred to a few thousand units.
We will apply the generalization to [0,1]-demand on the data provided by the CPG company in Section~\ref{sec::computational}.

\subsection{Differences between Our Static Policies and Existing Policies}
In this paper, we use the term ``static''
to describe a policy that prescribes a \textit{deterministic} pricing and/or assortment decision for each period $t$ at the very start of the time horizon.
Therefore, the decisions of the static policy must be independent of the sales that end up being realized.
Should an item that the calendar planned to offer be out of stock, we distinguish between two models for how customers behave.
\begin{enumerate}
\item \textbf{Static Substitution}: customers still see the same marked prices (and assortments), but if a customer would have chosen an out-of-stock item, then no sales are realized.
\item \textbf{Dynamic Substitution}: customers only see the calendar-planned items with remaining inventory at the time.  They never choose an out-of-stock item, and may or may not substitute to another in-stock item.
\end{enumerate}
This distinction is irrelevant for the dynamic policies previously studied in the literature, since they can be changed on-the-fly to never offer an out-of-stock item.



Our static policies are based on deterministic linear programs (see Section~\ref{sec::cdlp} for details),
which can be formulated for a given problem instance (items, inventory, prices, and demand distributions) in advance, and hence be used to derive static policies.
At a high level, the LPs use deterministic values to approximate the random execution of a policy, and we can use its optimal solution as a ``guide'' in designing actual policies.

Such an LP was first used for the single-item pricing problem under stationary demand in \citet{gallego1994optimal}, who show that the LP will suggest a single price to offer, and hence a static policy.
A recent paper by \citet{chen2018efficacy} also proposes a similar single-price policy in the face of strategic customers, that achieves the same $1-1/e$ guarantee.
However, this single-price policy requires the critical assumption that the demand, as a function over a continuous price range, is \textit{regular}\footnote{
The regularity assumptions (Assumptions 7.1, 7.2 and 7.3 in \citet{talluri2006theory}) require the demand function (as a function of price) to be strictly decreasing and continuously differentiable and require the revenue function (as a function of demand) to be concave. However, in practice, demand functions are usually not regular. See Figure 5.2 from \citet{talluri2006theory}.
}.
In the general setting with irregular demand or a demand function over discrete price points, the LP will suggest two prices, in which case \citet{gallego1994optimal} develop a policy that adaptively switches between them.

By contrast, we show that it is always better to switch from the higher suggested price to the lower suggested price, and furthermore, we show that a static \textit{switching point} can be computed in advance based on the LP.
The original dynamic pricing policy of \citet{gallego1994optimal} allows the two prices to be offered in either order, but we show that if the policy must be static, then only the high-to-low ordering of prices is effective.


Moving to non-stationary demand, we can no longer directly follow the LP solution.
In fact, we may want to modify certain decisions suggested by the LP to ensure that sufficient inventory is ``reserved'' for higher-revenue time periods (see Example~\ref{exa:LPsub} in Section~\ref{sec:thm:nonstationary:small}).
To accomplish this, we introduce a bid price $c_i$ for each item $i$, which can be interpreted as the opportunity cost of a unit of item $i$'s inventory.
Policies based on bid prices are common in revenue management, and bid prices which vary with the time $t$ can be derived using the approximate dynamic programming techniques in \citet{adelman2007dynamic,rusmevichientong2020dynamic}.
By contrast, our bid prices $c_i$ are time-invariant and reflect the aggregate value of item $i$ over the non-stationary time horizon.  This may be easier for managers to interpret, and also shows managers that an aggregate forecast of demand over the time horizon is sufficient for determining effective bid prices, if we translate those bid prices into a policy appropriately.

Our static policy is to take the LP solution, remove from the suggested assortments all instances where an item $i$ is offered at a price less than $c_i$, and then follow a de-randomized version of the modified solution.
We essentially treat $c_i$ as an acceptance threshold.
Our policy is similar to those of \citet{wang2015online, gallego2016online}, in that it imitates the LP solution and independently determines for each item $i$ when to discard it from the assortment.
However, our discarding rule is static and based on our fixed time-invariant bid prices $c_i$, whereas their discarding rule is dynamic and based on the realized inventory levels.

\subsection{Performance Guarantees and Analytical Techniques}\label{sec::comparisonOfBounds}

\begin{table}[!tb]
\TABLE
{Lower bounds on the performance of static and dynamic policies.  Our new results are bolded.
\label{tbl::outlineOfResults}}
{\begin{tabular}{c|c|c}
\updown & Dynamic Policies & \textbf{Static Policies} \\
\hline
\updown Stationary Demand & $1-1/e\ \xrightarrow[b\to\infty]{\text{w/ error rate }O(1/\sqrt{b})}\ 1$ & $1-1/e\ \xrightarrow[b\to\infty]{\text{w/ error rate }O(1/\sqrt{b})}\ 1$ \\
\hline
\updown \footnotesize Single-item Pricing/Assignment & [\citet{gallego1994optimal}] & [\textbf{Theorem~\ref{thm::assortmentStationary}}] \\
\hline
\updown \footnotesize Assortment (and Pricing) & \multicolumn{2}{c}{[\citet{liu2008choice}; \textbf{Theorem~\ref{thm::assortmentStationary}}]} \\
\hline\hline
\updown Non-stationary Demand & $1/2$ & $1/2$ \\
\hline
\updown \footnotesize Single-item Pricing/Assignment & [\citet{wang2015online}] & [\citet{rusmevichientong2020dynamic}; \\
\cline{1-2}
\updown \footnotesize Assortment (and Pricing) & [\citet{gallego2016online}] &  \textbf{Theorem~\ref{thm::assortmentNonstationary2}}] \\
\hline\hline
\updown Non-stationary Demand & $\xrightarrow[b\to\infty]{\text{w/ error rate }O(1/\sqrt{b})}\ 1$ & $\xrightarrow[b\to\infty]{\text{w/ error rate }O(\sqrt{\log b/b})}\ 1$ \\
\hline
\updown \footnotesize Single-item Pricing/Assignment & [\citet{wang2015online}] & [\citet{hajiaghayi2007automated}; \textbf{Theorem~\ref{thm:astmt:non:asy}}] \\
\hline
\updown \footnotesize Assortment (and Pricing) & [\citet{gallego2016online}] & [\textbf{Theorem~\ref{thm:astmt:non:asy}}] \\
\hline
\end{tabular}}
{Note: $b$ refers to the amount of starting inventory (or the smallest starting inventory, if there are multiple items).}
\end{table}
We establish performance guarantees for our static policies which, in many cases, improve existing guarantees even for dynamic policies.
All of our guarantees are ratios relative to the optimal LP objective value, which is an upper bound on the performance of any static or dynamic policy.
Generally, these LPs are useful because they portray a relaxation of the optimal policy, and hence an optimal LP solution can be used as a ``guide'' in designing a policy for the corresponding problem.
In this paper, we will focus on converting the LP solution into a static policy.

Our results are outlined in Table~\ref{tbl::outlineOfResults}.
The baseline performance ratio is $1-1/e$ for stationary demand and $1/2$ for non-stationary demand.
That is, our static policies always earn at least 50\% of the optimum in expectation, with the ratio improving to $\approx63.2\%$ if the given demand distributions are stationary.
Both of these ratios are tight.
The ratios also increase to 100\% as $b$, the starting inventory level when demand has been normalized to lie in [0,1] (or in the assortment setting, the minimum starting inventory among the items), increases to $\infty$.

In the stationary-demand pricing problem, 
\citet{gallego1994optimal} derived both the lower bound of $1-1/e$ and an asymptotic-optimality result.
However, their policy is in general dynamic, unless the demand function is \textit{regular} over a continuous interval -- the concavity assumption allows for a single price in the LP.
By contrast, we show that the same results can be obtained using our high-to-low static policy, regardless of demand regularity.
Additionally, in our analysis, we derive the tightest possible bound for every value of $b$ and $T$ (the number of time periods), which allows us to establish asymptotic optimality in only $b$ (instead of scaling both $T\to\infty$ and $b\to\infty$).

In the stationary-demand assortment problem, we analyze the policies originally proposed by \citet{liu2008choice} and obtain the same bounds as above that are tight in both $b$ and $T$.
To our knowledge, this type of result, which includes the baseline lower bound of $1-1/e$ when starting inventory is 1, has been previously unknown\footnote{
The results in \citet{golrezaei2014real} imply performance guarantees for our problem, but their ratios are smaller than ours, since they are designed to hold under the more general setting of \textit{adversarial} demand.  Under this demand model, they only obtain a $(1-1/e)$-guarantee under the additional assumptions that each item has a single price, and that starting inventories are asymptotically large.
} for the assortment problem.
Asymptotic optimality was previously derived by \citet{liu2008choice} when both $T\to\infty$ and $b\to\infty$.

Moving to non-stationary demand, the lower bound of 1/2 which improves to 1 as $b\to\infty$ has been previously established using dynamic policies,
in the assignment problem of \citet{wang2015online} and the more general assortment problem of \citet{gallego2016online}.
We establish the same bounds using static policies, with an extremely simple analysis based on \textit{prophet inequalities} from optimal stopping theory.
However, our convergence rate of $1-O(\sqrt{(\log b)/b})$ is worse than the rate of $1-O(1/\sqrt{b})$ achievable with their dynamic policies.

We should mention that the lower bound of 1/2 for static policies under non-stationary demand was also recently established by \citet{rusmevichientong2020dynamic}.
Their bound and analysis differ from ours in that theirs are relative to the optimal dynamic policy instead of the deterministic LP relaxation.
One benefit of using the LP is that it directly extends to the fractional-demand setting, since the LP does not change when demand can take any value in [0,1], which is our application of interest with the CPG company.
By contrast, their framework is designed for a very general setting where resources can be reused after a random amount of time.
We numerically compare the performance of their policy in Section~\ref{sec:synthetic}.




\subsection{Application on Data from CPG Company}\label{sec::intro:computational}
We use aggregated weekly sales data from a CPG company to validate our model, and test the performance of our proposed policies. We use random forest to build prediction models that suggest demand distributions (normalized to lie in [0,1], possibly fractional numbers) under different prices. Then we take these distributions as inputs, and numerically compare the performance of our policies to some basic benchmarks.
\begin{figure}[!htb]\centering
\includegraphics[width=0.7\textwidth,]{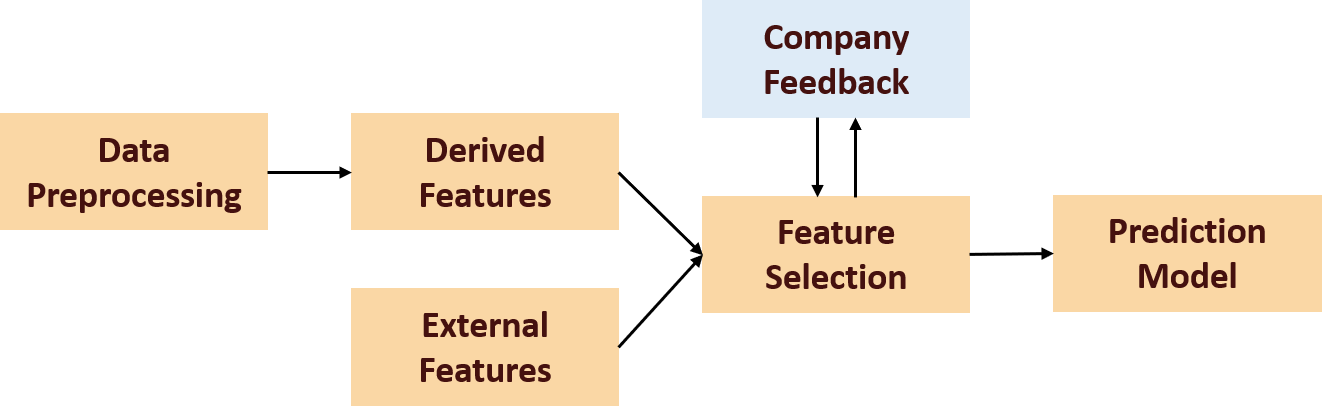}
\caption{Work flow: from data to prediction model}
\label{fig:workflow}
\end{figure}

Working together with the CPG company, we used the work flow depicted in Figure~\ref{fig:workflow} to build our demand model.
The average out-of-sample percent error in its sales predictions is 19.41\%.
It is worth highlighting the features selected by the random forest: the tagged price, external competitor prices, and some external features such as seasonality.
However, neither internal competitor prices (the prices of other SKUs of the CPG company) nor historical prices were selected.
This observation validates our model in the following two aspects:
internal competitor prices not being selected suggests that we can separately optimize the price calendar for each item;
historical prices not being selected suggests that demand can be modeled as independent over time.
The latter aspect is also validated by a stream of empirical literature on the ``pantry effect'' \citep{ailawadi1998effect, bell1999decomposition}, which observes for various consumable goods that if customers attempt to stockpile it when the price is low, then they will untimately consume it more quickly; hence, the low price did not necessarily cannibalize future demand.

Optimizing the price calendar based on our demand model, we find that for scenarios where the starting inventory is of moderate size compared to the total expected demand (i.e., for SKUs that were initially neither overstocked nor understocked),
our static policies outperform basic LP-based static policies by 5\% under stationarity and 1\% under non-stationarity.
Furthermore, our static policies lose at most 1\% under stationarity and 4\% under non-stationarity, compared to the optimal dynamic policies.

Both our theoretical guarantees and computational experiments suggest that static calendars perform nearly as well as their dynamic counterparts.
Managers will not lose much from planning a sequence of prices / assortments in advance.

Further details about our demand modeling and calendar optimization with data from the CPG company can be found in Sections~\ref{sec:predictionModel} and~\ref{sec:predictionModelEC}.

\subsection{Related Work}
%

Our 1/2 guarantee for the general assortment and pricing problem under non-stationary demand (and small inventory) is motivated by \textit{prophet inequalities}, which provide an elegant method for bounding the performance of online vs.\ offline algorithms (see \citet{samuel1984comparison,kleinberg2012matroid}).
The basic idea is to compute a threshold price for each item, based on the offline solution, such that either we are satisfied if an item sells out at its threshold price; or, if it does not, we are still satisfied from having had the opportunity to offer that item to every customer.
Using prophet inequalities, there are very general results for maximizing welfare in online combinatorial auction settings \citep{feldman2014combinatorial,dutting2017prophet}.

However, to the best of our knowledge, it is important for these techniques that the objective is \textit{welfare}, where the decision maker earns a reward equal to the sum of \textit{revenue} and \textit{customer surplus} generated.
When the objective is revenue alone, elegant connections have been made for the single-parameter domain \citep{chawla2010multi,correa2019pricing}, which hold under multiple settings including different arrival orderings.
Our work is the first to make the connection to revenue maximization for assortment optimization under substitutable choice models, which is essentially a multi-parameter domain.
The result is a simple and elegant 1/2 guarantee for assortment and pricing relative to the well-studied linear programming based upper bound, which holds without any assumptions on inventories being large or customers having identical willingness-to-pay distribution.
The fact that our thresholds and guarantee are relative to the upper bound is also novel.
The aforementioned literature has focused on comparing against the expected value of a ``prophet'' who knows the realized valuations in advance (a weaker benchmark than the LP).

We also compare our model in which demand is stochastic and distributionally known, to the models in which demand is completely unknown, or adversarial.
One of the benefits of the adversarial model is that it does not rely on correct ``forecasts'' of demand over time.
However, the drawback is that the resulting algorithm does not make use of forecasted demand information.
Starting with the online booking problem of \citet{ball2009toward} in which the decision concerns whether to accept or reject each customer, single-item pricing \citep{eren2010monopoly,ma2018policies}, assortment optimization \citep{golrezaei2014real}, and joint assortment and pricing \citep{ma2017algorithms} have all been studied under the adversarial demand model.

The guarantees relative to the optimum are worse than ours, because we have more demand information---in fact, in all of the settings except \citet{golrezaei2014real}, the papers resort to \textit{instance-dependent} competitive ratios because a universal non-zero guarantee is impossible when inventories can be depleted at multiple potential prices.
These papers have also tested the empirical performance of their algorithms on datasets and found that using a \textit{hybrid} strategy (as proposed by \citet{mahdian2007allocating}) which employs the algorithms from both the adversarial and stochastic demand models performs best.
This suggests that our algorithmic improvements in the stochastic demand model are valuable even if the given demand distributions are not 100\% correct.

\subsection{Outline}

In Section~\ref{sec:problemDefinitions} we define our basic problems and state the assumptions.
In Section~\ref{sec:ProblemGeneralizations} we discuss some generalizations and their required assumptions. 
In Sections~\ref{sec::stationary} -- \ref{sec::alternativePolicies}, we introduce randomized policies in stationary demand, non-stationary demand, and non-stationary demand with large inventory, respectively.
Then, in Sections~\ref{sec:derandomGeneric} and \ref{sec:structuralDeR}, we introduce general sampling-based de-randomization methods, and structural de-randomization methods, respectively.
These de-randomization methods yield deterministic calendars that (i) have the same theoretical guarantees, (ii) significantly improve computational performance, and (iii) are much easier for companies to accept.
Finally, in Sections~\ref{sec::computational} and \ref{sec:synthetic}, we conduct numerical experiments using real data provided by the CPG company, and synthetic data from the literature.
We also introduce how we estimate the demands from the data in Section~\ref{sec::computational}.

\section{Problem Definitions and Performance Guarantees via Randomized Static Policies} \label{sec:problems}

\subsection{Problem Definitions} \label{sec:problemDefinitions}

Let $\bN$ and $\bN_0$ denote the positive and non-negative integers, respectively.
For any positive integer $n\in\bN$, let $[n] = \{1,\ldots,n\}$.

A firm has $n\in\bN$ items to sell over a finite time horizon of $T\in\bN$ time periods.
Each item $i$ is endowed with $b_i\in\bN$ units of starting inventory, which is unreplenishable.
We assume that $b_i\le T$, which does not lose generality since at most one unit of any item $i$ can be sold during any time period.
Let $\bmin$ denote $\min_{i\in[n]} b_i$.

The firm can offer each item at one of $m\in\bN$ prices, $p_1,\ldots,p_m$, which are positive real numbers.
We will refer to each item-price combination $(i,j)\in[n]\times[m]$ as a \textit{product}, in which case the general assortment and pricing problem can be described as offering a set of products to each customer.

We let $\cS$ be any \textit{downward-closed}\footnote{A family of subsets $\cS$ is downward-closed if for any $S\in\cS$ and $S'\subseteq S$, we also have $S'\in\cS$.} family, which can be used to capture both physical constraints such as shelf-size limitations and business constraints whereby certain items cannot be offered at certain prices (or the same item cannot be simultaneously offered at multiple prices in the form of different products).
We allow for a constraint on the sets of products that can be feasibly offered, imposing that they must lie in some family $\cS$ of subsets of $\{(i,j):i\in[n],j\in[m]\}$.
We will refer to elements in $\cS$ as \textit{assortments}.

A \textit{static policy} is a calendar that must be fixed at the start, prescribing the assortments $S_1,\ldots,S_T\in\cS$ to offer over the time horizon.
In this section, we allow this calendar to be determined in a \textit{random} fashion at the start.
After this calendar has been fixed, sequentially over time $t=1,\ldots,T$, customer $t$ arrives and chooses to purchase at most one product from assortment $S_t$.
For the situation in which some of the products $(i,j)$ in the planned assortment $S_t$ have had their items $i$ stock out before time $t$, we distinguish between two models for how customers choose:
\begin{enumerate}
\item Customer $t$ always sees all of the products in $S_t$, and if her first choice from $S_t$ has stocked out, no sales are realized (\textit{static substitution});
\item Customer $t$ only sees the products in $S_t$ that are still in stock\footnote{Since $\cS$ is downward-closed, it is guaranteed that this selection of products seen by the customer is still feasible. Furthermore, $\emptyset \in \cS$, the empty set is always allowable.}, and chooses her favorite product from this selection (\textit{dynamic substitution}).
\end{enumerate}
We note that \textit{dynamic policies}, as traditionally studied in the literature, do not need to distinguish between static vs.\ dynamic substitution, since they can decide the assortments $S_t$ on-the-fly to never offer an out-of-stock item \citep{rusmevichientong2020dynamic}.
By contrast, for static policies, both models can be justified.
Static substitution occurs in parking systems, where customers are often shown ``phantom'' parking spots, only to drive there and discover that the spot is occupied \citep{owen2017price}.
On the other hand, dynamic substitution occurs if customers switch to a different product when they see that the shelf for their favorite product is empty \citep{anupindi1998estimation, mahajan2001stocking, honhon2010assortment, goyal2016near}.
One factor that possibly reduces the prevalence of dynamic substitution is brand loyalty, under which customers commit to a favorite brand at the supermarket and make no purchase if that brand has stocked out \citep{jacoby1973brand, amine1998consumers, roehm2002designing}.
For product categories where brand loyalty is less common, dynamic substitution is more common.

In either case, for all $t\in[T]$, $S\in\cS$, and $(i,j)\in S$, we let $q_t(i,j,S)$ be the probability of customer $t$ choosing product $(i,j)$ when she \textit{sees} assortment $S$.
Note that $\sum_{(i,j)\in S}q_t(i,j,S)\le1$ for all $t$ and $S$, where $1-\sum_{(i,j)\in S}q_t(i,j,S)$ denotes the probability of customer $t$ purchasing nothing when she sees assortment $S$.
If $(i,j) \notin S$, then $q_t(i,j,S) = 0$.
If the choice probabilities $q_t(i,j,S)$ are equal across $t=1,\ldots,T$, for all assortments $S$ and $(i,j)\in S$, then we say that demand is \textit{stationary}.
If so, we omit the subscript $t$ and refer to the choice probabilities as $q(i,j,S)$.

\subsubsection{Choice-based deterministic linear programs.}
\label{sec::cdlp}

Any (static or dynamic) policy for the assortment problem can be captured by the following LP: let $x_t(S), \forall t \in [T], \forall S \in \mathcal{S}$ represent the probability of offering assortment $S$ at time $t$.
\begin{align}
J^{CDLP-N} = \max\sum_{t=1}^T\sum_{S\in\cS}x_t(S)\sum_{(i,j)\in S}p_jq_t(i,j,S) &&& \label{eqn::obj::assortment} \\
\text{s.t.}\sum_{t=1}^T\sum_{S\in\cS}x_t(S)\sum_{j:(i,j)\in S}q_t(i,j,S) &\le b_i && \forall\ i=1,\ldots,n\label{eqn::constr::assortmentInventory} \\
\sum_{S\in\cS}x_t(S) &=1 && \forall\ t=1,\ldots,T \label{eqn::constr::assortmentCustomer} \\
x_t(S) &\ge0 && \forall\ t=1,\ldots,T;\ \forall\ S\in\cS \label{eqn::constr::assortmentNonneg}
\end{align}
Constraints~\eqref{eqn::constr::assortmentInventory} ensure that total units sold will not exceed the initial inventory, in expectation; constraints~\eqref{eqn::constr::assortmentCustomer} ensure that only one price can be chosen in each time period.
Note that we can assume equality in constraint~\eqref{eqn::constr::assortmentCustomer} because $\cS$ (being downward-closed) always contains the empty assortment $\emptyset$; hence, we can increase $x_t(\emptyset)$ until equality is achieved.

When demand is stationary, we let $x(S), \forall S \in \mathcal{S}$ represent the probability of offering assortment $S$ at any given time period.
Constraints~(\ref{eqn::constr::assortmentCustomer}) are then equivalent to the single constraint~(\ref{eqn::constr::assortmentStationary}).
\begin{align}
J^{CDLP-S} = \max\ T\cdot\sum_{S\in\cS}x(S)\sum_{(i,j)\in S}p_jq(i,j,S) && \label{eqn::obj::assortmentStationary} \\
\text{s.t. } T\cdot\sum_{S\in\cS}x(S)\sum_{j:(i,j)\in S} q(i,j,S) &\le b_i & \forall\ i=1,\ldots,n \label{eqn::constr::assortmentInventoryStationary} \\
\sum_{S\in\cS}x(S) &=1 & \label{eqn::constr::assortmentStationary} \\
x(S) &\ge0 & \forall\ S\in\cS \label{eqn::constr::assortmentNonnegStationary}
\end{align}

We derive performance guarantees for our policies, which are based on the deterministic LPs, relative to the optimal objective values of those LPs.
This also provides a performance guarantee relative to the revenue of any dynamic policy, which is upper-bounded by the LP objective value---this is a well-known type of result in revenue management.
\begin{lemma}[\citet{gallego1994optimal,gallego2004managing}]
The expected revenue of any (static or dynamic) policy for the assortment problem is upper-bounded by the optimal objective value of CDLP-N from \eqref{eqn::obj::assortment} when demand is non-stationary, and CDLP-S from \eqref{eqn::obj::assortmentStationary} when demand is stationary.
Analogously, the expected revenue of any policy for the single-item pricing problem is upper-bounded by the optimal objective value of DLP-S from \eqref{eqn::obj::pricing} in Section~\ref{sec:structuralDeR}.
\end{lemma}
Hereinafter, we will always use the LP objective value as our optimum and denote it using $\OPT$, where the distinction between the LPs will be clear from the context.

The following mild assumption is required for some of our results.  It originated from \citet{golrezaei2014real} and has been nearly omnipresent in the subsequent literature on inventory-constrained assortment optimization \citep{gallego2016online, chen2016assortment, ma2017algorithms, rusmevichientong2020dynamic, ma2018constant, cheung2018inventory}.
\begin{assumption}\label{ass::subs}
For all $t\in[T]$, $S\in\cS$, $S'\subseteq S$, and $(i,j)\in S'$, we have $q_t(i,j,S')\ge q_t(i,j,S)$.
\end{assumption}
This assumption states that the probability of selling a product $(i,j)$ can only be improved if it is offered as part of a smaller assortment $S'$ instead of a larger assortment $S$.
The condition on the choice probabilities is often called \textit{substitutability}.  It is implied by any random-utility choice model (e.g., the multinomial logit choice model used in our computational study in Section~\ref{sec:synthetic}), which treats the products as \textit{substitutes}.


\subsection{Generalized Model with Multi-Consumption and Fractional Consumption}
\label{sec:ProblemGeneralizations}

Before stating our results, we describe a generalized version of the model from Section~\ref{sec:problemDefinitions} that allows for multiple products, as well as fractional amounts of a product, to be consumed in the same time period.
This generalization is natural if we interpret each $t\in[T]$ as a larger-scale epoch (e.g., one week in the case of our CPG company) instead of the choice made by a single customer.

\subsubsection{Multiple purchases.}
We first consider the generalization where a customer can choose multiple products (but still demands exactly 1 unit of each product chosen).  This requires the following modifications to the model from Section~\ref{sec:problemDefinitions}.

We are now given the joint distribution for the \textit{set} of products demanded, when any assortment $S\in\cS$ is seen by any customer $t$.
We assume that the set demanded never contains two different products $(i,j_1),(i,j_2)$ corresponding to the same item $i$, so that even when there is only one unit of $i$ in stock, there is no ambiguity about whether $(i,j_1)$ or $(i,j_2)$ is purchased.

Note that this assumption can be enforced by restricting $\cS$ to not contain any assortment that simultaneously offers different products $(i,j_1),(i,j_2)$ corresponding to the same $i$.
Such a restriction is natural if the only difference between products $(i,j_1)$ and $(i,j_2)$ is in price ($p_{j_1}$ vs.\ $p_{j_2}$), in which case it is nonsensical to mark item $i$ at multiple prices.

The demand is said to be stationary if the distribution of the set of products demanded is identical across $t=1,\ldots,T$ for all assortments $S\in\cS$.
The definitions of Assumption~\ref{ass::subs} and the CDLPs remain unchanged if $q_t(i,j,S)$ now represents the \textit{marginal} probability of product $(i,j)$ being demanded when assortment $S$ is seen by customer $t$.

\subsubsection{Fractional demand consumption.}
The further generalization where demand can be fractional (for multiple products) requires the following further modifications to the model.

We are now given the joint distribution for the \textit{quantity} of each product demanded, when any assortment $S\in\cS$ is seen by any $t$.
We assume that this is a joint distribution over $[0,1]^{|S|}$, where we have normalized the sales of any item within a time period to lie in $[0,1]$ (and scaled its prices accordingly).
We now allow the starting inventories $b_1,\ldots,b_n$ to be any real numbers that are at least 1.  We can assume a lower bound of 1 because any item $i$ with $b_i<1$ can have its demand scaled up by $1/b_i$ so that the maximum possible sales during a time period is 1.

When demand can be fractional, we assume that the joint distribution of demand only depends on the assortment shown, not the exact quantity of each product available.
As a result, \textit{we assume static substitution when demand can be fractional}. Under static substitution, for each $t\in[T]$, $S\in\cS$, and $(i,j)\in S$, let $F_{t,(i,j,S)}(\cdot)$ be the CDF function for the quantity of product $(i,j)$ being demanded, should assortment $S$ be offered during time $t$. $F_{t,(i,j,S)}(\cdot)$ is given and known.
As before, demand is said to be stationary if the joint demand distribution is identical across time. In this case, $F_{t,(i,j,S)}(\cdot) = F_{(i,j,S)}(\cdot), \forall t \in [T], \forall S\in\cS$.
The definitions of Assumption~\ref{ass::subs} and the CDLPs remain unchanged if $q_t(i,j,S) = \bE_{Q\sim F_{t,(i,j,S)}}[Q]$ now represents the \textit{expected quantity} of product $(i,j)$ demanded when assortment $S$ is shown at time $t$.

We state one final assumption that is required for our result under stationary demand, \textit{only when demand can be fractional}.
This assumption is again very mild, in that it automatically holds for \{0,1\}-demand, which is the case studied in all of the existing literature.
We have to add it as a technical assumption in the setting of [0,1] demands.
Whereas we do not make any assumptions (e.g. concavity) on the demand distributions themselves, this assumption concerns the relationships between the different CDF's for the different prices.
\begin{assumption}\label{asp:sta2}
Let $F_{(i,j,S)}$ denote the marginal CDF for the quantity of product $(i,j)$ demanded when assortment $S$ is shown at time $t$.
For all items $i$, feasible assortments $S,S'\in\cS$, and prices $j,j'$ with $p_j > p_{j'}$, assume that for all $c\in[0,1]$,
\begin{align*}
\dfrac{\bE_{Q\sim F_{(i,j,S)}}[\min\{c,Q\}]}{\bE_{Q\sim F_{(i,j,S)}}[Q]}\ge\dfrac{\bE_{Q'\sim F_{(i,j',S')}}[\min\{c,Q'\}]}{\bE_{Q'\sim F_{(i,j',S')}}[Q']}.
\end{align*}
\end{assumption}
The intuitive explanation of Assumption~\ref{asp:sta2} is that for any amount of remaining inventory $c$ for item $i$, the fraction of un-truncated demand sold $\frac{\bE_{Q\sim F_{(i,j,S)}}[\min\{c,Q\}]}{\bE_{Q\sim F_{(i,j,S)}}[Q]}$ is greater at the higher price $j$.
Assumption~\ref{asp:sta2} can be seen as a weaker version of a stochastic dominance assumption on the \textit{hazard rates} of the distributions $F_{(i,j,S)}(x),F_{(i,j',S')}(x)$.
We provide examples and detailed discussions in Section~\ref{sec:assumptions:explanation} in the appendix.

\subsection{Stationary Demand}\label{sec::stationary}

\subsubsection{Statement of results.} \label{sec:stationary:statement}
Our assortment policy probabilistically follows the LP solution, without specifically re-ordering the decisions portrayed in the LP. 
\begin{algorithm}[htb]
\caption{Assortment (and pricing) policy when demand is stationary}
\label{def::pricingStationary}
\begin{algorithmic}[1]
\STATE{Solve CDLP-S, and let $\{\ssx(S):S\in\cS\}$ denote an optimal solution.}
\STATE{Independently for each time $t$, set the assortment $S_t$ to be $S$ with probability $\ssx(S)$, for all $S\in\cS$.}
\end{algorithmic}
\end{algorithm}

This policy that probabilistically imitates the LP was originally studied by \citet{gallego2004managing, liu2008choice}, where it was shown to be empirically effective and asymptotically optimal.
We now derive the first provable guarantees for it in the non-asymptotic setting, as well as a tight characterization of how the guarantee depends on both $T$ and $\underline{b}$.

\begin{theorem}\label{thm::assortmentStationary}
Under the static substitution model (with Assumption~\ref{asp:sta2} needed if demand is fractional),
for the assortment (and pricing) problem under stationary demand, if there are $T$ time periods and $\bmin = \min_{i\in[n]} b_i$, then Algorithm~\ref{def::pricingStationary} earns expected revenue of at least
\begin{align}
\frac{\bE[\min\{\Bin(T,\bmin / T),\bmin \}]}{\bmin}\cdot\OPT, \label{eqn::binomialBound}
\end{align}
where $\Bin(T,\bmin / T)$ denotes a Binomial random variable consisting of $T$ trials of probability $\bmin / T$.

If we let $\Delta^{APX}$ denote the term $\frac{\bE[\min\{\Bin(T,\bmin / T),\bmin\}]}{\bmin}$ from expression~(\ref{eqn::binomialBound}), then
\begin{align}
\Delta^{APX} \geq 1-\frac{\underline{b}^{\underline{b}}}{\underline{b}!}e^{-\underline{b}}, \label{eqn::poissonBound}
\end{align}
which states that $\Delta^{APX}=1-O(1/\sqrt{\underline{b}})$, and increases from $1-1/e$ to 1 as $\underline{b}\to\infty$ (regardless of $T$).
\end{theorem}

In Section~\ref{sec::bernoulliDemandExplanation} we sketch our proof technique for Theorem~\ref{thm::assortmentStationary}, and in Section~\ref{sec::tightness} we show that our approximation guarantee of $\Delta^{APX}=\frac{\bE[\min\{\Bin(T,\bmin / T),\bmin\}]}{\bmin}$ is tight for every value of $T$ and $\bmin$.
But first, we demonstrate why Theorem~\ref{thm::assortmentStationary} does not hold generally under the dynamic substitution model, and identify a special case when it does hold.

\begin{proposition}
\label{prop:StationaryBreaks}
Under the dynamic substitution model and Assumption~\ref{ass::subs} (substitutability), there is an instance of the assortment and pricing problem for which the expected revenue of Algorithm~\ref{def::pricingStationary} is strictly less than $(1-1/e) \cdot \OPT$.
\end{proposition}

The counterexample for Proposition~\ref{prop:StationaryBreaks} is detailed in Appendix~\ref{sec:NecessityOfAssumptions}.
Nonetheless, the counterexample requires both multiple prices (i.e., item 1 that can be sold at multiple prices) and multiple items (i.e., a second item that ``shields'' the first item from being sold at the lower price) to exist.
Theorem~\ref{thm::assortmentStationary} holds in both of the canonical cases of:
\begin{enumerate}
\item Single item, multiple prices (because with a single item, dynamic and static substitution are equivalent);
\item Multiple items, single price per item (this is the pure assortment problem without pricing, as stated next in Proposition~\ref{prop::assortmentStationary}).
\end{enumerate}

\begin{proposition}\label{prop::assortmentStationary}
Under the dynamic substitution model and Assumption~\ref{ass::subs} (substitutability), if each item has only one single price (pure assortment problem without pricing), then Algorithm~\ref{def::pricingStationary} earns expected revenue of at least $\Delta^{APX} \cdot\OPT. $
\end{proposition}
We prove Proposition~\ref{prop::assortmentStationary} and Theorem~\ref{thm::assortmentStationary} together in Section~\ref{sec:proof:assortmentStationary} in the appendix.
Moreover, for the joint assortment and pricing problem under dynamic substitution, a static calendar can still earn $\frac{1}{2}\cdot\OPT$, even when demand is non-stationary,
as we will show via Algorithm~\ref{def::assortmentNonstationary2} in Section~\ref{sec::nonstationary}.

\subsubsection{Two-step proof sketch of Theorem~\ref{thm::assortmentStationary}.}
\label{sec::bernoulliDemandExplanation}
The proof can be divided into two steps, which we will illustrate using the following example.
Consider a problem instance with a single item, time periods $T=3$ and starting inventory $b=2$. Suppose we have two prices. The higher price $p_\pH = 2$ earns a sale with probability $1/3$; the lower price $p_\pL = 1$ earns a sale with probability $1$, i.e. deterministically.
The optimal LP solution from \eqref{eqn::obj::assortmentStationary} -- \eqref{eqn::constr::assortmentNonnegStationary} suggests to offer a higher price $\pH$ for 1.5 time periods, and a lower price $\pL$ for 1.5 time periods.

Let $\bE[\Rev(0.5\pH , 0.5\pL;0.5\pH , 0.5\pL;0.5\pH , 0.5\pL)]$ denote the expected revenue of a randomized policy that offers
$\pH$ and $\pL$ each with probability one half in each period.
Suppose, for the purpose of analysis, that there existed a virtual price $p_\pC = (\bE[Q_\pH] p_\pH + \bE[Q_\pL] p_\pL)/(\bE[Q_\pH] + \bE[Q_\pL])$ with CDF $F_\pC(x) = 0.5 F_\pH(x) + 0.5 F_\pL(x), \forall x \in [0,1]$.
Note that $\bE[Q_\pC] = 2/3$.
We then establish the following sequence of two inequalities:
\begin{align}
\OPT\cdot\frac{\bE[\min\{\Bin(3,\frac{2}{3}),2\}]}{2}&\le \bE[\Rev(\pC;\pC;\pC)] \label{eqn::gen:robustArgument} \\
&\le \bE[\Rev(0.5\pH , 0.5\pL;0.5\pH , 0.5\pL;0.5\pH , 0.5\pL)] \label{eqn::gen:couplingArgument}
\end{align}

Inequality~(\ref{eqn::gen:robustArgument}) relates the LP optimum to the expected revenue of a virtual calendar that always offers $p_\pC$.
We interpret the LHS as the expectation of some Binomial random variable truncated by initial inventory and the RHS as the expectation of an identical-mean, smaller-variance random variable that is also truncated by initial inventory.
Although this virtual calendar cannot actually be offered (because the price $p_{\pC}$ never exists), it can bridge our analysis.

Inequality~(\ref{eqn::gen:couplingArgument}) is true under Assumption~\ref{asp:sta2}.
If the demand is never truncated by the amount of remaining inventory, then offering the virtual price $p_{\pC}$ is equivalent to randomly choosing prices $p_{\pH}$ and $p_{\pL}$ each with probability one-half.  However, if there is truncation, then Assumption~\ref{asp:sta2} guarantees that the revenue from randomly choosing between the real prices $p_{\pH}$ and $p_{\pL}$ cannot be less.

The formal, general proof of Theorem~\ref{thm::assortmentStationary} is deferred to Section~\ref{sec:proof:assortmentStationary} in the appendix.

\subsubsection{Tightness of results.}\label{sec::tightness}
We now show that the ratio produced in expression~(\ref{eqn::binomialBound}), which is dependent on $\underline{b}$ and $T$, is tight.
The proof of Proposition~\ref{prop::binomialBoundTight} can be found in Section~\ref{sec:binomialBoundTightProof} in the appendix.

\begin{proposition}\label{prop::binomialBoundTight}
For any positive integers $T$ and $b$, there exists an instance of the stationary-demand single-item pricing problem with $T$ time periods and $b$ starting inventory, for which the expected revenue of any policy is upper-bounded by expression~(\ref{eqn::binomialBound}).
\end{proposition}

\subsection{Non-stationary Demand with Small Inventory}\label{sec::nonstationary}

In this section, we present our results for non-stationary demand with small inventory.
Our results for non-stationary demand in the asymptotic regime will be discussed in Section~\ref{sec::alternativePolicies}.

\subsubsection{Statement of results.} \label{sec:thm:nonstationary:small}
In contrast to stationary demand, under the more general setting of non-stationary demand, following the LP solution may be undesirable, because it may be beneficial to ``reserve'' inventory for the highest-revenue time periods.  The following Example~\ref{exa:LPsub} demonstrates this idea.

\example
\label{exa:LPsub}
Let there be $T=2$ periods and $b = 1$ unit of initial inventory. Let $\epsilon \in (0,1)$ be some small positive number. Let there be two prices: $p_1 = 1/\epsilon^2, p_2 = 1$. Let random demands be Bernoulli random variables. During day $1$, the purchase probability of offering the higher price $p_1$ is $0$; and the purchase probability of offering the lower price $p_2$ is $1 - \epsilon$. During day $2$, the purchase probability of offering both prices is $\epsilon$.
\begin{table}[!h]
\centering
\begin{tabular}{c|cccccc}
Prices & Period 1 & Period 2 \\
\hline
$p_1$ & $0$ & $\epsilon$ \\
\hline
$p_2$ & $1-\epsilon$ & $\epsilon$
\end{tabular}
\end{table}
DLP-N suggests that we offer $p_2$ in the first period, then $p_1$ in the second period. The objective value of DLP-N is $(1 - \epsilon) + \epsilon \cdot \frac{1}{\epsilon^2} = 1-\epsilon + \frac{1}{\epsilon}$. By simply using the DLP-N solution as a calendar, the expected revenue is $(1 - \epsilon) + \epsilon \cdot \epsilon \cdot \frac{1}{\epsilon^2} = 2 - \epsilon$. We can pick $\epsilon$ to be arbitrarily small and thus directly using LP can be arbitrarily bad.
\Halmos \endexample

Nonetheless, we can still use the LP as a guide for our reservation policies.
\begin{algorithm}[htb]
\caption{Assortment (and pricing) policy when demand is non-stationary}
\label{def::assortmentNonstationary2}
\begin{algorithmic}[1]
\STATE{Solve CDLP-N, and let $\{\ssx_t(S):t\in[T],S\in\cS\}$ denote an optimal solution.}
\STATE{For each item $i$, let $\ssr_i=\sum_{t=1}^T\sum_{S\in\cS}\ssx_t(S)\sum_{j:(i,j)\in S}p_jq_t(i,j,S)$ be the contribution from item $i$ to the optimal objective value (note that $\OPT=\sum_{i=1}^n\ssr_i$).}
\STATE{Independently for each time $t$, first randomly select a $\tilde{S}_t$ to be equal to each $S\in\mathcal{S}$ with probability $x^*_t(S)$, which is a proper probability distribution by constraint \eqref{eqn::constr::assortmentCustomer}. If $\sum_S x^*_t(S) <1$, then select $\tilde{S}_t$ to be the empty set $\emptyset$ with the remaining probability, where $\emptyset \in \mathcal{S}$ is guaranteed by the downward-closed statement in Assumption~\ref{ass::subs}.}
\STATE{Define a discarding rule $D: \mathcal{S} \to \mathcal{S}$ to be
\begin{align} \label{eqn::policyAssortmentNonstationary2}
D(\tilde{S}_t) = \left\{ (i,j) \in \tilde{S}_t \left| p_j > \frac{\ssr_i}{2b_i} \right. \right\}.
\end{align}
After $\tilde{S}_t$ has been selected, set the final assortment to be $S_t = D(\tilde{S}_t) \subseteq \tilde{S}_t$, which is a feasible assortment to offer since $\mathcal{S}$ is downward-closed.}
\end{algorithmic}
\end{algorithm}

Our assortment policy under non-stationary demand uses each cost $\ssr_i/(2b_i)$ as an acceptance threshold.
We remove from the planned assortments all products of item $i$ being offered at prices below their thresholds.
It is probable that the final $S_t$ is an empty set $\emptyset$, even if $\tilde{S}_t$ is not empty, because we discard all the products from $\tilde{S}_t$.

\begin{theorem}\label{thm::assortmentNonstationary2}
Under Assumption~\ref{ass::subs} (substitutability), for the assortment (and pricing) problem where demand may be non-stationary, Algorithm~\ref{def::assortmentNonstationary2} earns expected revenue of at least $\OPT/2$.
\end{theorem}

\subsubsection{Proof sketch of Theorem~\ref{thm::assortmentNonstationary2}.}
By finding the assortment suggested by expression~\eqref{eqn::policyAssortmentNonstationary2}, each unit of item $i$ sold earns at least one-half of the per-inventory revenue of the corresponding $\ssr_i$, which is its contribution to the LP objective.
Thus, if inventory runs out during the horizon, then we have earned in total at least one-half of the LP upper bound.
If inventory never runs out, then the algorithm extracts the full ``opportunity'' from each time period which also results in at least one-half of the LP upper bound.
In other words, setting one-half of the per-inventory revenue as an acceptance threshold is neither too high nor too low, and results in a ``win-win'' situation.
This argument is based on the classical \textit{prophet inequalities} from \citet{krengel1977semiamarts, samuel1984comparison}, where we have modified their argument for optimal stopping to the pricing and assortment settings.

We outline two key steps here, and defer the details of our proof to Section~\ref{sec:proof:assortmentNonstationary2} in the appendix.
\begin{enumerate}
\item To evaluate Algorithm~\ref{def::assortmentNonstationary2}, we take out $\frac{r^*_i}{2b_i}$ revenue earned from each period for each product $(i,j)$. Since after the discarding rule, the prices should be no less than the threshold, i.e. $p_j \geq \frac{r^*_i}{2b_i}$. Thus, this difference should always be non-negative.
That is, $$\Rev \geq \sum_{t=1}^T \sum_{S \in \mathcal{S}} A_t(S) \sum_{(i,j)\in D(S)}(p_j-\frac{\ssr_i}{2b_i})\min\{B_{t-1}(i), Q_t(i,j, D(S))\}+\sum_{i=1}^n\frac{\ssr_i}{2b_i}(b_i-B_T(i)),$$
where $A_t(S)$ is an indicator if assortment $\tilde{S}_t = S$ was selected in period $t$, before the discarding rule from Algorithm~\ref{def::assortmentNonstationary2} was applied; the infimum between $B_{t-1}(i)$, the (random) remaining inventory of item $i$ at the end of period $t-1$, and $Q_t(i,j,D(S))$, the (random) quantity of product $(i,j)$ demanded, is the actual inventory of item $i$ sold in period $t$.
\item We relate the first triple summation term to CDLP-N, the deterministic linear program.
$$\sum_{t=1}^T \sum_{S \in \mathcal{S}} A_t(S) \sum_{(i,j)\in D(S)}(p_j-\frac{\ssr_i}{2b_i})\min\{B_{t-1}(i), Q_t(i,j, D(S))\} \geq \sum_{i=1}^n \frac{\bE[B_T(i)]}{b_i} \left( r_i^* - \frac{r_i^*}{2b_i}\cdot b_i \right).$$
Then after canceling and re-arranging terms we prove the desired result.
\end{enumerate}

\subsubsection{Tightness of results.}
We now show that the ratio in Theorem~\ref{thm::assortmentNonstationary2} is tight. It suffices to find an instance in the single-item pricing problem to show that the general result of assortments (and pricing) problem is tight.
The proof of Proposition~\ref{prop:nonstationaryTight} can be found in Section~\ref{sec:nonstationaryTightProof} in the appendix.
\begin{proposition}\label{prop:nonstationaryTight}
There exists an instance of the non-stationary demand single-item pricing problem for which the expected revenue of any policy is upper-bounded by $\OPT/2$.
\end{proposition}

\subsection{Non-stationary Demand with Large Inventory}\label{sec::alternativePolicies}

We present alternative policies for non-stationary demand that conduct ``reservation'' to a lesser degree than in Algorithm~\ref{def::assortmentNonstationary2}.
Our policies have better performance if starting inventory is large, where the law of large numbers reduces the necessity of reservation, even under non-stationary demand.

We propose a different asymptotic regime from the literature \citet{gallego1997multiproduct, talluri1998analysis, cooper2002asymptotic}, to name a few.
This is because traditional scaling requires $T$ and $b_i, \forall i \in [n]$ to scale up linearly, and under non-stationarity it is unclear how to scale the system.
Instead of letting all $T$ and $b_i, \forall i \in [n]$ to scale up linearly, we allow for arbitrary dependence among $T$ and $b_i, \forall i \in [n]$.
This asymptotic regime is more of theoretical interests, and is sometimes used in the theoretical CS literature.
Note that in practice, the number of initial inventory of different items may be significantly different, which might require some non-trivial normalization to fit into the standard asymptotic regime.

\subsubsection{Statement of results.}
\begin{algorithm}
\caption{Assortment (and pricing) policy when demand is non-stationary and inventory is large}
\label{def::assortmentNonstationary::asp}
\begin{algorithmic}[1]
\STATE{Solve CDLP-N, and let $\{\ssx_t(S):t\in[T],S\in\cS\}$ denote an optimal solution.}
\STATE{For each time $t$, offer each assortment $S \in \cS$ with probability $\ssx_t(S) \cdot (1-\delta)$, and offer $\emptyset\in\cS$ with probability $\delta$, where $\delta = \sqrt{\frac{3 \log{(\underline{b})}}{\underline{b}}}.$}
\end{algorithmic}
\end{algorithm}
In Algorithm~\ref{def::assortmentNonstationary::asp}, $\delta$ can be interpreted as the ``reservation'' probability, which decreases to zero as initial inventory increases.
We reserve inventory by offering the empty set, which is always available.
Note that Theorem~\ref{thm:astmt:non:asy} requires no assumption in the static substitution model.

\begin{theorem}
\label{thm:astmt:non:asy}
Under either the static substitution model or under Assumption~\ref{ass::subs} (substitutability), for the assortment (and pricing) policy where demand may be non-stationary, if $\underline{b} \geq 6$, then Algorithm~\ref{def::assortmentNonstationary::asp} earns revenue that is at least $\left(1 - \sqrt{\frac{3 \log{(\underline{b})}}{\underline{b}}} \right) \OPT$ in expectation.
In particular, Algorithm~\ref{def::assortmentNonstationary::asp} is asymptotically optimal as the starting inventories approach infinity.
\end{theorem}

\subsubsection{Proof sketch of Theorem~\ref{thm:astmt:non:asy}.}
Algorithm~\ref{def::assortmentNonstationary::asp} scales the LP solution by a factor of $1-\delta$, where $\delta$ is a small ``reservation'' probability.
$\delta$ is selected to balance two factors.
First, it is small enough such that if we never stock out, then earning $(1-\delta)\cdot\OPT$ is an asymptotically optimal ratio.
On the other hand, $\delta$ is large enough such that we stock out with probability at most $1/b$.
This intuition is motivated by a tutorial of Anupam Gupta \citep{guptaprophets}, where they introduced the original work of \citet{hajiaghayi2007automated}.
We improve the bounds in the original paper, so that our bound only depends on $b$, but not on $T$. We also generalize to assortment and pricing problems with fractional-demand consumptions.

We outline two key steps here and defer the details of our proof to Section~\ref{sec:proof:thm:astmt:non:asy} in the appendix.
The intuition is as follows: conditioning on the event that ``inventory never runs out'', the expected revenue is at least $1-\delta$ fraction of the LP objective.
Then, we show that this event happens with high probability.
\begin{enumerate}
\item Lower bound the expected revenue by a multiplicative factor of the LP objective, i.e. $\bE[\Rev] \geq \Pr[B_T > 0](1-\delta)\OPT$, where $\delta$ is as defined in Algorithm~\ref{def::assortmentNonstationary::asp}.
\item Using concentration of inequality, lower bound the probability that inventory never runs out, i.e. $\Pr[B_T > 0] \geq 1 - \Pr\left[\sum_{t=1}^T \left\{ Q_t - \bE[Q_t] \right\} \geq \delta b \right] \geq 1 - e^{- \frac{(\delta b)^2}{2 \text{Var}(\sum_{t=1}^T Q_t) + 2/3 \delta b}} \geq 1 - e^{- \frac{\delta^2 b}{2}} = 1 - \frac{1}{b}$.
\end{enumerate}

\section{De-randomization Methods}

In this section we introduce de-randomization methods, which yield deterministic calendars that (i) have the same theoretical guarantees, (ii) significantly improve computational performance (See Section
~\ref{sec:synthetic}), and (iii) are much easier to accept in practice.

Specifically, we first introduce a general de-randomization method in Section~\ref{sec:derandomGeneric} that applies to any
randomized static policy for joint assortment and pricing that we proposed
in Section~\ref{sec:problems}.
Then we introduce two specialized de-randomization methods for Algorithms~\ref{def::pricingStationary} and \ref{def::assortmentNonstationary2} when there is only a single item.
These methods take advantage of the structural properties in the single-item pricing problem.
The de-randomization methods are summarized in Table~\ref{tbl::DeRMethods}.
\begin{table}[!tb]
\TABLE
{A summary of de-randomization methods
\label{tbl::DeRMethods}}
{\begin{tabular}{|l|c|c|}
\hline
\updown  & \footnotesize Single-item Pricing &  \footnotesize Multi-item Joint Assortment and Pricing \\
\hline
\updown  \footnotesize Stationary Demand &  \textbf{Theorem~\ref{thm::pricingStationary} (or~\ref{thm:derandomize})} & \textbf{Theorem~\ref{thm:derandomize}} \\
\hline
\updown  \footnotesize Non-Stationary Demand &  \textbf{Theorem~\ref{thm::pricingNonstationary} (or~\ref{thm:derandomize})} & \textbf{Theorem~\ref{thm:derandomize}} \\
\hline
\updown  \footnotesize Non-Stationary Demand with Large Inventory & \textbf{Theorem~\ref{thm:derandomize}} & \textbf{Theorem~\ref{thm:derandomize}} \\
\hline
\end{tabular}}{}
\end{table}

\subsection{General De-randomization Methods} \label{sec:derandomGeneric}


In this section, we introduce a general simulation-based de-randomization method that achieves the same guarantee as any policy suggested by Algorithms~\ref{def::pricingStationary}--\ref{def::assortmentNonstationary::asp} does.
We consider the general assortment (and pricing) problem under non-stationary demand, which captures single-item pricing and stationary demand as special cases.

Any policy as suggested by Algorithms~\ref{def::pricingStationary}--\ref{def::assortmentNonstationary::asp} that \textit{independently} chooses the assortment in each time period, implies a distribution over static calendars, which can be characterized by the following vector: $$\bm{z} = (z_t(S) : t \in [T], S \in \mathcal{S}) \in [0,1]^{T \cdot \left|\mathcal{S}\right|},$$ where $z_t(S) = \Pr\{S_t=S\}$ is the probability that we offer $S$ in period $t$.
Note there might exist $t$ such that $z_t(\emptyset) > 0$ because our policies could possibly suggest offering nothing in some periods.
We will use the following example to illustrate our de-randomization procedure.


\example
\label{exa:ThreePeriod}
Consider a three-period problem with three options $\cS = \{S_A, S_B, \emptyset\}$.
Note that $\emptyset$ is always available.
Suppose that our randomized policy (possibly from Algorithm~\ref{def::assortmentNonstationary2}) is characterized by
\begin{align*}
\bm{z} = ( & z_1(S_A) = 0.5, & & z_1(S_B) = 0.5, & & z_1(\emptyset) = 0;\\
& z_2(S_A) = 0.5, & & z_2(S_B) = 0, & & z_2(\emptyset) = 0.5;\\
& z_3(S_A) = 0, & & z_3(S_B) = 1, & & z_3(\emptyset) = 0)
\end{align*}
This policy implies a distribution over static calendars such that it takes $(S_A, S_A, S_B)$ with prob. $1/4$, $(S_A, \emptyset, S_B)$ with prob. $1/4$, $(S_B, S_A, S_B)$ with prob. $1/4$, and $(S_B, \emptyset, S_B)$ with prob. $1/4$ (because the assortments for $t=1$ and $t=2$ are drawn independently).
The idea of our de-randomization method is to select one of them that garners the same expected revenue as the distribution of calendars does.
\Halmos \endexample

In general, this distribution over static calendars has a finite but exponentially large support; computing the expected revenue of each calendar using brute force is impossible.
Instead, our method identifies the best assortment to offer iteratively over $t=1,\ldots,T$, using simulation.
Our method requires a simulator $\nu(\bm{z}, \xi): [0,1]^{T \cdot \left|\mathcal{S}\right|} \times \Xi \to \mathbb{R}$, 
whose source of randomness (e.g. the random seed) is characterized by $\xi \in \Xi$.
In each single run of the simulator, it randomly generates (i) a calendar of assortments $S_1, S_2, ..., S_T$, based on the probabilities suggested by $\bm{z}$, and (ii) a sequence of demands based on the choice models and the assortments on the calendar.
Finally, the simulator calculates revenue based on the simulated assortments and demands.
The simulator generates revenue from a bounded interval $[0, (b_1+\cdots+b_n)p_{\mathrm{max}}]$ where $p_{\mathrm{max}}=\max_{j\in[m]} p_j$ is given.

If we query this simulator $K$ times, then we obtain an estimator $\hat{\mu}_K(\bm{z}) = \frac{1}{K}\sum_{k=1}^K \nu(\bm{z}, \xi_k)$ of the expected revenue of policy $\bm{z}$.
We can select $K$ to be a large number such that $\hat{\mu}_K(\bm{z})$ is close to $\bE[\Rev(\bm{z})]$ via a concentration inequality.
We specify our de-randomization method in Algorithm~\ref{def::derandomize}.
\begin{algorithm}[htb]
\caption{Simulation-based de-randomization method for Algorithms~\ref{def::pricingStationary}--\ref{def::assortmentNonstationary::asp}}
\label{def::derandomize}
\begin{algorithmic}[1]
\STATE{
Initialize $\bm{z}$ to be the distribution over calendars suggested by any of Algorithms~\ref{def::pricingStationary}--\ref{def::assortmentNonstationary::asp}.}
\STATE{Fix $K=\lceil \frac{T^2 (b_1+\cdots+b_n)^2 p_{\mathrm{max}}^2}{\OPT^2} \cdot \frac{1}{\epsilon^2} \cdot (\log{n} + \log{T}) \rceil$, the number of samples to average in each period.}
\FOR{$t=1,2,...,T$}
\STATE{
For each assortment $S\in\cS$ with $z_t(S)>0$, estimate the revenue of the randomized calendar $\bm{z}$ with the assortment $S_t$ to offer at time $t$ deterministically set to $S$.
Let $\hat{\mu}_K(\bm{z} \left| S_t = S \right)$ denote the estimate realized (taking the average of $K$ simulation runs), for each $S\in\cS$.
Select $$\hS_t\in\argmax_{S \in \{S \in \mathcal{S} | z_t(S)>0 .\}} \hat{\mu}_K(\bm{z} | S_t = S ).$$
}
\STATE{Update $\bm{z}$ such that $z_t(\hS_t) = 1$, and $z_t(S) = 0$ for all $S \ne \hS_t$.}
\STATE{Offer assortment $\hS_t$ at time $t$.}
\ENDFOR
\end{algorithmic}
\end{algorithm}

\example{\textit{(Example~\ref{exa:ThreePeriod} Continued.)}}
Let $(0.5 S_A, 0.5 S_B; 0.5 S_A, 0.5 \emptyset; S_B)$ denote a randomized policy that
offers $S_A$ and $S_B$ each with probability one half in the first period,
offers $S_A$ and $\emptyset$ each with half-probability in the second period,
and finally offers $S_B$ in the third period.
Algorithm~\ref{def::derandomize} first finds the better one between $(S_A; 0.5 S_A, 0.5 \emptyset; S_B)$ and $(S_B; 0.5 S_A, 0.5 \emptyset; S_B)$.
If the latter is better, in the second iteration it finds the better one between $(S_B; S_A; S_B)$ and $(S_B; \emptyset; S_B)$.
\Halmos \endexample

This idea of iterative de-randomization, when the support of the randomized solution is exponentially sized, has commonly appeared in the computer science literature \citep{motwani1995randomized}.
However, the need for a simulator to evaluate the assortments at each iteration and the analysis of how many samples are needed to lose at most $\epsilon$ in the final de-randomized solution are new to our paper, to the best of our knowledge.

We prove the following result, the proof of which is deferred to Section~\ref{sec:proof:thm:derandomize} in the appendix.
\begin{theorem}
\label{thm:derandomize}
If any policy from Algorithms~\ref{def::pricingStationary} -- \ref{def::assortmentNonstationary::asp} earns expected revenue of at least $\alpha \cdot \OPT$, then the static calendar suggested by Algorithm~\ref{def::derandomize} earns expected revenue of at least $(\alpha - \epsilon) \cdot \OPT$.

The time complexity of Algorithm~\ref{def::derandomize} is $O(\frac{(b_1+\cdots+b_n)^2 p_{\mathrm{max}}^2}{\OPT^2} \cdot \frac{1}{\epsilon^2} n^2 T^4 (\log{n} + \log{T}))$.
\end{theorem}

\subsection{Single-Item Stationary Demand} \label{sec:structuralDeR}
We now introduce a specific de-randomization method for Algorithm~\ref{def::pricingStationary} (the algorithm for stationary demand) in the special case of single-item pricing.  While the generic de-randomization method from Section~\ref{sec:derandomGeneric} will also suffice, the one presented here contains additional structural insights about the de-randomized calendar.
Note that with a single item, dynamic and static substitution are equivalent.

We establish a structural property in Section~\ref{sec:structuralProperty} 
which shows that sorting the static calendar in order of high-to-low prices is dominating.
We show that sorting in the opposite order (low-to-high) earns strictly less expected revenue, in Example~\ref{eg::highToLowNecessary} in Section~\ref{sec:HtoLnecessary}.
We also show that using only one price (without regularity assumptions) earns strictly less expected revenue, in Proposition~\ref{prop:singlePriceInfeasible} in Section~\ref{sec:HtoLnecessary}.

In the single-item pricing problem, we have $n=1$, and we will omit index $i$.
To be able to handle [0,1]-demand (instead of just \{0,1\}-demand), we require the following assumption.
\begin{assumption}\label{asp:sta1}
For any $j,j'\in[m]$, either $F_j(x)\ge F_{j'}(x)$, or $F_j(x)\le F_{j'}(x)$, for all $x\in[0,1]$.
\end{assumption}

When demand is stationary, we have the following LP.
For simplicity, we omit the infinite price, and we put an inequality instead of an equality in \eqref{eqn::constr::pricingStationary}.
\begin{align}
J^{DLP-S} = \max T \cdot \sum_{j=1}^m p_j q_{j} x_{j} && \label{eqn::obj::pricing} \\
\text{s.t.} T \cdot \sum_{j=1}^m q_{j} x_{j} &\le b & \label{eqn::constr::pricingInventory} \\
\sum_{j=1}^m x_j &\le 1 & \label{eqn::constr::pricingStationary}\\
x_{j} &\ge 0 & \forall \ j=1,\ldots,m \label{eqn::constr::pricingNonneg}
\end{align}

The LP for pricing under stationary demand has the following structure.
\begin{lemma}[\citet{gallego1994optimal}]\label{lem::twoPrices}
The DLP-S defined by (\ref{eqn::obj::pricing}), (\ref{eqn::constr::pricingInventory}), (\ref{eqn::constr::pricingStationary}), and (\ref{eqn::constr::pricingNonneg}) has a basic optimal solution $(\ssx_j)_{j=1}^m$ with at most two non-zeros in its support, which we will denote using $\ssx_{\pH}$ (``Higher price'') and $\ssx_{\pL}$ (``Lower price''), with $p_{\pH}\ge p_{\pL}$
\end{lemma}

Based on the above LP and the optimal structure from Lemma~\ref{lem::twoPrices}, we devise the following policy.
\begin{algorithm}
\caption{Single-item pricing policy when demand is stationary}
\label{def::singleStationary}
\begin{algorithmic}[1]
\STATE{Solve DLP-S, and let $p_{\pH},p_{\pL},\ssx_{\pH},\ssx_{\pL}$ correspond to an optimal solution as described in Lemma~\ref{lem::twoPrices}. Denote $s_\pH = T \cdot x^*_\pH / (x^*_\pH + x^*_\pL)$.}
\STATE{Set the price to be $p_{\pH}$ for $t=1,\ldots,\sss$ and $p_{\pL}$ for $t=\sss+1,\ldots,T$. Here $\sss$, the duration for which the higher price is offered, is either $\lfloor\sss_{\pH}\rfloor$ or $\lceil\sss_{\pH}\rceil$.}
\end{algorithmic}
\end{algorithm}
Our policy offers the prices in a \textit{high-to-low} order, with a static switching point.
Intuitively, the high-to-low ordering is desirable, because should we stock out early from higher-than-expected demand realizations, we would rather lose low-priced sales at the end.

\begin{theorem}\label{thm::pricingStationary}
Under Assumption~\ref{asp:sta1}, for the single-item pricing problem under stationary demand with $b$ units of inventory to sell over $T$ periods, Algorithm~\ref{def::singleStationary} earns expected revenue at least
\begin{align*}
\frac{\bE[\min\{\Bin(T,b/T),b\}]}{b}\cdot\OPT,
\end{align*}
\end{theorem}

We prove Theorem~\ref{thm::pricingStationary} in the next section.

\subsubsection{Structural property: monotonicity.} \label{sec:structuralProperty}
We begin by quickly establishing a structural property, Lemma~\ref{lem:structure1}, as a warm-up to the proof of Lemma~\ref{lem::convexity}, which is the key to the de-randomization in the single-item stationary demand setting.
Let $v_t \in [m]$ denote the price index for time $t$ in a calendar, and $v^*_t \in [m]$ denote the optimal price index in a revenue-maximizing calendar. We use $\bm{v}$ to describe the calendar, a vector of price indices. The structural property states the following:

\begin{lemma}
\label{lem:structure1}
In any calendar $\bm{v}$, if two consecutive price indices $v_t,v_{t+1}$ are such that $p_{v_t}<p_{v_{t+1}}$, then indices $v_t$ and $v_{t+1}$ can be exchanged in the calendar without decreasing its expected revenue.
\end{lemma}

The proof of Lemma~\ref{lem:structure1} is deferred to Section~\ref{sec::proof:thm::gen:pricingStationary} in the appendix.
From this Lemma, we know that there exists an optimal static calendar, the prices of which are non-increasing over time.

Now we strengthen the monotonicity property in Lemma~\ref{lem:structure1}.
Consider a problem instance with $T=3$ time periods and starting inventory $b=2$. Suppose we have two prices. The higher price of $2$ earns a sale with probability $1/3$; the lower price of $1$ earns a sale with probability $1$, i.e. deterministically.
The optimal LP solution (according to Lemma~\ref{lem::twoPrices}) suggests offering a higher price index $\pH$ for 1.5 time periods, and a lower price index $\pL$ for 1.5 time periods.

We let $\bE[\Rev(\pH; 0.5\pH , 0.5\pL; \pL)]$ denote the expected revenue of a randomized policy that offers $\pH$ in the first period,
offers $\pH$ and $\pL$ each with probability one half in the second period,
and offers $\pL$ in the third period.
We define $\bE[\Rev(0.5\pH , 0.5\pL;0.5\pH , 0.5\pL;0.5\pH , 0.5\pL)]$ analogously.

The structural property states that if there is a positive probability that one policy offers a lower price before a higher price, then this policy can be improved. For example, 
\begin{align}
\bE[\Rev(0.5\pH , 0.5\pL;0.5\pH , 0.5\pL;0.5\pH , 0.5\pL)] \leq \bE[\Rev(\pH;0.5\pH,0.5\pL;\pL)], \label{eqn::convexityArgument}
\end{align}
There is a positive probability that the policy $(0.5\pH , 0.5\pL;0.5\pH , 0.5\pL;0.5\pH , 0.5\pL)$ does so, since there is already a 1/4 chance that it offers $\pL$ in period 1 and $\pH$ in period 2.
However, the randomized policy $(\pH;0.5\pH , 0.5\pL;\pL)$ could only lead to the calendars $(\pH;\pH;\pL)$ or $(\pH;\pL;\pL)$; in either case it always offers higher prices before lower prices.
The conclusion of inequality~(\ref{eqn::convexityArgument}) is that the first policy can be changed to the second policy without reducing revenue; note that the total expected number of periods that both $\pH$ and $\pL$ are offered is still the same (1.5 periods each).

We now formalize inequality~(\ref{eqn::convexityArgument}).
Let $\{x\} = x - \lfloor x \rfloor$ be the fractional part of a real number $x$.

\begin{lemma}
\label{lem::convexity}
Consider the following two policies:
\begin{enumerate}
\item A policy that offers in each period the same probabilistic mixture of two prices, i.e. a probability $\alpha$ of offering the higher price and a probability $1 - \alpha$ of offering the lower price;
\item A policy that starts by deterministically offering the higher price for $\lfloor \alpha \cdot T \rfloor$ periods, then in the next period offers the higher price with probability $\{\alpha\cdot T\}$ and the lower price with probability $1-\{\alpha \cdot T\}$, and finally switches to offering the lower price in the last $\lceil (1-\alpha) \cdot T \rceil -1$ periods.
\end{enumerate}
The expected revenue of the second policy is no less than the expected revenue of the first policy.
\end{lemma}

The proof of Lemma~\ref{lem::convexity} can be found in Section~\ref{sec::proof:thm::gen:pricingStationary}.

\proof{Proof of Theorem~\ref{thm::pricingStationary}.}
Observe that Algorithm~\ref{def::pricingStationary} suggests the first calendar in Lemma~\ref{lem::convexity}; Algorithm~\ref{def::singleStationary} suggests the second calendar in Lemma~\ref{lem::convexity}.
Theorem~\ref{thm::pricingStationary} holds by invoking Theorem~\ref{thm::assortmentStationary}.
\Halmos\endproof

\subsubsection{Switching from high to low is necessary.}\label{sec:HtoLnecessary}
We show that switching from a higher price to a lower price is necessary, in the sense that if we switch from a lower price to a higher price, we may fail to achieve the bound by expression~(\ref{eqn::binomialBound}).

\example
\label{eg::highToLowNecessary}
Let there be $T=2$ periods and $b=1$ unit of initial inventory. Let there be two prices: $p_1 = 8, p_2 = 1$. The corresponding purchase probabilities are $q_1=0.1, q_2=0.9$. The LP suggests that we offer both $p_1$ and $p_2$ for exactly one period. The LP objective is $\OPT = p_1 q_1 \cdot 1 + p_2 q_2 \cdot 1 = 1.7$

We calculate the bound in expression~(\ref{eqn::binomialBound}): it suggests a $\bE[\min\{\Bin(T,b),b\}] / b = 75\%$ guarantee.

If we offer $p_2$ in period $1$ and then $p_1$ in period $2$, this earns an expected revenue of $p_2 q_2 + (1-q_2) p_1 q_1 = 0.98$, which is $0.98 / 1.7 \approx 57.6\%$ of the LP upper bound.

If we offer $p_1$ in period $1$ and then $p_2$ in period $2$, this earns an expected revenue of $p_1 q_1 + (1-q_1) p_2 q_2 = 1.61$, which is $1.61 / 1.7 \approx 94.7\%$ of the LP upper bound.
\Halmos \endexample

This example demonstrates that switching from a lower price to a higher price is worse than the bound by expression~(\ref{eqn::binomialBound}), and it is even worse than the $1-1/e \approx 63.2\%$ ratio. On the other hand, switching from a higher price to a lower price performs much better.

We also show that two prices are needed to obtain our results in Theorem~\ref{thm::assortmentStationary}.
This is because we do not assume regularity assumptions.
We state Proposition~\ref{prop:singlePriceInfeasible} here and defer its proof to Section~\ref{sec:proof:singlePriceInfeasible} in the appendix.

\begin{proposition}
\label{prop:singlePriceInfeasible}
There exists an instance of the stationary-demand single-item pricing problem for which the expected revenue of any single price policy is strictly smaller than expression~(\ref{eqn::binomialBound}).
\end{proposition}

\subsection{Single-Item Non-Stationary Demand}
Analogous to Section~\ref{sec:structuralDeR}, we now introduce a specific de-randomization method for Algorithm~\ref{def::assortmentNonstationary2} (the algorithm for non-stationary demand) in the special case of single-item pricing.

When demand is non-stationary, we have the following LP.
Let $x_{tj}, \forall t \in [T], j \in [m]$ be the probability that we offer price $j$ in time $t$.
\begin{align*}
J^{DLP-N} = \max\sum_{t=1}^T\sum_{j=1}^mp_jq_{tj}x_{tj} & && \\
\text{s.t.}\sum_{t=1}^T\sum_{j=1}^mq_{tj}x_{tj} &\le b && \\
\sum_{j=1}^{m}x_{tj} &\le1 && \forall\ t=1,\ldots,T \\
x_{tj} &\ge0 && \forall\ t=1,\ldots,T;\ \forall\ j=1,\ldots,m 
\end{align*}

\begin{algorithm}
\caption{Single-item pricing policy when demand is non-stationary}
\label{def::pricingNonstationary}
\begin{algorithmic}[1]
\STATE{Solve DLP-N, and let $\ssr$ denote the optimal objective value.}
\STATE{For each time $t$, set the price to be $p_{j_t}$, where
\begin{align}
j_t\in\argmax_{j}(p_j-\frac{\ssr}{2b})q_{tj}. \label{eqn::policyPricingNonstationary}
\end{align}}
\end{algorithmic}
\end{algorithm}
In (\ref{eqn::policyPricingNonstationary}), $\ssr/b$ can be interpreted as the \textit{per-inventory revenue of the LP}.
Algorithm~\ref{def::pricingNonstationary} guarantees to sell inventory for at least half of this value, since at each time $t$, it maximizes the expected profit with a \textit{bid price} (opportunity cost) of $\ssr/(2b)$.
The intuition is that when there is only one item, we can treat the threshold as a bid price and maximize with respect to it to obtain a deterministic calendar.
This de-randomization is because we use maximization instead of a discarding rule.

\begin{theorem}\label{thm::pricingNonstationary}
For the single-item pricing problem where demand may be non-stationary, Algorithm~\ref{def::pricingNonstationary} earns expected revenue at least $\OPT/2$.
\end{theorem}

We defer the proof of Theorem~\ref{thm::pricingNonstationary} to Section~\ref{sec:proof:thm::pricingNonstationary} in the appendix.

\section{Computational Study}

\subsection{Computational Study: Using Real Data from A CPG Company}\label{sec::computational}

We first describe the business model. Then in Section~\ref{sec:predictionModel}, we explain how we develop the prediction model from data. We discuss the details of feature selection in Section~\ref{sec:selectedFeatures} and justify the motivation of our dynamic pricing model. In Section~\ref{sec:curveFitting}, we explain how we adapt the prediction model such that its output is consistent with managerial intuition and statistically effective. Finally, in Section~\ref{sec:numericalPerformance}, we explain the numerical performance under our proposed policies.

At the end of each year, the CPG company requires a price calendar to be planned for the next year.
This calendar contains 52 weekly prices for each SKU. The CPG company then brings this calendar to its channels (e.g. supermarkets) to negotiate the price-to-customers (PTC).
We assume that they are the same, since the CPG company has full bargaining power.
After the calendar is delivered to channels, the channels decide their yearly advertising strategy,  produce flyers, and make price tags.
These are the reasons (e.g., long lead time on flyers) why we need to plan a calendar in advance.
Customers will not see the prices until the channels release their prices, so there is no anticipatory behavior.

\subsubsection{The random forest model.}\label{sec:predictionModel}

In this section we explain in detail how we develop the prediction model from the data. We will follow the workflow shown in Figure~\ref{fig:workflow} from Section~\ref{sec::intro:computational}.

We begin with weekly sales data in the past 3 years. 
After cleaning the missing data, we select SKUs that generated 90\% of the revenue in the past three years and eliminate the rest.
We also eliminate SKUs that were newly introduced in the most recent year.
Some SKUs are already grouped together by the company.
They are similar brands sold at similar pack sizes.
The company requires that all SKUs in the same group be sold at the same price.
There are 52 distinct groups in total.
We build group-specific prediction models with the same combination of features, i.e., all SKUs use the feature ``tagged price'', but it refers to a different tagged price for each SKU.

We derive a list of features from the data that will be used to predict demand at each time step.
These features include the price that this group is tagged at, its internal competitor prices, its external competitor prices, and its history prices.
The internal competitor prices are the prices of the brands owned by the same company. The external competitor prices are the prices of its true competitors, owned by its rival companies.
The features of history prices are take from the past week to the past 3 weeks, as 3 different features.

The external features include industry seasonal trend (after applying moving average), total number of stores in the district, festivals and sports events.
The first two features are provided by the company, and the rest are obtained by scripting from the Internet.
We create dummy variables for festivals and sports events to characterize categorical data.
\begin{table}[t]
\centering
\TABLE{Different combinations of features, and the resulting out-of-sample error rates.
\label{tbl::featureSelection}}
{\begin{tabular}{|l|c|c|c|c|c|c|}
\hline
Tagged price                                   &       & \checkmark   & \checkmark   & \checkmark   & \checkmark   & \checkmark   \\ \hline
Seasonal industry trend (after moving average) & \checkmark   & \checkmark   & \checkmark   & \checkmark   & \checkmark   & \checkmark   \\ \hline
Total number of stores in the district         &       & \checkmark   & \checkmark   & \checkmark   & \checkmark   & \checkmark   \\ \hline
Festivals and sports events                    &       & \checkmark   & \checkmark   & \checkmark   & \checkmark   &       \\ \hline
External competitor prices                     &       & \checkmark   & \checkmark   & \checkmark   & \checkmark   & \checkmark   \\ \hline
Internal competitor prices (within brand)      &       & \checkmark   &       & \checkmark   &       & \checkmark   \\ \hline
History prices                                 &       & \checkmark   & \checkmark   &       &       &       \\ \hline
Error rate (average over time periods and different SKUs)                   & 32.21 & 21.07 & 21.09 & 19.47 & \textbf{19.41} & 19.80 \\ \hline
\end{tabular}}{}
\end{table}

We tested a few algorithms and finally choose to use random forest (\citet{liaw2002classification, ferreira2015analytics}) as the prediction model.
In Section~\ref{sec:curveFitting} we will discuss an important challenge associated with random forest prediction and how it is addressed.
We aggregate all the features together, then simultaneously perform feature selection and parameter tuning, by using a 5-fold cross-validation.
Finally, the average prediction error is reported as 19.41\%.
This demonstrates a very good prediction model, compared with the number in \citet{ferreira2015analytics}.

\subsubsection{Feature selection.}\label{sec:selectedFeatures}
In this section we explain how we select features.
We validated our model based on the selected features, because this provides indications of the existence (or non-existence) of different sources of cannibalization effects.
For example, if we identified a temporal cannibalization effect, which states that a promotion given today will take away future sales, then our model should address it.
Fortunately, there are no such complications, as supported by the data.

In cross-validation, we evaluated each feature combination based on its median absolute percentage error (MdAPE) on the validation set.
During this procedure, we engaged in rounds of discussions with the company to ensure that the features selected are interpretable.
There are some sub-optimal combinations that the company believed would make more practical sense, and we followed their advice.

These features were both approved by the CPG company's management as consistent with their expedience and also resulted in the lowest out-of-sample prediction errors --- see Table~\ref{tbl::featureSelection} for the reported error rates.
Each column depicts a combination of features, and the corresponding numbers are prediction errors under this feature combination.
The first column serves as a benchmark.
We omit some trivial duplicates of the same feature, but note that some rows represent many features, e.g., festivals and sports events.

The features that were ultimately selected include: the tagged price, external competitor prices\footnote{We do not know the true competitor prices, but we can use ARIMA \citep{hyndman2007automatic, hyndman2020package}, a time series model, to predict competitor prices; by substituting true competitor prices with predicted competitor prices we find the prediction accuracy evaluated on the testing set remains almost unchanged. Thus, we use predicted competitor prices instead of true competitor prices.}, and some external features.
Note that this list includes neither internal competitor prices (from other SKUs of the CPG company) nor historical prices.
We validate our model with the following two observations, which suggest that our model captures the real retail dynamics:

Cross-product cannibalization is not significant. A probable reason is that we have already grouped similar brands sold at similar pack sizes. This suggests that we can employ single-item calendar pricing without considering the joint optimization of simultaneously deciding all SKUs' calendars.

Inter-temporal cannibalization is also not significant. This suggests that demands are not correlated across time, in alignment with our theoretical model. A good explanation of this observation is due to the ``pantry effect'' \citep{ailawadi1998effect, bell1999decomposition}. Even if customers stockpile a product during promotions, this accelerates their consumption rate. Thus, they ultimately purchase as much of the product in subsequent time periods as they otherwise would have, especially for products such as carbonated beverages and ice cream. Therefore, we model demands as independent across time.

\subsubsection{Monotone demand curve.} \label{sec:curveFitting}
\begin{figure}[!htb]
\begin{minipage}[b]{0.49\textwidth}
\includegraphics[width=\textwidth,trim={0 0.5 0 0},clip]{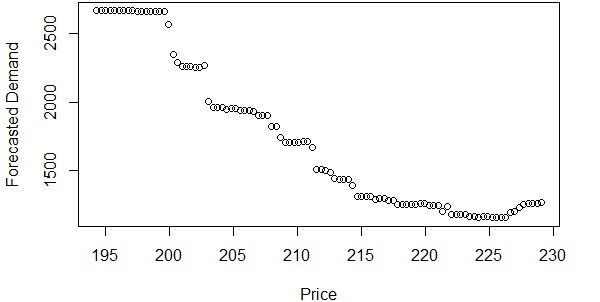}
\end{minipage}
\begin{minipage}[b]{0.49\textwidth}
\includegraphics[width=\textwidth,trim={0 0.5 0 0},clip]{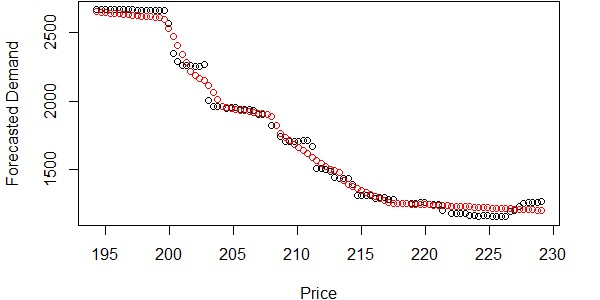}
\end{minipage}
\caption{Predicted price-demand relationship, before and after curve fitting (prices are in industry units)}
\label{fig:PDRelationship}
\end{figure}

If we focus on the price-demand relationship, we observe that the direct output of random forest yields a non-monotone prediction.
In Figure \ref{fig:PDRelationship}, the black dots show the predicted price-demand curve for one SKU in one week.
There are occasions when the predicted demand has positive price elasticity, e.g. when price is between $225$ and $230$.
Since we are selling consumer packaged goods, there is no conspicuous leisure \citep{veblen2017theory}, and price elasticity should be negative.
A non-decreasing demand curve is not acceptable to the CPG company from a managerial perspective.
Theoretically, a non-decreasing demand curve might also (but not necessarily) violate Assumption~\ref{asp:sta1}.

To solve this problem, we introduce curve fitting. First uniformly draw $100$ samples $(P_i, D_i)_{i=1}^{100}$ from the demand curve. Then fit a piecewise linear function, written as
$$f_{\Theta}(x) = \sum_{i = 1}^{d-1} \bI_{\{x_i < x < x_{i+1}\}} \left( \frac{y_{i+1} - y_{i}}{x_{i+1} - x_{i}} (x - x_i) + y_i \right) + \bI_{\{x < x_1\}} \left(a_l (x - x_1) + y_1 \right) + \bI_{\{x > x_d\}} \left(a_r (x - x_d) + y_d \right)$$parameterized by $\Theta = \{x_1, ..., x_d, y_1, ..., y_d, a_l, a_r\} \in \mathrm{R}_+^{2d+2}$, where $d$ is the number of breakpoints, and $\bI_{\{x>a\}}$ are indicator functions equal to one if $x > a$, zero if $x \leq a$. We arbitrarily selected $d$ to be $10$. Finally, we minimize the mean squared error over these $100$ sample points, with shape constraints enforcing a monotonic decreasing function.
\begin{align*}
\min_{\Theta \in \mathcal{R}^{2d+2}} \quad & \sum_{i=1}^{100} (Y_i - f_{\Theta}(X_i))^2 \\
s.t. \quad & \min_j{X_j} \leq x_1 < x_2 < ... < x_d \leq \max_j{X_j}\\
& \max_j{Y_j} \geq y_1 > y_2 > ... > y_d \geq \min_j{Y_j}\\
& a_l, a_r < 0
\end{align*}
We solve the above program using heuristics. The curve-fitting output is depicted in Figure \ref{fig:PDRelationship} as the red dots. The accuracy is slightly improved from 19.41\% to 18.66\%. We do not view this improvement as tremendous, but we have built a model that is in greater agreement with managerial suggestions.
Finally, a monotone demand curve\footnote{To be precise, we derive several scenarios for a specific SKU at a specific price to obtain a discrete distribution, and this suggests a monotone demand curve under each scenario.} satisfies Assumption~\ref{asp:sta1}, as addressed in Section~\ref{sec:structuralDeR}.

\subsubsection{Computational performance of policies.} \label{sec:numericalPerformance}
In this section, we take distributions obtained from the above sections as inputs and compare the performance of our policies to selected benchmarks.
We fix the feasible price set for each SKU to be the prices from its historical data.
The planning horizon is one year, $52$ weeks.
We normalize demands to take $[0,1]$ values by dividing the predicted demands by the highest predicted demand.
We consider different scenarios in which the starting inventory ranges from $1$ unit to $52$ units and analyze both stationary and non-stationary demand models.

\begin{figure}[!htb]\centering
\includegraphics[width=0.8\textwidth,]{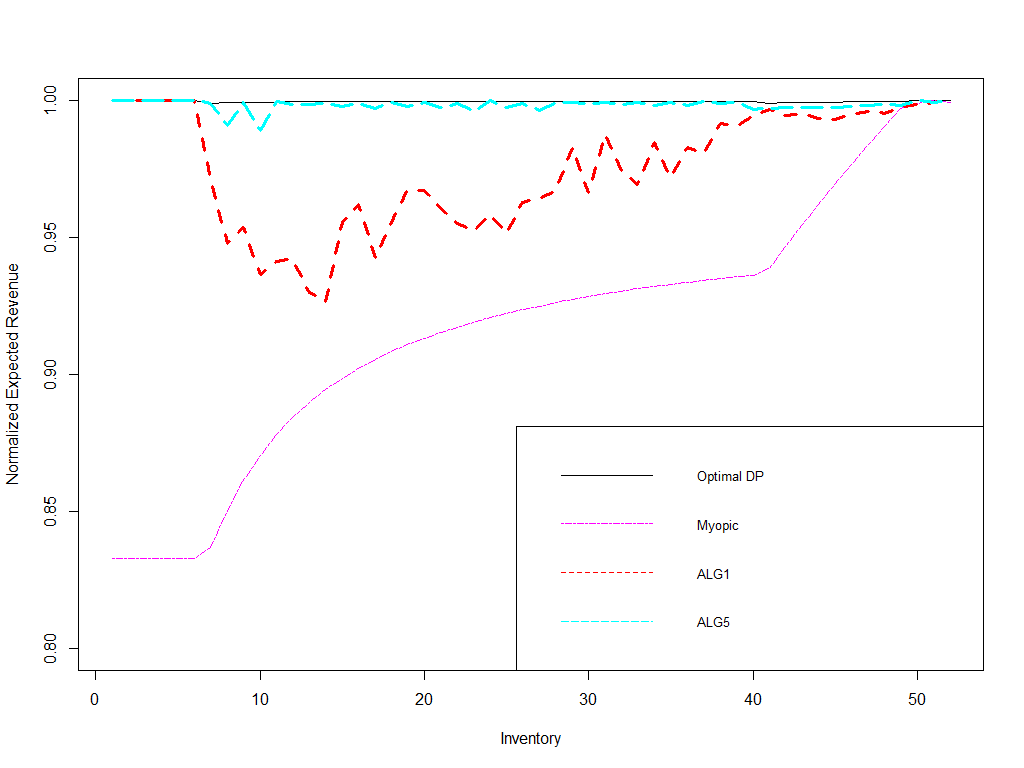}
\caption{Computational performance of polocies under stationary demand}
\label{fig:Stationary}
\end{figure}

\begin{figure}[!htb]\centering
\includegraphics[width=0.8\textwidth,]{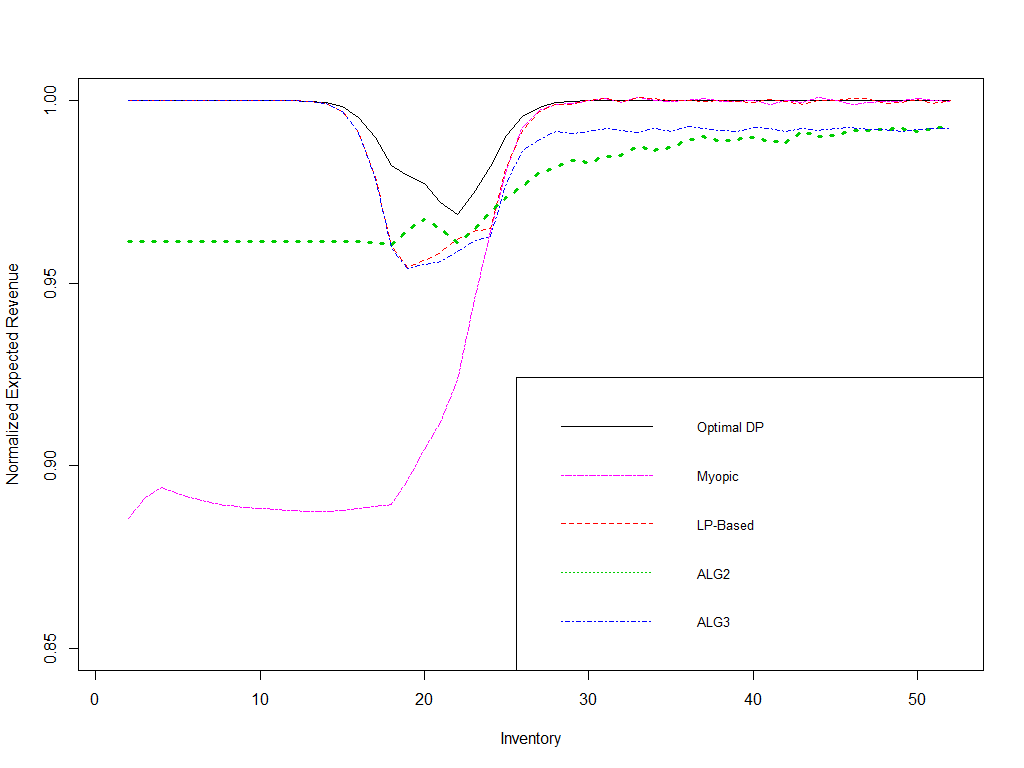}
\caption{Computational performance of polocies under non-stationary demand}
\label{fig:NonStationary}
\end{figure}

We compute the expected revenue from our proposed policies (\textbf{ALG1} from Algorithm~\ref{def::pricingStationary} under stationarity, \textbf{ALG2} from Algorithm~\ref{def::assortmentNonstationary2} under non-stationarity, \textbf{ALG3} from Algorithm~\ref{def::assortmentNonstationary::asp} under non-stationarity, and \textbf{ALG5} from Algorithm~\ref{def::singleStationary} the deterministic policy under stationarity), the LP upper bound, the \textbf{Optimal DP} for the policy solving the optimal dynamic program, the \textbf{Myopic} policy as one benchmark, and an \textbf{LP-Based} randomized policy as another benchmark.
The results are shown in Figures~\ref{fig:Stationary} and~\ref{fig:NonStationary}, where we have divided all numbers by the corresponding LP upper bound, meaning that the performance ratio is always between $0$ and $1$, with higher ratios indicating better performance.

For scenarios in which the starting inventory is of moderate size compared to the total expected demand (i.e.\ for SKUs that were initially neither overstocked nor understocked),
our static policies outperform basic LP-based policies by 5\% under stationarity and 1\% under non-stationarity.
Furthermore, our static policies lose at most 1\% under stationarity and 4\% under non-stationarity, compared to the optimal dynamic policies.

Note that in practice, it is rare for the initial inventory level to be very small or very large, since it would have been pre-optimized\footnote{Pre-optimized by some higher level managers, \citet{talluri2006theory}.} to sell out exactly.
When inventory is of moderate size, our policies outperform the existing benchmarks under both stationary and non-stationary demand settings.

In fact, if we consider the prediction model, the expected demand is approximately $0.2 \sim 0.6$ on each day under different prices, which corresponds to the dip
when inventory $b$ is around $10 \sim 30$.
If we divide them by the time horizon of $T=52$ weeks, we see that the expected units sold per week $b/T$ roughly meets the expected (normalized) demand of $0.2 \sim 0.6$.
This is the region where the pricing problem is non-trivial in theory, and most common in practice.
When the inventory level is such that the problem falls into degenerate cases, 
all the curves are close to the LP upper bound.
This moderate inventory size corresponds to the moderate load scaling factor of $0.6 \sim 1.4$, which is the ratio between initial inventory and mean demand in the admission control problems originated from \citet{zhang2005revenue}.

\subsection{Computational Study: Using Synthetic Data from Literature} \label{sec:synthetic}

In this section, we study the joint assortment and pricing problem, using synthetic data that are commonly adopted in the choice-based deterministic linear program literature, such as \citet{zhang2005revenue, liu2008choice, gallego2016online}.

We closely follow their numerical setup. Let there be three items, each of which can be consumed fractionally.
The items have initial inventory proportional to $\bm{b} = (3,5,4)$.
We will later normalize initial inventory by a load scaling factor $\alpha \in \{0.6, 0.8, 1.0, 1.2, 1.4\}$.
Each item has two prices to be offered, high and low.
We fix the low prices $\bm{p}_\mathrm{L} = (400, 500, 300)$ and let the high price to change from $\bm{p}_\mathrm{H} \in \{(800, 1000, 600), (8000, 10000, 6000)\}$.
We also denote these prices as $\mathrm{L}_1, \mathrm{L}_2, \mathrm{L}_3$ and $\mathrm{H}_1, \mathrm{H}_2, \mathrm{H}_3$.

We specify a choice model to be adopted in our computational study.
It is an adaption from a mixture of MNL models (see, e.g., \citet{li2018product, mcfadden2000mixed}).
However, we interpret the choice probabilities as purchasing market shares, which are between [0,1].
The choice model can be explicitly written as follows
$$Q_t(i,j,S) = A_t^\mathrm{L} \frac{\sum_{(l,\mathrm{L}_l) \in S} \bI_{(l, \mathrm{L}_l) = (i,j)} \cdot v_{l,\mathrm{L}_l}}{v^L_{\emptyset} + \sum_{(l,\mathrm{L}_l) \in S} v_{l,\mathrm{L}_l}} + A_t^\mathrm{H} \frac{\sum_{(l,\mathrm{H}_l) \in S} \bI_{(l, \mathrm{H}_l) = (i,j)} \cdot v_{l,\mathrm{H}_l}}{v^H_{\emptyset} + \sum_{(l,\mathrm{H}_l) \in S} v_{l,\mathrm{H}_l}},$$
where we adopt the fashion that $\frac{0}{0} = 0$.
The randomness in this model comes from the coefficients in the front of each single MNL model.
We let $A_t^\mathrm{L}$ and $A_t^\mathrm{H}$ be Bernoulli random variables.
The mean values $\bE[A_t^\mathrm{L}]$ and $\bE[A_t^\mathrm{H}]$ are given for any $t \in [T]$.

In our computational study, we distinguish between a stationary setting and a non-stationary setting.
In both settings, $T=20$.
In the stationary setting, 
$\bE[A_t^\mathrm{L}] = 0.3, \bE[A_t^\mathrm{H}] = 0.2, \forall t \in [T]$.
In the non-stationary setting,
$\bE[A_t^\mathrm{L}] = 0.8, \bE[A_t^\mathrm{H}] = 0, \forall t \leq 12; \bE[A_t^\mathrm{L}] = 0.2, \bE[A_t^\mathrm{H}] = 0.2, \forall t \geq 13$.
We also specify the attractiveness vectors as $\bm{v}_{\mathrm{L}} = (5, 1, 10), \bm{v}_{\mathrm{H}} = (5, 10, 1)$. We test four different no-purchase vectors to be $(v^\mathrm{L}_\emptyset, v^\mathrm{H}_\emptyset) \in \{(0,0), (1,5), (5,10), (10,20)\}$.

In Tables~\ref{tbl:Sta}--\ref{tbl:Non.BIG}, \textbf{LP UB} stands for the corresponding DLP upper bounds; \textbf{Myopic} stands for the myopic policy that offers the assortment that gives the highest expected revenue, regardless of inventory; \textbf{LP-Sol} stands for the CDLP benchmark, which first solves the LP, then directly uses the optimal solution to implement a (randomized) policy; \textbf{ALG2} and \textbf{ALG3} stand for the policy suggested by Algorithm~\ref{def::assortmentNonstationary2} and Algorithm~\ref{def::assortmentNonstationary::asp}, respectively; \textbf{RST17} stands for the static policy suggested by \citet{rusmevichientong2020dynamic}, by considering the extension in their Section 5.2; and \textbf{DeRLP}, \textbf{DeR2}, and \textbf{DeR3} stand for our de-randomization method from Algorithm~\ref{def::derandomize} applied to the LP-Sol, ALG2, and ALG3, respectively.
All of the percentages are relative to the LP upper bound.

\begin{table}[!htb]
\TABLE
{Computational performance in the stationary setting
\label{tbl:Sta}}
{\footnotesize
\begin{tabular}{c|ccccc|ccccc}
\hline
         & LP UB  & Myopic  & LP-Sol  & RST17   & DeRLP   & LP-UB  & Myopic  & LP-Sol  & RST17   & DeRLP  \\ \hline
$\alpha$ & \multicolumn{5}{c|}{(0,0)}           & \multicolumn{5}{c}{(1,5)}         \\ \hline
0.6      & 4300.0 & 67.76\% & 74.51\% & 81.79\% & \textbf{82.19\%} & 3800.0 & 71.15\% & 80.15\% & \textbf{86.62\%} & 85.50\% \\
0.8      & 5200.0 & 71.28\% & 80.27\% & 83.55\% & \textbf{83.90\%} & 4266.7 & 75.93\% & 87.41\% & 88.23\% & \textbf{90.11\%} \\
1        & 6050.0 & 71.95\% & 82.77\% & 81.27\% & \textbf{83.22\%} & 4566.7 & 78.22\% & \textbf{92.74\%} & 89.74\% & 92.53\% \\
1.2      & 6100.0 & 79.70\% & 90.54\% & 89.17\% & \textbf{90.88\%} & 4586.7 & 83.13\% & 95.04\% & 93.77\% & \textbf{95.87\%} \\
1.4      & 6150.0 & 84.42\% & 93.46\% & 92.13\% & \textbf{94.45\%} & 4606.7 & 86.61\% & 95.26\% & 95.85\% & \textbf{97.63\%} \\ \hline
$\alpha$ & \multicolumn{5}{c|}{(5,10)}          & \multicolumn{5}{c}{(10,20)}           \\ \hline
0.6      & 3200.0 & \textbf{91.67\%} & 87.92\% & 89.54\% & 91.23\% & 2468.9 & \textbf{94.46\%}  & 92.80\% & 91.05\% & 90.81\% \\
0.8      & 3466.7 & 94.28\% & 94.20\% & 90.97\% & \textbf{93.32\%} & 2533.3 & \textbf{97.47\%}  & 97.33\% & 94.78\% & 97.33\% \\
1        & 3500.0 & 97.37\% & 97.46\% & 95.29\% & \textbf{97.47\%} & 2533.3 & 99.29\%  & \textbf{99.35\%} & 98.23\% & \textbf{99.35\%} \\
1.2      & 3500.0 & \textbf{99.14\%} & 99.13\% & 98.00\% & 99.09\% & 2533.3 & 99.76\%  & 99.81\% & 99.41\% & \textbf{99.87\%} \\
1.4      & 3500.0 & 99.86\% & 99.81\% & 99.28\% & \textbf{99.89\%} & 2533.3 & \textbf{100.06\%} & 99.92\% & 99.94\% & 99.96\% \\ \hline
\end{tabular}}{}
\end{table}

In the stationary setting, we observe from Table~\ref{tbl:Sta} that Myopic performs the best when inventory is too much, which is not surprising.
LP-Sol and RST17 have similar performance.
The de-randomization method from Algorithm~\ref{def::derandomize} uniformly improves ($7.68\% \sim -1.99\%$) on the randomized policy in most scenarios.
It also has better performance than RST17 in most scenarios.

\begin{table}[!htb]
\TABLE
{Computational performance in the non-stationary setting when the price difference is small
\label{tbl:Non.SMALL}}
{\footnotesize
\begin{tabular}{c|ccccccccc}
\hline
         & LP UB  & Myopic  & LP-Sol  & ALG2    & ALG3    & RST17    & DeRLP   & DeR2    & DeR3   \\ \hline
$\alpha$ & \multicolumn{8}{c}{(0,0)} \\ \hline
0.6      & 3936.0 & 53.76\% & 81.42\% & 81.37\% & 79.29\% & \textbf{84.13\%} & \textbf{83.47\%} & \textbf{83.40\%} & \textbf{83.51\%} \\
0.8      & 4981.3 & 54.43\% & 84.65\% & 84.61\% & 82.34\% & \textbf{86.26\%} & \textbf{86.80\%} & \textbf{86.82\%} & 85.39\% \\
1        & 6026.7 & 54.50\% & \textbf{86.39\%} & \textbf{86.41\%} & 84.23\% & 84.16\% & \textbf{87.34\%} & \textbf{87.59\%} & \textbf{87.31\%} \\
1.2      & 6304.0 & 60.83\% & 87.17\% & 87.22\% & 84.94\% & \textbf{90.08\%} & \textbf{89.72\%} & \textbf{89.67\%} & \textbf{89.60\%} \\
1.4      & 6581.3 & 66.59\% & 87.33\% & 87.38\% & 85.37\% & \textbf{90.58\%} & \textbf{89.43\%} & \textbf{89.37\%} & \textbf{89.42\%} \\ \hline
$\alpha$ & \multicolumn{8}{c}{(1,5)} \\ \hline
0.6      & 3696.0 & 65.96\% & 84.81\% & 84.75\% & 82.85\% & \textbf{87.85\%} & \textbf{86.03\%} & \textbf{85.93\%} & \textbf{85.93\%} \\
0.8      & 4396.3 & 62.98\% & \textbf{91.80\%} & \textbf{91.84\%} & 88.75\% & 87.00\% & 91.67\% & \textbf{92.28\%} & \textbf{92.23\%} \\
1        & 4535.0 & 66.91\% & \textbf{95.15\%} & \textbf{95.16\%} & 91.14\% & 92.00\% & \textbf{95.95\%} & \textbf{96.02\%} & \textbf{96.10\%} \\
1.2      & 4673.7 & 70.39\% & \textbf{94.91\%} & \textbf{94.93\%} & 91.38\% & 94.69\% & \textbf{96.12\%} & \textbf{96.11\%} & \textbf{96.11\%} \\
1.4      & 4765.1 & 74.26\% & \textbf{97.00\%} & \textbf{96.87\%} & 92.86\% & 95.20\% & \textbf{97.13\%} & \textbf{97.21\%} & \textbf{97.22\%} \\ \hline
$\alpha$ & \multicolumn{8}{c}{(5,10)} \\ \hline
0.6      & 2862.7 & 84.17\% & \textbf{94.29\%} & \textbf{94.16\%} & 90.32\% & 85.83\% & \textbf{95.22\%} & \textbf{95.19\%} & \textbf{95.13\%} \\
0.8      & 3250.2 & 90.20\% & \textbf{94.56\%} & \textbf{94.47\%} & 90.77\% & 91.11\% & \textbf{95.50\%} & \textbf{95.54\%} & \textbf{95.45\%} \\
1        & 3633.9 & 90.88\% & \textbf{95.24\%} & \textbf{95.17\%} & 91.67\% & 93.20\% & \textbf{95.17\%} & \textbf{95.14\%} & \textbf{95.16\%} \\
1.2      & 3696.0 & 95.20\% & \textbf{97.20\%} & \textbf{97.09\%} & 92.92\% & \textbf{97.16\%} & \textbf{97.50\%} & \textbf{97.56\%} & \textbf{97.55\%} \\
1.4      & 3730.3 & \textbf{97.96\%} & \textbf{97.97\%} & \textbf{97.92\%} & 93.55\% & \textbf{98.16\%} & \textbf{97.97\%} & \textbf{97.85\%} & \textbf{97.87\%} \\ \hline
$\alpha$ & \multicolumn{8}{c}{(10,20)} \\ \hline
0.6      & 2364.1 & 92.34\% & 91.70\% & 91.80\% & 88.57\% & \textbf{92.83\%} & \textbf{93.37\%} & \textbf{93.37\%} & \textbf{93.43\%} \\
0.8      & 2755.7 & 93.34\% & 94.63\% & \textbf{94.67\%} & 91.17\% & 93.46\% & \textbf{95.07\%} & \textbf{95.11\%} & \textbf{95.08\%} \\
1        & 2878.3 & \textbf{96.58\%} & \textbf{96.82\%} & \textbf{96.83\%} & 92.70\% & \textbf{96.52\%} & \textbf{96.74\%} & \textbf{96.65\%} & \textbf{96.79\%} \\
1.2      & 2910.8 & \textbf{98.94\%} & \textbf{98.97\%} & \textbf{99.00\%} & 94.40\% & \textbf{99.05\%} & \textbf{98.98\%} & \textbf{99.07\%} & \textbf{98.99\%} \\
1.4      & 2910.8 & \textbf{99.92\%} & \textbf{99.87\%} & \textbf{99.87\%} & 94.95\% & \textbf{99.88\%} & \textbf{99.87\%} & \textbf{99.90\%} & \textbf{99.89\%} \\ \hline
\end{tabular}}{}

\TABLE
{Computational performance in the non-stationary setting when the price difference is large
\label{tbl:Non.BIG}}
{\footnotesize
\begin{tabular}{c|ccccccccc}
\hline
         & LP UB   & Myopic  & LP-Sol  & ALG2    & ALG3    & RST17    & DeRLP   & DeR2    & DeR3    \\ \hline
$\alpha$ & \multicolumn{8}{c}{(0,0)} \\ \hline
0.6      & 48034.0 & 11.54\% & 70.38\% & \textbf{92.04\%} & 69.82\% & 87.43\% & 84.06\% & \textbf{90.88\%} & \textbf{90.49\%} \\
0.8      & 50745.3 & 13.99\% & 71.22\% & \textbf{91.36\%} & 71.16\% & 87.03\% & 86.35\% & \textbf{91.99\%} & 89.14\% \\
1        & 53456.7 & 16.15\% & 71.47\% & \textbf{89.96\%} & 72.49\% & 85.30\% & 74.28\% & \textbf{89.91\%} & 88.36\% \\
1.2      & 54176.0 & 18.91\% & 69.99\% & 87.80\% & 71.65\% & \textbf{92.07\%} & 82.56\% & 87.56\% & \textbf{88.78\%} \\
1.4      & 54895.3 & 22.73\% & 69.25\% & 84.80\% & 71.21\% & \textbf{92.41\%} & 74.46\% & 84.25\% & \textbf{91.03\%} \\ \hline
$\alpha$ & \multicolumn{8}{c}{(1,5)} \\ \hline
0.6      & 36064.0 & 70.34\% & 67.26\%  & 85.89\% & 68.89\% & \textbf{87.00\%} & 71.36\%  & 85.82\% & \textbf{87.97\%}  \\
0.8      & 37835.7 & 70.06\% & \textbf{84.14\%}  & 83.66\% & 82.85\% & \textbf{84.41\%} & 87.79\%  & 83.59\% & \textbf{92.26\%}  \\
1        & 38195.4 & 71.51\% & \textbf{87.89\%}  & 84.81\% & 85.54\% & 85.50\% & 91.34\%  & 84.36\% & \textbf{93.34\%}  \\
1.2      & 38555.0 & 72.62\% & \textbf{88.64\%}  & 86.90\% & 86.62\% & 86.30\% & \textbf{94.42\%}  & 86.22\% & 93.61\%  \\
1.4      & 38831.1 & 73.61\% & \textbf{88.88\%}  & 87.01\% & 87.21\% & 87.44\% & 91.94\%  & 87.12\% & \textbf{94.16\%}  \\ \hline
$\alpha$ & \multicolumn{8}{c}{(5,10)} \\ \hline
0.6      & 28569.7 & 76.22\% & 88.32\% & 88.41\% & 85.89\% & \textbf{89.11\%} & 90.74\% & 89.03\% & \textbf{93.34\%} \\
0.8      & 29574.9 & 87.57\% & 88.32\% & \textbf{88.39\%} & 86.34\% & \textbf{88.44\%} & 94.87\% & 88.65\% & \textbf{96.43\%} \\
1        & 30580.1 & \textbf{96.76\%} & 88.94\% & 87.70\% & 87.03\% & 87.55\% & \textbf{95.67\%} & 87.58\% & 95.06\% \\
1.2      & 30758.5 & 98.52\% & \textbf{99.06\%} & 89.08\% & 94.47\% & 89.45\% & \textbf{98.79\%} & 89.46\% & 92.86\% \\
1.4      & 30855.2 & 98.95\% & \textbf{99.30\%} & \textbf{99.26\%} & 94.26\% & 90.88\% & 99.25\% & \textbf{99.41\%} & 97.83\% \\ \hline
$\alpha$ & \multicolumn{8}{c}{(10,20)} \\ \hline
0.6      & 21379.9 & 83.74\% & \textbf{88.02\%}  & 86.44\% & 85.86\% & \textbf{87.89\%} & \textbf{94.24\%}  & 86.18\% & 91.16\%  \\
0.8      & 22275.8 & \textbf{97.58\%} & 92.11\%  & 86.74\% & 90.23\% & 87.44\% & 93.75\%  & 87.05\% & \textbf{95.77\%}  \\
1        & 22594.0 & \textbf{98.92\%} & 98.81\%  & 91.26\% & 94.00\% & 89.01\% & \textbf{99.02\%}  & 91.38\% & 95.73\%  \\
1.2      & 22676.2 & \textbf{99.80\%} & 99.61\%  & 99.56\% & 95.03\% & 90.80\% & \textbf{99.71\%}  & \textbf{99.77\%} & 95.90\%  \\
1.4      & 22676.2 & \textbf{99.96\%} & \textbf{100.02\%} & 99.87\% & 95.20\% & 92.55\% & \textbf{100.00\%} & \textbf{99.79\%} & \textbf{100.13\%} \\ \hline
\end{tabular}}{}
\end{table}

Moving to non-stationary setting, we observe from Table~\ref{tbl:Non.SMALL} that when the price difference (between $\bm{p}_\mathrm{L}$ and $\bm{p}_\mathrm{H}$) is small, our ALG2 is identical to CDLP and performs well ($81.37\% \sim 99.87\%$).
This is because in many cases, our virtual cost is not large enough to discard any products from the assortment.
ALG2 is among the best algorithms in many of the simulation scenarios $(v_\emptyset^L, v_\emptyset^L) \in \{ (1,5), (5,10), (10,20) \}.$
RST17 also performs well ($84.13\% \sim 99.88\%$) and is the best algorithm in some of the simulation scenarios $(v_\emptyset^L, v_\emptyset^L) \in \{ (0,0), (10,20) \}.$
The de-randomization method from Algorithm~\ref{def::derandomize} also uniformly improves on the corresponding randomized policy in almost all the scenarios and has better performance than RST17.

In Table~\ref{tbl:Non.BIG}, when the price difference is large, our algorithm performs uniformly well ($84.80\% \sim 99.87\%$), while CDLP does not ($67.26\% \sim 100.02\%$).
Note that when $\alpha$ is large and $(v_\emptyset^L, v_\emptyset^L)$ are large, this corresponds to scenarios when inventory is too much.
Myopic performs the best when inventory is too high, which is not surprising.
For the rest of the scenarios, ALG2 and RST17 have the best performance, and sometimes one has better performance than the other.
While it is difficult to say which algorithm performs the best, we remark that RST17 requires more information than ALG2, because RST17 needs to know the exact order in which customers $t$ arrive to solve the DP, while ALG2 only needs to know the universe of customers to solve the LP.
Again, the de-randomization method from Algorithm~\ref{def::derandomize} uniformly improves on the corresponding randomized policy in almost all the scenarios and has better performance than RST17.

\section{Conclusions}
We proposed and analyzed a calendar pricing problem that a consumer packaged goods company favors given its operational convenience.
We considered both single-item pricing and assortment (and pricing) controls.
We showed that our policies are within 1-1/e (approximately 0.63) of the optimum under stationary demand and 1/2 of the optimum under non-stationary demand, with both guarantees approaching 1 if the starting inventory is large.
Our techniques to analyze the best-possible performance guarantees are of theoretical interest per se.
Finally, we fitted the real problem faced by the CPG company into the fractional demand setting of our model and demonstrated using data provided by the CPG company that our simple price calendars are effective.
We also tested our simple policies and literature benchmarks on synthetic data, using the same numerical setup as in the literature.

There are three streams of possible variations on this paper.
First, if we were introducing new products to the market or if there were very little data to generate accurate demand distributions, then we should treat demand distributions as unknown. One interesting question is to address the unknown demand distributions.
The second possible variant is to consider strategic customers.
If we observed significant inter-temporal cannibalization, it would be interesting to develop a strategic customer model.
The literature \citep{gul1986foundations, chen2018efficacy} suggests that commitment power gains an advantage in the presence of strategic customers.
Calendar pricing is naturally such a commitment policy.
A third possible variant is to consider oligopoly pricing.
If we observed cross-SKU cannibalization, then it would be worthwhile to consider how the incentives of different agents align with one another and even with the retailers that carry all competitors' products.

\section*{Acknowledgments}

The authors would like to thank Huseyin Topaloglu for discussions and suggestions during the course of this research.
The authors also thank the department editor Hamid Nazerzadeh, the anonymous associate editor, and three anonymous referees whose comments improved the manuscript.



\begingroup
\OneAndAHalfSpacedXI
\bibliographystyle{informs2014} 
\bibliography{bibliography} 
\endgroup

\ECSwitch


\ECHead{E-Companion}
\section{Building the Random Forest Model}\label{sec:predictionModelEC}

In this section we explain in detail how we built the prediction model from the data.
We follow the workflow shown in Figure~\ref{fig:workflow}.

We begin with weekly sales data in the past 3 years.
After cleaning the missing data, we select SKUs that generated 90\% of the revenue in the past three years and eliminate the rest.
We also eliminate SKUs that were newly introduced in the most recent year.
Some SKUs are already grouped together by the company.
They are similar brands sold at similar pack sizes.
The company requires that all SKUs in the same group be sold at the same price.
There are 52 distinct groups in total.
We build group-specific prediction models with the same combination of features, i.e., all SKUs use the feature ``tagged price'', but it refers to a different tagged price for each SKU.

We derive a list of features from the data that will be used to predict demand at each time step.
These features include the price that this group is tagged at, its internal competitor prices, its external competitor prices, and its history prices.
The internal competitor prices are the prices of the brands owned by the same company. The external competitor prices are the prices of its true competitors, owned by its rival companies.
The features of history prices are take from the past week to the past 3 weeks, as 3 different features.

The external features include industry seasonal trend (after applying moving average), total number of stores in the district, festivals and sports events.
The first two features are provided by the company, and the rest are obtained by scripting from the Internet.
We create dummy variables for festivals and sports events to characterize categorical data.

We tested a few algorithms and finally choose to use random forest (\citet{liaw2002classification, ferreira2015analytics}) as the prediction model.
Random forest provides us the flexibility to use piecewise constant functions to approximate any true demand function, possibly nonlinear functions.
Random forest provides us better performance than simpler models such as linear regression.
On the other hand, it preserves some interpretability of the features, compared to more advanced methods such as neural networks.

Then we aggregate all the features together and simultaneously perform feature selection and parameter tuning by using a 5-fold cross-validation.
We use stepwise backward selection to select features.
In the cross-validation, we evaluate each combination based on its performance on the validation set.

During this procedure, we engaged in rounds of discussions with the company to ensure that the features selected are interpretable.
There are some sub-optimal combinations that the company believed would make more practical sense, and we followed their advice.
These features were both approved by the CPG company's management as consistent with their expedience and also resulted in the lowest out-of-sample prediction errors --- see Table~\ref{tbl::featureSelection} for the reported error rates.
Each column depicts a combination of features, and the corresponding numbers are prediction errors under this feature combination.
The first column serves as a benchmark.
We omit some trivial duplicates of the same feature, but note that some rows represent many features, e.g., festivals and sports events.

The average prediction error is reported as 19.41\%.

\section{Further Justification of Assumptions~\ref{asp:sta2} and \ref{asp:sta1}} \label{sec:assumptions:explanation}

In this section we provide further explanation of Assumptions~\ref{asp:sta2} and \ref{asp:sta1}. Throughout this section we assume that the prices are sorted in decreasing order, i.e. $p_1 > p_2 > ... > p_m$. This is without any loss of generality.


\subsection{Two Examples}
We provide two examples here to illustrate our assumptions.
\begin{figure}[!htb]\centering
   \begin{subfigure}{0.49\textwidth}
     \includegraphics[width=\linewidth,trim={5 15 12 56},clip]{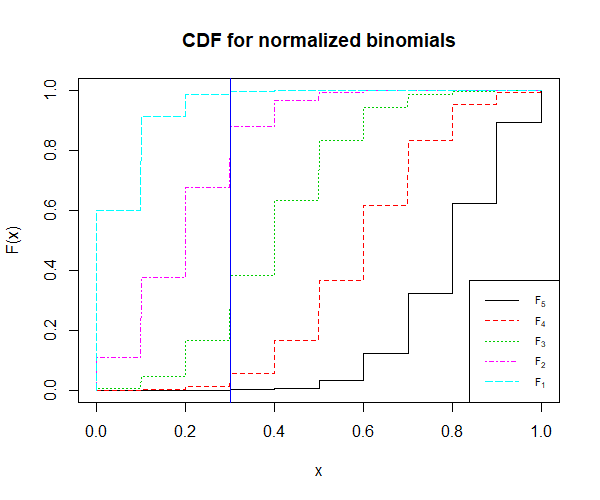} \caption{\footnotesize Example \ref{exa:for:gen:asp2.2}: CDF for normalized binomials}
   \end{subfigure}
   \begin{subfigure}{0.49\textwidth}
     \includegraphics[width=\linewidth,trim={5 15 12 56},clip]{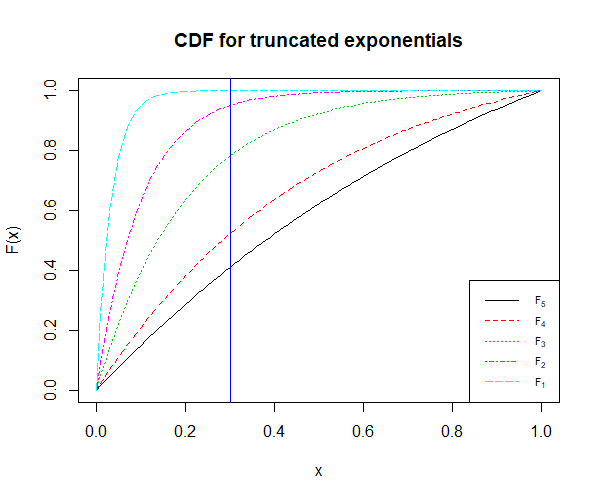} \caption{\footnotesize Example \ref{exa:for:gen:asp2.1}: CDF for truncated exponentials}
   \end{subfigure}
\end{figure}

\example \textit{Normalized binomial distributions.}
\label{exa:for:gen:asp2.2}
Let us restrict ourselves to normalized binomial distributions that have the same number of coin flips, i.e. $\Bin(N, \beta_j) / N$, where $N$ denotes the total number of coin flips, and $\beta_j$ denotes the probability of head-ups. We normalize it by $N$ so that this is a proper distribution with bounded support within $[0,1]$.

Assumption~\ref{asp:sta1} is naturally satisfied.
From Lemma~\ref{lem:StochasticDominance} we know that $\forall j < j', \beta_j \leq \beta_{j'}$ ensures Assumption~\ref{asp:sta2} to hold.
\Halmos \endexample

\example \textit{Truncated exponential distributions.}
\label{exa:for:gen:asp2.1}
Let us restrict ourselves to truncated exponential distributions with bounded support on $[0,1]$, whose CDF can be written as $$F(x) = \frac{1 - e^{-\lambda x}}{1 - e^{-\lambda}}, \forall x \in [0,1].$$

Again Assumption~\ref{asp:sta1} is naturally satisfied.

Assumption~\ref{asp:sta2} states that $\forall j < j', c \in [0,1], \frac{\bE[Q_j]}{\bE[Q_{j'}]} \leq \frac{\bE[\min\{c,Q_j\}]}{\bE[\min\{c,Q_{j'}\}]}.$ Notice that $\bE[X] = \frac{1}{\lambda}, \bE[\min\{c,X\}] = \frac{1 - e^{- \lambda c}}{\lambda (1 - e^{-\lambda})}$. So we have $$\frac{\bE[Q_j]}{\bE[Q_{j'}]} \leq \frac{\bE[\min\{c,Q_j\}]}{\bE[\min\{c,Q_{j'}\}]} \Longleftrightarrow \frac{\lambda_{j'}}{\lambda_{j}} \leq \frac{\lambda_{j'}}{\lambda_{j}} \frac{\frac{1 - e^{- \lambda_{j} c}}{\lambda_{j} (1 - e^{-\lambda_{j}})}}{\frac{1 - e^{- \lambda_{j'} c}}{\lambda_{j'} (1 - e^{-\lambda_{j'}})}} \Longleftrightarrow \lambda_{j} \geq \lambda_{j'}$$ So Assumption~\ref{asp:sta2} holds if and only if $\forall j < j', \lambda_{j} \geq \lambda_{j'}$.
\Halmos \endexample


\subsection{Necessity of Assumption~\ref{asp:sta2}}
Notice that inequality~\eqref{eqn::gen:couplingArgument} holds if and only if Assumption~\ref{asp:sta2} holds. So if Assumption~\ref{asp:sta2} does not hold then inequality~\eqref{eqn::gen:couplingArgument} and Theorem~\ref{thm::assortmentStationary} break down.

\subsection{Necessity of Assumption~\ref{asp:sta1} Through Examples}
We show an example that does not satisfy Assumption~\ref{asp:sta1} and breaks Lemma~\ref{lem::convexity}.
\example
\label{exa:assumption:fail}
Let there be $T=2$ periods and $b = 1$ unit of initial inventory. Let $\epsilon \in (0,1)$ be some small positive number. Let there be two prices: $p_1 = 1+\epsilon, p_2 = 1$. Demand at the higher price $p_1$ is deterministically $1/2 - \epsilon$; and demand at the lower price $p_2$ is $1$ with probability $1/2$, and $2 \epsilon$ with probability $1/2$.

DLP-S suggests that we offer both prices $p_1$ and $p_2$ for one period, since that uses up the $b=1$ inventory exactly in expectation. Indeed, if we ignore the $\epsilon$ terms, $\Rev(\pH; \pL) \approx 3/4$; and $\Rev(0.5\pH, 0.5\pL; 0.5\pH 0.5\pL) \approx 13/16 > 3/4$. So Inequality~\eqref{eqn::convexityArgument} and Lemma~\ref{lem::convexity} break down.
\Halmos \endexample

\section{Inequalities Involving Truncations}
\begin{lemma}
\label{lem:trc:ineq}
$\forall c,x,y \geq 0$, $$\min\{c,x\} + \min\{c,y\} \geq \min\{c,x+y\}$$
\end{lemma}

\proof{Proof.}
We prove by discussing all the possibilities. If $c \leq \min\{x,y\}$, then following from $c \geq 0$ we know that $\min\{c,x\} + \min\{c,y\} = c + c \geq c = \min\{c,x+y\}$.

If $\min\{x,y\} \leq c \leq \max\{x,y\}$, then following from $x,y \geq 0$ we know that $\min\{c,x\} + \min\{c,y\} = \min\{x,y\} + c \geq c = \min\{c,x+y\}$.

If $c \geq \max\{x,y\}$ then we know that $\min\{c,x\} + \min\{c,y\} = x + y \geq \min\{c,x+y\}$.
\Halmos\endproof

\begin{lemma}
\label{lem:trc:ineq2}
For any $p > 0, c \geq 0, x \geq w \geq 0, z \geq y \geq 0$, if $J$ is differentiable, $J(0) = 0, 0 \leq J'(u) \leq p$, and $J$ is a concave function, then the following holds:
\begin{align*}
0 \leq & - p \min\{c, w + y\} + p \min\{c, w + z\} \\
& + p \min\{c, x + y\} - p \min\{c, x + z\} \\
& - J((c - w - y)^+) + J((c - w - z)^+)\\
& + J((c - x - y)^+) - J((c - x - z)^+)
\end{align*}
\end{lemma}

\proof{Proof.}
We prove by enumerating all the possibilities. If $c \leq w+y$, then $0 \leq 0$ we are done.

If $w+y \leq c \leq \min\{w+z, x+y\}$, then it suffices to show that $0 \leq p(c-w-y) - J(c - w - y)$, which is proved by $J(c-w-y) = \int_0^{c-w-y} J'(u) \mathrm{d} u \leq \int_0^{c-w-y} p \mathrm{d} u = p (c-w-y)$.

If $\min\{w+z, x+y\} \leq c \leq \max\{w+z, x+y\}$, without loss of generality we assume $w+z \leq x+y$. So it suffices to show that $0 \leq p(z-y) + J(c-w-z) - J(c-w-y)$, which is proved by $J(c-w-y) = \int_0^{c-w-y} J'(u) \mathrm{d} u \leq \int_0^{c-w-z} J'(u) \mathrm{d} u + \int_{c-w-z}^{c-w-y} p \mathrm{d} u = J(c-w-z) + p(z-y)$.

If $\max\{w+z, x+y\} \leq c \leq x+z$, it suffices to show that $J(c-w-z) + J(c-x-y) + p(x+z-c) \geq J(c-w-y)$. Since $J$ is concave, $J'$ is non-increasing.
\begin{align*}
& J(c-w-z) + J(c-x-y) + p(x+z-c) \\
= & \int_0^{c-w-z} J'(u) \mathrm{d}u + \int_0^{c-x-y} J'(u) \mathrm{d}u + \int_{2c-w-x-y-z}^{c-w-y} p \mathrm{d}u \\
\geq & \int_0^{c-w-z} J'(u) \mathrm{d}u + \int_0^{c-x-y} J'(u + (c-w-z)) \mathrm{d}u + \int_{2c-w-x-y-z}^{c-w-y} p \mathrm{d}u \\
= & \int_0^{c-w-z} J'(u) \mathrm{d}u + \int_{c-w-z}^{2c-w-x-y-z} J'(u) \mathrm{d}u + \int_{2c-w-x-y-z}^{c-w-y} p \mathrm{d}u\\
\geq & \int_0^{c-w-z} J'(u) \mathrm{d}u + \int_{c-w-z}^{2c-w-x-y-z} J'(u) \mathrm{d}u + \int_{2c-w-x-y-z}^{c-w-y} J'(u) \mathrm{d}u \\
= & J(c-w-y)
\end{align*}
where the first inequality is due to concavity of $J$; second inequality due to $J'(u) \leq p$.

Finally if $c \geq x+z$, it suffices to show that $0 \leq - J(c - w - y) + J(c - w - z) + J(c - x - y) - J(c - x - z),$ which is due to concavity of $J$.
\Halmos\endproof

\section{Lemmas for the proof of Theorem~\ref{thm::assortmentStationary}} \label{sec:thm1Lemmas}

These Lemmas are of independent interests.
We state them here, and prove them one by one.

\begin{lemma}
\label{lem:gen:robust}
Let $c \in \mathrm{R}^+$ be any positive real number, $T \in \mathrm{N}$ any positive integer, and $p \in [0,1]$ be any positive fractional number.
Let $\{X_t\}, t = 1,2,...,T$ be i.i.d. random variables with bounded support over $[0,1]$, such that $\bE[X_t] = p, \forall t = 1,2,...,T.$
Let $\{Y_t\}, t = 1,2,...,T$ be i.i.d. Bernoulli random variables, such that with probability $p$, $Y_t =1 , \forall t = 1,2,...,T$.
Then we have $$\bE\left[ \min\{c, \sum_{t=1}^T X_t\} \right] \geq \bE\left[ \min\{c, \sum_{t=1}^T Y_t\} \right].$$
\end{lemma}

\begin{lemma}
\label{lem:SaltyWater}
Suppose $a_i, b_i >0, \forall i \in [n]$, and $\frac{a_1}{b_1} \geq \frac{a_2}{b_2} \geq ... \geq \frac{a_n}{b_n}$; suppose $\beta_1 \geq \beta_2 \geq ... \geq \beta_n \geq 0$. Then we have $$\frac{\sum_{i \in [n]} \beta_i a_i}{\sum_{i \in [n]} \beta_i b_i} \geq \frac{\sum_{i \in [n]}a_i}{\sum_{i \in [n]}b_i}$$
\end{lemma}

\begin{lemma}
\label{lem:common}
Let $c \in \mathrm{R}^+$ be any positive real number, $T \in \mathrm{N}$ any positive integer, the following function $f: (0,1] \to \mathrm{R}^+$ is non-increasing in $x$. $$f(x) = \frac{\bE\left[ \min\{c, \Bin(T,x)\} \right]}{Tx}.$$ 
\end{lemma}

\begin{lemma}
\label{lem:StochasticDominance}
Let $T \in \mathrm{N}$ be any positive integer. For any positive real numbers $x,y$ such that $T \geq x \geq y > 0$, we have $$\frac{\bE[\min\{\Bin(T, x / T),x\}]}{x} \geq \frac{\bE[\min\{\Bin(T, y / T),y\}]}{y}.$$ 
\end{lemma}

\subsection{Proof of Lemma~\ref{lem:gen:robust}}

\proof{Proof of Lemma~\ref{lem:gen:robust}.}
Let $F_X(\cdot)$ and $F_Y(\cdot)$ denote the CDF of $X_1$ and $Y_1$, respectively. First we show $\forall c \geq 0$,
\begin{equation}
\label{eq:x1y1}
\bE\left[ \min\{c, X_1\} \right] \geq \bE\left[ \min\{c,  Y_1\} \right].
\end{equation}
This is obvious when $c \geq 1$, and now we focus on $c < 1$ case.
\begin{align*}
\bE\left[ \min\{c, X_1\} \right] = & \int_{[0,1]} \min\{c, x\} \mathrm{d} F_X(x) \\
= & \int_{[0,1]} x \mathrm{d} F_X(x) - \int_{[c,1]} (x - c) \mathrm{d} F_X(x) \\
= & \bE[X_1] - \left\{ \int_{[c,1]} x \mathrm{d} F_X(x) - c(1 - F_X(c)) \right\} \\
= & \bE[X_1] - \left\{ 1 - c F_X(c) - \int_{[c,1]} F_X(x-0^+) \mathrm{d} x - c(1 - F_X(c)) \right\} \\
= & \bE[X_1] - (1 - c) + \int_{[c,1]} F_X(x) \mathrm{d} x
\end{align*}
where the fourth equality is due to integration by part, as a corollary of Fubini's theorem.
Due to similar analysis, $\bE\left[ \min\{c, Y_1\} \right] = \bE[Y_1] - (1 - c) + \int_{[c,1]} F_Y(x) \mathrm{d} x$. So it suffices to show $\int_{[c,1]} F_X(x) \mathrm{d} x \geq \int_{[c,1]} F_Y(x) \mathrm{d} x$.

Note that $\int_{[0,1]} F_X(x) \mathrm{d} x = \int_{[0,1]} F_Y(x) \mathrm{d} x$, that $F_Y(x) = p, \forall x \in [0,1)$ is a constant, and that $F_X(x)$ is non-decreasing over $x \in [0,1)$. Denote $x_0$ to be the smallest number from $x_0 = \arg \min_{x \in [0,1)} \left\{x \left| F_X(x) \geq p, \lim_{u \to x^-} F_X(u) \leq p \right. \right\}$. Since $F_X(\cdot)$ is right-continuous, $\lim_{u \to x^+} F_X(u) = F_X(x)$.

We distinguish the following two cases. When $c \geq x_0$, $\forall x \in [c,1], F_X(x) \geq F_Y(x).$ So we know $\int_{[c,1]} F_X(x) \mathrm{d} x \geq \int_{[c,1]} F_Y(x) \mathrm{d} x$. When $c \leq x_0$, $\forall x \in [0,c), F_X(x) \leq F_Y(x).$ So we know $\int_{[c,1]} F_X(x) \mathrm{d} x = \int_{[0,1]} F_X(x) \mathrm{d} x - \int_{[0,c)} F_X(x) \mathrm{d} x \geq \int_{[0,1]} F_Y(x) \mathrm{d} x - \int_{[0,c)} F_Y(x) \mathrm{d} x = \int_{[c,1]} F_Y(x) \mathrm{d} x$. In both cases, we have shown that $\forall c \geq 0, \bE\left[ \min\{c, X_1\} \right] \geq \bE\left[ \min\{c,  Y_1\} \right]$.

Then we prove the desired conclusion by pairwise switching $X_t$ into $Y_t$. Fix $\tau$. For any realization of random variables except $X_\tau, Y_\tau$, i.e., for any realization of $X_t = x_t, Y_t = y_t, \forall t \ne \tau$, we have the following:
\begin{align*}
& \bE\left[ \min\{c, \sum_{t=1}^{\tau} X_t + \sum_{t=\tau+1}^{T} Y_t\} \right] \\
= & \bE\left[ \min\{c, \sum_{t=1}^{\tau - 1} x_t + \sum_{t=\tau+1}^{T} y_t + X_\tau\} \right] \\
= & \bE\left[ \min \left\{ \left(c - \sum_{t=1}^{\tau - 1} x_t - \sum_{t=\tau+1}^{T} y_t \right)^+, X_\tau \right\} \right] + \min \left\{c, \sum_{t=1}^{\tau - 1} x_t + \sum_{t=\tau+1}^{T} y_t \right\}\\
\geq & \bE\left[ \min \left\{ \left(c - \sum_{t=1}^{\tau - 1} x_t - \sum_{t=\tau+1}^{T} y_t \right)^+, Y_\tau \right\} \right] + \min \left\{c, \sum_{t=1}^{\tau - 1} x_t + \sum_{t=\tau+1}^{T} y_t \right\}\\
= & \bE\left[ \min\{c, \sum_{t=1}^{\tau-1} X_t + \sum_{t=\tau}^{T} Y_t\} \right]
\end{align*}
where the inequality is due to \eqref{eq:x1y1}.

Repeatedly applying the above inequality, $\bE\left[ \min\{c, \sum_{t=1}^T X_t\} \right] \geq \bE\left[ \min\{c, \sum_{t=1}^T Y_t\} \right], \forall c \geq 0$.
\Halmos\endproof

\subsection{Proof of Lemma~\ref{lem:SaltyWater}}

\proof{Proof of Lemma~\ref{lem:SaltyWater}.}
Since $\frac{a_1}{b_1} \geq \frac{a_2}{b_2} \geq ... \geq \frac{a_n}{b_n}$, we have $\forall i \in [n-1]$, $$\frac{\sum_{j \in [i]}a_j}{\sum_{j \in [i]}b_j} \geq \frac{\sum_{j \in [n]}a_j}{\sum_{j \in [n]}b_j}.$$ Then we plug it into the following fraction, and use the fact that $\beta_i \geq \beta_{i+1}, \forall i \in [n-1]$:
\begin{align*}
\frac{\sum_{i \in [n]} \beta_i a_i}{\sum_{i \in [n]} \beta_i b_i} = & \frac{\sum_{i \in [n-1]} (\beta_i - \beta_{i+1}) \cdot \sum_{j \in [i]}a_j + \beta_n \cdot \sum_{j \in [n]} a_j}{\sum_{i \in [n-1]} (\beta_i - \beta_{i+1}) \cdot \sum_{j \in [i]}b_j + \beta_n \cdot \sum_{j \in [n]} b_j} \\
\geq & \frac{\sum_{i \in [n-1]} (\beta_i - \beta_{i+1}) \cdot \sum_{j \in [n]}a_j + \beta_n \cdot \sum_{j \in [n]} a_j}{\sum_{i \in [n-1]} (\beta_i - \beta_{i+1}) \cdot \sum_{j \in [n]}b_j + \beta_n \cdot \sum_{j \in [n]} b_j} \\
= & \frac{\sum_{i \in [n]}a_i}{\sum_{i \in [n]}b_i}
\end{align*}
\Halmos\endproof

\subsection{Proof of Lemma~\ref{lem:common}} \label{sec:prop1}

Similar to Lemma~\ref{lem:SaltyWater}, we can show the following
\begin{lemma}
\label{lem:SaltyWaterOtherDirection}
Suppose $a_i, b_i >0, \forall i \in [n]$, and $\frac{a_0}{b_0} \geq \frac{a_1}{b_1} \geq ... \geq \frac{a_n}{b_n}$; suppose $0 \leq \beta_0 \leq \beta_1 \leq ... \leq \beta_n$. Then we have $$ \frac{\sum_{i = 0}^n a_i}{\sum_{i = 0}^n b_i} \geq \frac{\sum_{i = 0}^n \beta_i a_i}{\sum_{i = 0}^n \beta_i b_i}.$$
\end{lemma}

The proof is the same as the proof of Lemma~\ref{lem:SaltyWater}.

\proof{Proof of Lemma~\ref{lem:common}.}
Observe that Binomial distribution is a discrete distribution.
It only suffices to prove Lemma~\ref{lem:common} in the case when $c \in \mathrm{N}$ is an integer.

Let $C^l_T$ be $T$ choose $l$.

Take any $x,y \in [0,1]$ such that $x<y$.
Notice that Binomial distributions $\Bin(T,x)$ and $\Bin(T,y)$ only have finite supports over $\{0,1,...,T\}.$
It is trivial when $c \geq T$ because the truncation does not take effect and both fractions equal to 1.
Take any $z \in \{0,1,...,T-1\}$.
Since both enumerators are linear in $c \in (z,z+1)$, it only suffices to check for $c \in \{0,1,...,T-1\}$, i.e. when $c$ is an integer.

First notice that $\frac{x (1 - y)}{y (1 - x)} \leq 1$.
$\forall l \in \{0,1,...,T-1\}$, $$\left( \frac{x (1 - y)}{y (1 - x)} \right)^l \geq \left( \frac{x (1 - y)}{y (1 - x)} \right)^{l+1}.$$
Both multiply by $\left(\frac{1-x}{1-y}\right)^T$ we have $\forall l \in \{0,1,...,T-1\}$, $$\frac{x^l (1 - x)^{T-l}}{y^l (1 - y)^{T-l}} \geq \frac{x^{l+1} (1 - x)^{T-l-1}}{y^{l+1} (1 - y)^{T-l-1}}.$$
Further multiply by some constants to both enumerators and denominators the inequality still holds.
$$\frac{C_T^l x^l (1 - x)^{T-l} \cdot l}{C_T^l y^l (1 - y)^{T-l} \cdot l} \geq \frac{C_T^{l+1} x^{l+1} (1 - x)^{T-l-1} \cdot (l+1)}{C_T^{l+1} y^{l+1} (1 - y)^{T-l-1} \cdot (l+1)}.$$

$\forall l \in \{0,1,...,T\}$, let $a_l = C_T^l x^l (1 - x)^{T-l} \cdot l; b_l = C_T^l y^l (1 - y)^{T-l} \cdot l$.

$\forall i \in \{0,1,...,c\}$, let $\beta_i = 0$; $\forall i \in \{c+1,c+2,...,T\}$, $\beta_i = (i-c)/i$. It is easy to verify that $0 \leq \beta_0 \leq \beta_1 \leq ... \leq T$.

Invoking Lemma~\ref{lem:SaltyWaterOtherDirection}, we have $$\frac{\sum_{l=0}^T C_T^l x^l (1 - x)^{T-l} \cdot l}{\sum_{l=0}^T C_T^l y^l (1 - y)^{T-l} \cdot l} \geq \frac{\sum_{l=c}^T C_T^l x^l (1 - x)^{T-l} \cdot (l-c)}{\sum_{l=c}^T C_T^l y^l (1 - y)^{T-l} \cdot (l-c)}.$$

Re-arranging terms, $$\frac{\sum_{l=0}^{c-1} C_T^l x^l (1 - x)^{T-l} \cdot l + \sum_{l=c}^T C_T^l x^l (1 - x)^{T-l} \cdot c}{\sum_{l=0}^T C_T^l x^l (1 - x)^{T-l} \cdot l} \geq \frac{\sum_{l=0}^{c-1} C_T^l y^l (1 - y)^{T-l} \cdot l + \sum_{l=c}^T C_T^l y^l (1 - y)^{T-l} \cdot c}{\sum_{l=0}^T C_T^l y^l (1 - y)^{T-l} \cdot l}.$$

Equivalently, $$\frac{\bE\left[ \min\{c, \Bin(T,x)\} \right]}{Tx} \geq \frac{\bE\left[ \min\{c, \Bin(T,y)\} \right]}{Ty}$$
\Halmos\endproof



\subsection{Proof of Lemma~\ref{lem:StochasticDominance}}

\proof{Proof of Lemma~\ref{lem:StochasticDominance}.}
It suffices to prove the following
$$\bE[\min\{\frac{y}{x} \cdot \Bin(T, \frac{x}{T}),y\}] \geq \bE[\min\{\Bin(T, \frac{y}{T}),y\}]$$

For any $t\in[T]$, denote $X_t = \frac{y}{x} \cdot \Ber(T, \frac{x}{T})$ as a Bernoulli random variable such that $\bE[X_t] = y/T$, and $X_t \in [0,1]$ has bounded support between $[0,1]$;
For any $t\in[T]$, denote $Y_t = \Ber(T, \frac{y}{T})$ as a Bernoulli random variable such that with probability $y/T$, $Y_t=1$.
Pick $c = y$ to be a positive real number.

From Lemma~\ref{lem:gen:robust} we have
$$\bE[\min\{\sum_{t\in[T]} X_t,y\}] \geq \bE[\min\{\sum_{t\in[T]} Y_t,y\}],$$
which finishes the proof.
\Halmos \endproof

\section{Proof of Theorem~\ref{thm::assortmentStationary} and Proposition~\ref{prop::assortmentStationary}} \label{sec:proof:assortmentStationary}

We write the most general proof precisely by combining Theorem~\ref{thm::assortmentStationary} and Proposition~\ref{prop::assortmentStationary}.

\begin{theorem}\label{thm::appendixMerged}
Under one of the following three conditions:
\begin{enumerate}[label=(\roman*)]
\item the static substitution model with integral demand;
\item the static substitution model with fractional demand and Assumption~\ref{asp:sta2};
\item the dynamic substitution model with integral demand and Assumptions~\ref{ass::subs} (substitutability), and when one item has only one single price (pure assortment problem without pricing);
\end{enumerate}
for the assortment (and pricing) problem under stationary demand, if there are $T$ time periods and $\bmin = \min_{i\in[n]} b_i$, then Algorithm~\ref{def::pricingStationary} earns expected revenue of at least
\begin{align}
\frac{\bE[\min\{\Bin(T,\bmin / T),\bmin \}]}{\bmin}\cdot\OPT,
\end{align}
where $\Bin(T,\bmin / T)$ denotes a Binomial random variable consisting of $T$ trials of probability $\bmin / T$.

If we let $\Delta^{APX}$ denote the term $\frac{\bE[\min\{\Bin(T,\bmin / T),\bmin\}]}{\bmin}$ from expression~(\ref{eqn::binomialBound}), then
\begin{align}
\Delta^{APX} \geq 1-\frac{\underline{b}^{\underline{b}}}{\underline{b}!}e^{-\underline{b}},
\end{align}
which states that $\Delta^{APX}=1-O(1/\sqrt{\underline{b}})$, and increases from $1-1/e$ to 1 as $\underline{b}\to\infty$ (regardless of $T$).
\end{theorem}

It is easy to see that Theorem~\ref{thm::assortmentStationary} corresponds to (i) and (ii) of Theorem~\ref{thm::appendixMerged}, and Proposition~\ref{prop::assortmentStationary} corresponds to (iii) of Theorem~\ref{thm::appendixMerged}
Now we prove Theorem~\ref{thm::appendixMerged}.

\proof{Proof of Theorem~\ref{thm::appendixMerged}.}
This proof consists of two steps.
In the first step, we lower bound the performance of Algorithm~\ref{def::pricingStationary}, which is a randomized policy, by the performance of a virtual calendar.
We define the choice model of this virtual calendar to have \textit{only} static substitution, yet it is a lower bound to the performance of our Algorithm~\ref{def::pricingStationary}, under both static and dynamic substitution.
In the second step, we lower bound the performance of this virtual calendar by $\Delta^{APX} \cdot \OPT$.
Under all three conditions as stated in Theorem~\ref{thm::assortmentStationary}, the virtual calendar is the same.
We state this virtual calendar in its most general form as in Step 0, and illustrate how the three conditions simplifies to this most general form.
Under three conditions, the first step that lower bounds the performance of Algorithm~\ref{def::pricingStationary} to the performance of this virtual calendar may be different, as we shall see in Step 1.
Under all three conditions, the second step is the same, as we prove in Step 2.

Now we introduce the following random variables, which depict a run of our assortment policy.
Let $S_t$ be the assortment that we select to offer in period $t$.
Let $B_t(i)$ be the remaining inventory of item $i$ at the \textit{end} of time $t$.
We have $B_0(i) = b_i$.
Under all three conditions, let $R_t(i,j)$ be the amount of sales that a customer chooses product $(i,j)$ during period $t$.
We will always specify the distribution of $R_t(i,j)$, by using a conditional probability.
For example, we will use $\bE[R_t(i,j) \left| S_t = S, \bm{B}_t = \bm{B}\right.]$, for the expected sales that a customer chooses product $(i,j)$ during period $t$, when we plan to offer assortment $S$, and when the remaining inventory level for each resource is $\bm{B} = (B_1, B_2, ..., B_n)$.

Under condition (i), under static substitution and when demand is integral, $R_t(i,j) \in \{0,1\}$ is binary.
Conditional on any $S \in \cS, \bm{B} \in \bN_0^n$, $R_t(i,j) = \bI_{\{B_i > 0\}} Q$, where $Q$ is a binary random variable, which takes $1$ with probability $q(i,j,S)$.
Under condition (ii) under static substitution and when demand is fractional, $R_t(i,j) \in [0,1]$ is continuous.
Conditional on any $S \in \cS, \bm{B} \in \bR_+^n$, $R_t(i,j) = \min \{B_i, Q\}$, where $Q$ is a random variable whose CDF is $F_{(i,j,S)} (\cdot)$.
Under condition (iii) under dynamic substitution and when demand is integral, $R_t(i,j) \in \{0,1\}$ is binary.
Conditional on any $S \in \cS, \bm{B} \in \bN_0^n$, $R_t(i,j)$ takes $1$ with probability $q(i,j,\bS)$.
Here we define $\bS = \{(i,j) \in S \left| B_i > 0 \right.\}$ to be a function of $S$.

Under condition (iii), Assumption~\ref{ass::subs} suggests that $q(i,j,\bS) \geq q(i,j,S), \forall (i,j) \in \bS$, because $\bS \subseteq S$.
The demand that originally would have chosen the stocked out items would go to their substitutes (as well as leaving, in which case the inequality takes equality).
On the other hand, $q(i, j, \bS) = 0, \forall (i,j) \notin \bS$.
The demand for any stocked out item is zero.

We can use indicator variables to write the above inequalities in a compact form
\begin{align}
\label{eqn:DynamicSubSubAsp}
q(i,j, \bS) \geq \bI_{\{B_{t-1}(i) > 0\}} q(i,j,S)
\end{align}

For any period $t$, given the remaining inventory from the last period to be $\bm{B}_{t-1}$, conditional on any $S \in \cS$, the remaining inventory updates in the following fashion,
$$B_t(i) = B_{t-1}(i) - R_t(i,j), \forall i.$$
Note that no item can be offered multiple times at different prices in one assortment.
Also note that we have defined $R_t(i,j)$ as the amount of sales, so $R_t(i,j)$ can never go beyond $B_{t-1}(i)$.

\noindent \textbf{Step 0 Statement of the virtual calendar.}

Consider Algorithm~\ref{def::pricingStationary} that offers each assortment randomly.
Define the set of items $I_{\bm{x}^*(S)} = \left\{ i \in [n] \left| \exists j \in [m], \exists S \in \mathcal{S}, s.t. (i,j)\in S, x^*(S) > 0 \right. \right\}$.
These are the items that are relevant to the probablistic offering of assortments from Algorithm~\ref{def::pricingStationary}.
In other words, $I_{\bm{x}^*(S)}$ is the set of items such that there is a positive probability that Algorithm~\ref{def::pricingStationary} suggests an assortment that contains item $i$.

Associated with each item in $I_{\bm{x}^*(S)}$, there is a unique price $p_{\pC, i}$.
For any $i \in I_{\bm{x}^*(S)}$, define $$p_{\pC,i} = \dfrac{\sum_{S \in \mathcal{S}} \ssx(S) \sum_{j: (i,j) \in S} p_j q(i,j,S)}{\sum_{S \in \mathcal{S}} \ssx(S) \sum_{j: (i,j) \in S} q(i,j,S)}.$$ 
We have $\sum_{S \in \mathcal{S}} \ssx(S) = 1$, and $\ssx(S) \geq 0, \forall S \in \mathcal{S}$ due to constraints \eqref{eqn::constr::assortmentCustomer} and \eqref{eqn::constr::assortmentNonneg}.
Notice that, here we only use $q(i,j,S)$.

Under conditions (i) and (ii), $p_{\pC, i}$ cannot be simplified.
Under condition (iii), since each product has only one single price, we can define $j^i, \forall i \in [n]$ to be the price index that item $i$ can be offered.
The price offered is simply $p_{\pC, i} = p_{j^i}$.

Now in each period suppose we had an option to offer a deterministic assortment $S_\pC$ that consists of the products $S_\pC = \left\{ (i,p_{\pC,i}) \left| \forall i \in I_{\bm{x}^*(S)} \right. \right\}$.

Associated with each product $(i,p_{\pC,i}), \forall i \in I_{\bm{x}^*(S)}$, we prescribe a choice model.
Under conditions (i) and (ii), we use the following notation.
Let $Q(i,p_{\pC,i},S_\pC), \forall i \in I_{\bm{x}^*(S)}$ be a \textit{random variable} for the quantity that customers attempt to purchase product $(i,p_{\pC,i})$, should assortment $S_\pC$ be \textit{offered}.
Here we directly define the choice model to be under static substitution, for this virtual calendar.
Nonetheless, as we will show in Step 1, the performance of this virtual calendar is a lower bound to the performance of our Algorithm~\ref{def::pricingStationary}, under both static and dynamic substitutions.
We define the CDF function of $Q(i,p_{\pC,i},S_\pC)$ to be $F_{(i,p_{\pC,i},S_\pC)}(\cdot) = \sum_{S \in \mathcal{S}} \ssx(S) \sum_{j:(i,j)\in S} F_{(i,j,S)}(\cdot)$.
Notice that $j:(i,j) \in S$ selects only one price, because no item can be offered multiple times at different prices in one assortment.

Under condition (iii), since we only consider dynamic substitution with integral demand, we can simplify the notations.
Let $q(i,j^i,S_\pC), \forall i \in I_{\bm{x}^*(S)}$ be the probability that product $(i,j^i)$ is demanded, should assortment $S_\pC$ be offered.
And assortment $S_\pC$ is under static substitution -- $q(i,j^i,S_\pC)$ is unchanged even if some items from the assortment stocks out.
Denote $q(i,j^i,S_\pC) = \sum_{S \in \mathcal{S}} x^*(S) q(i,j^i,S)$.

We wish to show that the expected revenue earned from this deterministic assortment is a lower bound to the probablistic offering of assortments from Algorithm~\ref{def::pricingStationary}.
Denote $J_\pC(\bm{b},t)$ as the \textit{expected} revenue earned from always offering the deterministic assortment $S_\pC$, if at the beginning of period $t$ we are endowed with $\bm{b}$ units of inventory---this corresponds to the expression in line~\eqref{eqn::gen:robustArgument}.
The expectation has only one source of randomness, which comes from the random demand.
Denote $\Rev(\bm{b},t)$ as the \textit{expected} revenue earned by the policy from Algorithm~\ref{def::pricingStationary}, if at the beginning of period $t$ we are endowed with $\bm{b}$ units of inventory---this corresponds to the expression in line~\eqref{eqn::gen:couplingArgument}.
The expectation has two sources of randomness, which come from both the random demand, and the randomization from Algorithm~\ref{def::pricingStationary}.

\noindent \textbf{Step 1 Lower bounding the performance of Algorithm~\ref{def::pricingStationary}.}

In this step, we distinguish the following two cases: static substitution and dynamic substitution.
Under static substitution, we talk about conditions (i) and (ii); under dynamic substitution, we talk about condition (iii).

\textbf{Case 1} Static substitution

This proof unifies conditions (i) and (ii).
Note that condition (i) is integral, Bernoulli demand, which naturally satisfies Assumption~\ref{asp:sta2}.

Assumption~\ref{asp:sta2} suggests that $\forall i \in [n], \forall S, S' \in \mathcal{S}$, for all $j,j'\in[m]$ such that $p_j > p_{j'}$, we have $\forall c\in[0,1]$,
\begin{align*}
\frac{\ssx(S) \bE_{Q\sim F_{(i,j,S)}}[\min\{c,Q\}]}{\ssx(S) \bE_{Q\sim F_{(i,j,S)}}[Q]}\geq\frac{\ssx(S')\bE_{Q\sim F_{(i,j',S')}}[\min\{c,Q\}]}{\ssx(S')\bE_{Q\sim F_{(i,j',S')}}[Q]}.
\end{align*}

From Lemma~\ref{lem:SaltyWater}, if we treat $p_j \geq p_{j'}$ as $\beta$'s, then we have $\forall i \in [n], \forall c \in [0,1]$,
\begin{align*}
\frac{\sum_{S \in \mathcal{S}} \ssx(S) \sum_{j: (i,j) \in S} p_j \bE_{Q\sim F_{(i,j,S)}}[\min\{c,Q\}]}{\sum_{S \in \mathcal{S}} \ssx(S) \sum_{j: (i,j) \in S} p_j \bE_{Q\sim F_{(i,j,S)}}[Q]} \geq \frac{\sum_{S \in \mathcal{S}} \ssx(S) \sum_{j: (i,j) \in S} \bE_{Q\sim F_{(i,j,S)}}[\min\{c,Q\}]}{\sum_{S \in \mathcal{S}} \ssx(S) \sum_{j: (i,j) \in S} \bE_{Q\sim F_{(i,j,S)}}[Q]}
\end{align*}
which simplifies to
\begin{align}
\label{eq:avgprice}
\frac{\sum_{S \in \mathcal{S}} \ssx(S) \sum_{j: (i,j) \in S} p_j \bE_{Q\sim F_{(i,j,S)}}[\min\{c,Q\}]}{\sum_{S \in \mathcal{S}} \ssx(S) \sum_{j: (i,j) \in S} \bE_{Q\sim F_{(i,j,S)}}[\min\{c,Q\}]} \geq p_{\pC, i} = \frac{\sum_{S \in \mathcal{S}} \ssx(S) \sum_{j: (i,j) \in S} p_j \bE_{Q\sim F_{(i,j,S)}}[Q]}{\sum_{S \in \mathcal{S}} \ssx(S) \sum_{j: (i,j) \in S} \bE_{Q\sim F_{(i,j,S)}}[Q]}
\end{align}

Now we prove by backward induction on $t$. In the last period $T, \forall \bm{c} \geq \bm{0},$
\begin{align}
\Rev(\bm{c},T) & = \sum_{i \in [n]} \sum_{S \in \mathcal{S}} \ssx(S) \sum_{j: (i,j) \in S} p_j \bE_{Q\sim F_{(i,j,S)}}[\min\{c_i,Q\}] \nonumber \\
& \geq \sum_{i \in [n]} p_{\pC, i} \sum_{S \in \mathcal{S}} \ssx(S) \sum_{j: (i,j) \in S} \bE_{Q\sim F_{(i,j,S)}}[\min\{c_i,Q\}] = J_\pC(\bm{c},T) \label{eqn:DynamicSubBreaksHere}
\end{align}
The first equality is because each item can only be offered at one price in each assortment.
So when truncation happens, there is no ambiguity which price of demand is lost.
And the inequality is due to \eqref{eq:avgprice}.

To continue the induction, if we can show $\Rev(\bm{c},t+1) \geq J_\pC(\bm{c},t+1), \forall \bm{c} \geq \bm{0}$, then we can show:
\begin{align*}
& \Rev(\bm{c},t) \\
= & \sum_{i \in [n]} \sum_{S \in \mathcal{S}} \ssx(S) \sum_{j: (i,j) \in S} p_j \bE_{Q\sim F_{(i,j,S)}}[\min\{c_i,Q\}] + \sum_{S \in \mathcal{S}} \ssx(S) \bE_{\bm{Q}}[\Rev(\max\{\bm{0}, \bm{c} - \bm{Q}\}, t+1)] \\
\geq & \sum_{i \in [n]} p_{\pC, i} \sum_{S \in \mathcal{S}} \ssx(S) \sum_{j: (i,j) \in S} \bE_{Q\sim F_{(i,j,S)}}[\min\{c_i,Q\}] + \sum_{S \in \mathcal{S}} \ssx(S) \bE_{\bm{Q}}[J_\pC(\max\{\bm{0}, \bm{c} - \bm{Q}\}, t+1)] \\
= & J_\pC(\bm{c},t)
\end{align*}
where we use $\bm{Q} = (Q_1, ..., Q_n)$ as a vector form to stand for the joint distribution. Specifically, $Q_i \sim F_{(i,j^i,S)}, \forall i \in [n]$ specifies the distribution of demand associated with $i \in [n]$, and $j^i: (i,j) \in S$ specifies the price that is uniquely determined.

Now we explain this block of inequalities.
The inequality is due to \eqref{eq:avgprice} and due to induction hypothesis.
By induction on $t$ we show $\Rev(\bm{b},t) \geq J_\pC(\bm{b},t)$. That is, the expected revenue earned from deterministic assortment $S_\pC$ is a lower bound to the probabilistic offering of assortments from Algorithm~\ref{def::pricingStationary}.

\textbf{Case 2} Dynamic substitution

This proof is for condition (iii), for integral demand.
We prove by backward induction on $t$.
In the last period $T, \forall \bm{c} \geq \bm{0},$
\begin{align*}
\Rev(\bm{c},T) & = \sum_{i \in [n]} \sum_{S \in \mathcal{S}} x^*(S) p_{j^i} q(i,j^i,\bS) \\
& \geq \sum_{i \in [n]} p_{j^i} \bI_{\{c_i > 0\}} \sum_{S \in \mathcal{S}} x^*(S) q(i,j^i,S) \\
& = \sum_{i \in [n]} p_{j^i} \bI_{\{c_i > 0\}} q(i,j^i,S_\pC) = J_\pC(\bm{c},T)
\end{align*}
The inequality is due to \eqref{eqn:DynamicSubSubAsp}.

To continue the induction, if we can show $\Rev(c,t+1) \geq J_\pC(c,t+1), \forall c \geq 0$, then we can show:
\begin{align*}
\Rev(\bm{c},t) = & \sum_{i \in [n]} \sum_{S \in \mathcal{S}} x^*(S) p_{j^i} q(i,j^i,\bS) + \sum_{S \in \mathcal{S}} x^*(S) \left\{ \sum_{i \in [n]} q(i,j^i,\bS) \cdot \Rev(\max\{\bm{0}, \bm{c} - \bm{e}_i\}, t+1) \right. \\
& \qquad \qquad \qquad \qquad \qquad \qquad \qquad \qquad \quad \left. + (1-\sum_{i \in [n]}q(i,j^i,\bS)) \cdot \Rev(\bm{c}, t+1) \right\} \\
\geq & \sum_{i \in [n]} \sum_{S \in \mathcal{S}} x^*(S) p_{j^i} q(i,j^i,\bS) + \sum_{S \in \mathcal{S}} x^*(S) \left\{ \sum_{i \in [n]} q(i,j^i,\bS) \cdot J_\pC(\max\{\bm{0}, \bm{c} - \bm{e}_i\}, t+1) \right. \\
& \qquad \qquad \qquad \qquad \qquad \qquad \qquad \qquad \quad \left. + (1-\sum_{i \in [n]}q(i,j^i,\bS)) \cdot J_\pC(\bm{c}, t+1) \right\} \\
= & \sum_{S \in \mathcal{S}} x^*(S) \left\{ \sum_{i \in [n]} q(i,j^i,\bS) \left\{ p_{j^i} + J_\pC(\max\{\bm{0}, \bm{c} - \bm{e}_i\}, t+1) - J_\pC(\bm{c}, t+1) \right\}\right\} + J_\pC(\bm{c}, t+1) \\
\geq & \sum_{S \in \mathcal{S}} x^*(S) \left\{ \sum_{i \in [n]} q(i,j^i,S) \left\{ p_{j^i} + J_\pC(\max\{\bm{0}, \bm{c} - \bm{e}_i\}, t+1) - J_\pC(\bm{c}, t+1) \right\}\right\} + J_\pC(\bm{c}, t+1) \\
= & J_\pC(\bm{c},t)
\end{align*}
where the first inequality is due to induction hypothesis;
the second equality is taking out $J_\pC(\bm{c}, t+1)$ and re-arranging terms;
the second inequality is because the marginal revenue of one extra unit of resource $i$ is bounded by $p_{j^i}$, and because $q(i,j^i,\bS) \geq q(i,j^i,S)$.
By induction on $t$ we show $\Rev(\bm{b},t) \geq J_\pC(\bm{b},t)$. That is, the expected revenue earned from deterministic assortment $S_\pC$ is a lower bound to the probablistic offering of assortments from Algorithm~\ref{def::pricingStationary}.

\noindent \textbf{Step 2 Lower bounding the performance of the virtual calendar.}

Now we further lower bound the expected revenue earned from deterministic assortment $S_\pC$.
Since we have defined the choice model of the virtual calendar to be under static substitution, in all the remaining proof, we will only use notations like $Q(i,j,S)$, for the random quantity that customers attempt to purchase product $(i,j)$, should assortment $S$ be offered, no matter if any of the items from the assortment is stocked out.
We do this because now we are under static substitution.
The total quantity of demands attempting to consume inventory $i$ is
\begin{align}
Q_i:=\sum_{S\in\cS}\ssx(S)\sum_{j:(i,j)\in S}Q(i,j,S) \label{eqn::rho}
\end{align}

Note that $Q_i$ does not depend on $t$.
By the independence of both policy decisions and customer decisions across time, the total sales of inventory $i$ is a sum of $T$ trials of independent random variables $Q_i$, truncated by the starting inventory $b_i$.
To summarize, the expected consumption of item $i$, regardless of inventory availability, is $\bE[\min\{\sum_{t\in[T]} Q_i(\xi_t),b_i\}]$, with $Q_i$ defined as in \eqref{eqn::rho}, for all $i\in[n]$.
We have used $\xi_t$ to emphasize the randomness in each trial.

Now denote $\rho_i:=\sum_{S\in\cS} \ssx(S)\sum_{j:(i,j)\in S}q(i,j,S)$.
Due to Lemma~\ref{lem:gen:robust}, we have $\bE[\min\{\sum_{t\in[T]} Q_i(\xi_t),b_i\}] \geq \bE[\min\{\Bin(T,\rho_i),b_i\}]$.
Summing over all $i\in[n]$, the expected revenue of the policy is
\begin{align*}
\sum_{i=1}^n \bE[\min\{\Bin(T,\rho_i),b_i\}] \cdot p_{\pC,i} = & \sum_{i=1}^n\frac{\bE[\min\{\Bin(T,\rho_i),b_i\}]}{T \rho_i}\sum_{S\in\cS}\ssx(S)\sum_{j:(i,j)\in S}p_jq(i,j,S) \\
\geq & \sum_{i=1}^n\frac{\bE[\min\{\Bin(T, b_i / T),b_i\}]}{T \cdot (b_i/T)}\sum_{S\in\cS}\ssx(S)\sum_{j:(i,j)\in S}p_jq(i,j,S) \\
\geq & \frac{\bE[\min\{\Bin(T, \bmin / T),\bmin\}]}{\bmin}\sum_{i=1}^n\sum_{S\in\cS}\ssx(S)\sum_{j:(i,j)\in S}p_jq(i,j,S).
\end{align*}
The first inequality follows from Lemma~\ref{lem:common}, because $\rho_i\le b_i/T$ for all $i$, since the LP solution $\ssx(S)$ satisfies constraints~(\ref{eqn::constr::assortmentInventory}).
The second inequality is due to Lemma~\ref{lem:StochasticDominance}.
\Halmos\endproof

\begin{proposition}
\label{prop:bound}
For all $b \leq T$, $\Delta^{APX} = \bE[\min\{\Bin(T,b/T),b\}] / b$ is non-increasing in $T$, and
\begin{align*}
\lim_{T \to +\infty} \Delta^{APX} = \left(1 - \frac{\lfloor b \rfloor^{\lfloor b \rfloor}}{{\lfloor b \rfloor}!}e^{-\lfloor b \rfloor}\right) \geq 1 - 1/e
\end{align*}
\end{proposition}

\proof{Proof of Proposition~\ref{prop:bound}.}
Since $T \in [N]$, and $T \geq b \geq \lfloor b \rfloor > 0$,
from Lemma~\ref{lem:StochasticDominance} we have
$$\Delta^{APX} = \frac{\bE[\min\{\Bin(T, b / T),b\}]}{b} \geq \frac{\bE[\min\{\Bin(T, \lfloor b \rfloor / T),\lfloor b \rfloor\}]}{\lfloor b \rfloor}.$$ 

Since we have normalized $b$ such that $b \geq 1$, we know that $\lfloor b \rfloor \geq 1$.
Now we wish to prove Proposition~\ref{prop:bound} when $b \in [N]$ is any positive integer, i.e. we wish to show that for any positive integer $b \leq T$,
\begin{align*}
\lim_{T \to +\infty} \Delta^{APX} = \left(1 - \frac{b^b}{b!}e^{-b}\right) \geq 1 - 1/e
\end{align*}

Denote $a = b / T$.
We prove the first equality by telescoping.
Denote $C_n^m = \frac{n!}{m!(n-m)!}$ as $n$ choose $m$.
\begin{align*}
b \Delta^{APX} = & E \left[ \min \left\{Bin(T,b/T), b\right\} \right] \\
= & Ta - \sum_{i = b+1}^T C^i_T (a^i (1-a)^{T-i})(i - b)\\
= & Ta - \sum_{i = b+1}^T \frac{T!}{(i-1)!(T-i)!} a^i (1-a)^{T-i} + \sum_{i = b+1}^T \frac{b T!}{i! (T-i)!} a^i (1-a)^{T-i}\\
= & Ta - \frac{T!}{b!(T-b-1)!} a^{b+1} (1-a)^{T-b-1} + \frac{b T!}{T! 0!} a^T (1-a)^0\\
& \quad -\sum_{i = b+1}^{T-1} \frac{T!}{i!(T-i-1)!} a^{i+1} (1-a)^{T-i-1} + \sum_{i = b+1}^{T-1} \frac{b T!}{i! (T-i)!} a^i (1-a)^{T-i}\\
= & Ta - \frac{T!}{b!(T-b-1)!} a^{b+1} (1-a)^{T-b-1} + b a^T \\
& \quad + \sum_{i = b+1}^{T-1} \frac{T!}{i!(T-i)!} a^{i} (1-a)^{T-i-1} \underbrace{\left( b(1-a) - a(T-i)\right)}_{= a (i-b) + (b-aT) \geq 0}\\
\geq & b \left( 1 + a^T - C^b_{T-1} a^b (1-a)^{T-b-1} \right)\\
= & b \left( 1 + a^T - C^b_T a^b (1-a)^{T-b} \right)
\end{align*}
where the fourth equality follows from telescoping. Then we can take $T \to +\infty$ and use Stirling's formula:
\begin{align*}
\lim_{T \to +\infty} \Delta^{APX} = & \lim_{T \to +\infty} 1 + a^T - C^b_T a^b (1-a)^{T-b}\\
= & 1 + \lim_{T \to +\infty} \left( \frac{b}{T} \right)^T - \lim_{T \to +\infty} \frac{T! b^b (T-b)^{T-b}}{b! (T-b)! T^b T^{T-b}}\\
= & 1 + 0 - \lim_{T \to +\infty} \frac{\sqrt{2 \pi T} \frac{T^T}{e^T} (T-b)^{T-b}}{\sqrt{2 \pi (T-b)} \frac{(T-b)^{T-b}}{e^{T-b}} T^T} \frac{b^b}{b!}\\
= & 1 - \frac{b^b}{b!}e^{-b}
\end{align*}
This term is increasing in $b$. So it obtains minimum when $b=1$: $$\lim_{T \to +\infty} \Delta^{APX} = 1 - \frac{b^b}{b!}e^{-b} \geq 1 - \frac{1}{e}$$
\Halmos \endproof

\section{Tightness of Theorem~\ref{thm::assortmentStationary}: Proof of Proposition~\ref{prop::binomialBoundTight}}
\label{sec:binomialBoundTightProof}

\proof{Proof of Proposition~\ref{prop::binomialBoundTight}.}
Construct the following instance. There is only one price option, i.e. $m=1$. So there only exists one calendar to sell at this single price everyday. For any given $b$ and $T$, the only price option has a purchase probability of $b/T$, and earns $1$ unit revenue.

The LP upper bound suggests a total of $b$ units of revenue. And the only calendar earns $\bE \left[ \min \left\{Bin(T,b/T), b\right\} \right]$ units of revenue. So the expected revenue of the only policy is exactly expression~(\ref{eqn::binomialBound}), which finishes the tightness proof.
\Halmos\endproof

\section{Necessity of Assumptions for Theorem~\ref{thm::assortmentStationary}: Proof of Proposition~\ref{prop:StationaryBreaks}}
\label{sec:NecessityOfAssumptions}
\proof{Proof of Proposition~\ref{prop:StationaryBreaks}.}
Consider the following problem with $T=5$ periods.
There are $3$ products, $A$, $B$, and $C$.
Products $A$ and $B$ use the first resource, and product $C$ uses the second resource.
Both resources have initial inventory $1$.
Product $A$ is sold at $p_\pH = 1$, and products $B$ and $C$ are sold at $p_\pL = \epsilon$.

The choice model is a distribution of ordinal preferences:
it takes $C\succ B\succ \emptyset$ with probability $1/5$, $A\succ \emptyset$ with probability $1/5-\epsilon$, $B\succ \emptyset$ with probability $\epsilon$, and $\emptyset$ with probability $3/5$.
Since this choice model is prescribed by a distribution of ordinal preferences, it satisfies Assumption~\ref{ass::subs}, the substitutability assumption.
Since consumptions from this choice model are binary, it satisfies Assumption~\ref{asp:sta2}.

The optimal solution from the LP is to offer assortment $\{A, B, C\}$ in all $5$ periods.
And the LP objective value is $1 + O(\epsilon)$.
With some calculation, the actual expected revenue is $1959/3125 + O(\epsilon)$.
Taking $\epsilon \to 0^+$ we have $1959/3125 \approx 0.6269 < 0.6321 \approx 1-1/e$.
\Halmos \endproof

\section{Proof of Theorem~\ref{thm::assortmentNonstationary2}}
\label{sec:proof:assortmentNonstationary2}

\proof{Proof of Theorem~\ref{thm::assortmentNonstationary2}.}
Denote the following random variables, which depict a run of our assortment policy from Algorithm~\ref{def::assortmentNonstationary2}.
Let $A_t(S)$ be the indicator random variable for $\tilde{S}_t = S$, where $\tilde{S}_t$ was the assortment selected before discarding in \eqref{eqn::policyAssortmentNonstationary2} was applied.
Let $B_t(i)$ be the remaining inventory of item $i$ at the \textit{end} of time $t$.  Defined for all $i\in[n]$ and $t=0,\ldots,T$, where $B_0(i)=b_i$ for all $i$.

Under either static or dynamic substitution, let $R_t(i,j)$ be the amount of sales that a customer chooses product $(i,j)$ during period $t$.
We will always specify the distribution of $R_t(i,j)$, by using a conditional probability.
For example, we will use $\bE[R_t(i,j) \left| S_t = S, \bm{B}_t = \bm{B}\right.]$, for the expected sales that a customer chooses product $(i,j)$ during period $t$, when we plan to offer assortment $S$, and when the remaining inventory level for each resource is $\bm{B} = (B_1, B_2, ..., B_n)$.

Under static substitution, conditional on any $S \in \cS, \bm{B} \in \bR_+^n$, $R_t(i,j) = \min \{B_i, Q\}$, where $Q$ is a random variable whose CDF is $F_{t,(i,j,S)} (\cdot)$.
Under dynamic substitution, conditional on any $S \in \cS, \bm{B} \in \bN_0^n$, $R_t(i,j)$ takes $1$ with probability $q_t(i,j,\bS)$.
Here we define $\bS = \{(i,j) \in S \left| B_i > 0 \right.\}$ to be a function of $S$.

Under dynamic substitution, Assumption~\ref{ass::subs} suggests that $q_t(i,j,\bS) \geq q_t(i,j,S), \forall (i,j) \in \bS$, because $\bS \subseteq S$.
The demand that originally would have chosen the stocked out items would go to their substitutes (as well as leaving, in which case the inequality takes equality).
On the other hand, $q_t(i, j, \bS) = 0, \forall (i,j) \notin \bS$.
The demand for any stocked out item is zero.
We can use indicator variables to write the above inequalities in a compact form $q_t(i,j, \bS) \geq \min \{B_{t-1}(i), q_t(i,j,S)\}$.

In all, we have
\begin{align}
\label{eqn:DynamicSubSubAsp2}
\bE[R_t(i,j) \left| S_t = S, \bm{B}_t = \bm{B}\right.] \geq \bE_{Q\sim F_{t,(i,j,S)}}[\min \{B_{t-1}(i), Q\}]
\end{align}
where $F_{t,(i,j,S)}$ may prescribe a Bernoulli distribution, e.g. under dynamic substitution.

For any period $t$, given the remaining inventory from the last period to be $\bm{B}_{t-1}$, conditional on any $S \in \cS$, the remaining inventory updates in the following fashion,
$$B_t(i) = B_{t-1}(i) - R_t(i,j), \forall i.$$
Note that no item can be offered multiple times at different prices in one assortment.
Also note that we have defined $R_t(i,j)$ as the amount of sales, so $R_t(i,j)$ can never go beyond $B_{t-1}(i)$.

Following each sample path, we let $\Rev$ denote the revenue earned by the policy suggested in Algorithm~\ref{def::assortmentNonstationary2}.
\begin{align*}
\bE[\Rev]=&\sum_{t=1}^T \sum_{S \in \mathcal{S}} A_t(S) \sum_{ (i,j)\in D(S) }p_j \bE[R_t(i,j) \left| S_t = D(S), \bm{B}_t = \bm{B}_{t-1}\right.] \\
\geq &\sum_{t=1}^T \sum_{S \in \mathcal{S}} A_t(S) \sum_{ (i,j)\in D(S) }p_j \bE_{Q \sim F_{t, (i,j, D(S))}}[\min\{B_{t-1}(i), Q\}] \\
=&\sum_{t=1}^T \sum_{S \in \mathcal{S}} A_t(S) \sum_{(i,j)\in D(S)}(p_j-\frac{\ssr_i}{2b_i})\bE_{Q \sim F_{t, (i,j, D(S))}}[\min\{B_{t-1}(i), Q\}] \\
&+\sum_{i=1}^n\frac{\ssr_i}{2b_i}\sum_{t=1}^T\sum_{S \in \mathcal{S}}A_t(S)\sum_{j:(i,j)\in D(S)}\bE_{Q \sim F_{t, (i,j, D(S))}}[\min\{B_{t-1}(i), Q\}] \\
=&\sum_{t=1}^T \sum_{S \in \mathcal{S}} A_t(S) \sum_{(i,j)\in D(S)}(p_j-\frac{\ssr_i}{2b_i})\bE_{Q \sim F_{t, (i,j, D(S))}}[\min\{B_{t-1}(i), Q\}]+\sum_{i=1}^n\frac{\ssr_i}{2b_i}(b_i-\bE_{\bm{Q}}[B_T(i)])
\end{align*}
where the first inequality is due to \eqref{eqn:DynamicSubSubAsp2}; the second equality is due to our discarding rule from \eqref{eqn::policyAssortmentNonstationary2};
the last equality is counting how much inventory has been sold throughout the horizon, where we implicitly use $\bm{Q}$ to stand for the randomness from all periods.

We further observe that, $\forall t \in [T], S \in \mathcal{S}, (i,j) \in S$,
\begin{equation}
\label{eqn:expectations}
\begin{split}
\bE_{Q \sim F_{t, (i,j, S)}; B_{t-1}(i)}[\min\{B_{t-1}(i), Q_t(i,j,S)\}] = & \bE_{Q \sim F_{t, (i,j, S)}; B_{t-1}(i)}\left[\frac{B_{t-1}(i) \cdot Q}{\max\{B_{t-1}(i), Q\}}\right] \\
\geq & \bE_{Q \sim F_{t, (i,j, S)}, B_{t-1}(i)}\left[\frac{B_{t-1}(i) \cdot Q}{b_i}\right] \\
= & \frac{\bE_{B_{t-1}(i)}[B_{t-1}(i)]}{b_i} \cdot \bE_{Q \sim F_{t, (i,j, S)}}[Q] \\
\geq & \frac{\bE[B_{T}(i)]}{b_i} \cdot q_t(i,j,S)
\end{split}
\end{equation}
where the first inequality is due to $B_{t-1}(i) \leq b_i$ inventory is no more than initial inventory level, and $Q \leq b_i, \forall t \in [T]$ demand is smaller than initial inventory; the second equality is because $B_{t-1}(i)$ is independent to any consumption that occurs at time $t$; the last inequality is because $B_t(i)$ is non-increasing in $t$, and evaluating the expectation of $\bE_{Q \sim F_{t, (i,j, S)}}[Q] = q_t(i,j,S)$.

Finally, the expectation of the our policy's revenue can be decomposed as
\begin{align*}
\bE[\Rev] & = \sum_{i=1}^n\sum_{t=1}^T\sum_{S \in \mathcal{S}} \bE[A_t(S)] \sum_{j:(i,j)\in D(S)}(p_j-\frac{\ssr_i}{2b_i})\bE_{Q \sim F_{t, (i,j, D(S))}}[\min\{B_{t-1}(i), Q\}] +\sum_{i=1}^n\frac{\ssr_i}{2b_i}(b_i-\bE[B_T(i)])\\
& \geq \sum_{i=1}^n \frac{\bE[B_T(i)]}{b_i} \sum_{t=1}^T\sum_{S \in \mathcal{S}} x^*_t(S) \sum_{j:(i,j)\in D(S)}(p_j-\frac{\ssr_i}{2b_i})q_t(i,j,D(S)) +\sum_{i=1}^n\frac{\ssr_i}{2b_i}(b_i-\bE[B_T(i)])\\
& \geq \sum_{i=1}^n \frac{\bE[B_T(i)]}{b_i} \sum_{t=1}^T\sum_{S \in \mathcal{S}} x^*_t(S) \sum_{j:(i,j)\in S}(p_j-\frac{\ssr_i}{2b_i})q_t(i,j,S) +\sum_{i=1}^n\frac{\ssr_i}{2b_i}(b_i-\bE[B_T(i)])\\
& \geq \sum_{i=1}^n \frac{\bE[B_T(i)]}{b_i} \left( r_i^* - \frac{r_i^*}{2b_i}\cdot b_i \right) +\sum_{i=1}^n\frac{\ssr_i}{2b_i}(b_i-\bE[B_T(i)])\\
& = \sum_{i=1}^n\frac{\ssr_i}{2}
\end{align*}
where the first equality is because the selection of set $\tilde{S}_t$ is independent to the remaining inventory $B_{t-1}(i)$, and also independent to the consumption $Q \sim F_{t,(i,j, D(S))}$ in period $t$;
the first inequality is because of \eqref{eqn:expectations} and evaluating the expectation of $\bE[A_t(S)] = x^*_t(S)$;
the second inequality is because (i) we include non-positive terms into the summation, and (ii) $q_t(i,j,D(S)) \geq q_t(i,j,S)$ since $D(S) \subseteq S$, due to Assumption~\ref{ass::subs}.
the third inequality is due to the definition of $\ssr_i$ in the policy, and the fact that the inventory constraint~(\ref{eqn::constr::assortmentInventory}) in the LP is satisfied.
Thus we complete the proof of the theorem.
\Halmos\endproof

\section{Tighness of Theorem~\ref{thm::assortmentNonstationary2}: Proof of Proposition~\ref{prop:nonstationaryTight}}
\label{sec:nonstationaryTightProof}
\proof{Proof of Proposition~\ref{prop:nonstationaryTight}.}
We use the problem instance suggested by Example~\ref{exa:LPsub}.
Observe that any revenue-maximizing policy offers $p_1$ in Day 2. So the only decision to make is in Day 1. Since offering $p_1$ in Day 1 sells nothing, the outside option of selling nothing on Day 1 is simply the same as offering $p_1$ in Day 1.

Now we parameterize any revenue-maximizing policy by $x$, its probability in offering $p_1$ in Day 1; and $1-x$ is the probability in offering $p_2$ in Day 1. In expectation, this policy earns $\bE[\Rev] = x \cdot (0+\epsilon \cdot 1/\epsilon) + (1-x) \cdot ((1-\epsilon) \cdot 1 + \epsilon \cdot 1/\epsilon) = 1$ unit of revenue. But the LP objective is $2-\epsilon$. So taking $\epsilon \to 0^+$, the expected revenue of any policy is upper-bounded by $\OPT/2$. 
\Halmos\endproof

\section{Proof of Theorem~\ref{thm:astmt:non:asy}} \label{sec:proof:thm:astmt:non:asy}
\proof{Proof of Theorem~\ref{thm:astmt:non:asy}.}
Denote $\hat{\delta} = \sqrt{\frac{2 \log{(\underline{b})}}{\underline{b}}}.$
And note that $\delta b_i -1 \geq \hat{\delta} b_i, \forall i \in [n]$, because $\underline{b} \geq 6$.

Denote the following random variables, which depict a run of our assortment policy from Algorithm~\ref{def::assortmentNonstationary2}.
Let $A_t(S)$ be the indicator random variable for $\tilde{S}_t = S$, where $\tilde{S}_t$ was the assortment selected before discarding in \eqref{eqn::policyAssortmentNonstationary2} was applied.
Let $B_t(i)$ be the remaining inventory of item $i$ at the \textit{end} of time $t$.  Defined for all $i\in[n]$ and $t=0,\ldots,T$, where $B_0(i)=b_i$ for all $i$.

Under either static or dynamic substitution, let $R_t(i,j)$ be the amount of sales that a customer chooses product $(i,j)$ during period $t$.
We will always specify the distribution of $R_t(i,j)$, by using a conditional probability.
For example, we will use $\bE[R_t(i,j) \left| S_t = S, \bm{B}_t = \bm{B}\right.]$, for the expected sales that a customer chooses product $(i,j)$ during period $t$, when we plan to offer assortment $S$, and when the remaining inventory level for each resource is $\bm{B} = (B_1, B_2, ..., B_n)$.

Under static substitution, conditional on any $S \in \cS, \bm{B} \in \bR_+^n$, $R_t(i,j) = \min \{B_i, Q\}$, where $Q$ is a random variable whose CDF is $F_{t,(i,j,S)} (\cdot)$.
Under dynamic substitution, conditional on any $S \in \cS, \bm{B} \in \bN_0^n$, $R_t(i,j)$ takes $1$ with probability $q_t(i,j,\bS)$.
Here we define $\bS = \{(i,j) \in S \left| B_i > 0 \right.\}$ to be a function of $S$.

Under dynamic substitution, Assumption~\ref{ass::subs} suggests that $q_t(i,j,\bS) \geq q_t(i,j,S), \forall (i,j) \in \bS$, because $\bS \subseteq S$.
The demand that originally would have chosen the stocked out items would go to their substitutes (as well as leaving, in which case the inequality takes equality).
On the other hand, $q_t(i, j, \bS) = 0, \forall (i,j) \notin \bS$.
The demand for any stocked out item is zero.
We can use indicator variables to write the above inequalities in a compact form $q_t(i,j, \bS) \geq \min \{B_{t-1}(i), q_t(i,j,S)\}$.

In all, we have
\begin{align}
\bE[R_t(i,j) \left| S_t = S, \bm{B}_t = \bm{B}\right.] \geq \bE_{Q\sim F_{t,(i,j,S)}}[\min \{B_{t-1}(i), Q\}]
\end{align}
where $F_{t,(i,j,S)}$ may prescribe a Bernoulli distribution, e.g. under dynamic substitution.

For any period $t$, given the remaining inventory from the last period to be $\bm{B}_{t-1}$, conditional on any $S \in \cS$, the remaining inventory updates in the following fashion,
$$B_t(i) = B_{t-1}(i) - R_t(i,j), \forall i.$$
Note that no item can be offered multiple times at different prices in one assortment.
Also note that we have defined $R_t(i,j)$ as the amount of sales, so $R_t(i,j)$ can never go beyond $B_{t-1}(i)$.

Following each sample path, we let $\Rev$ denote the revenue earned by the policy suggested in Algorithm~\ref{def::assortmentNonstationary::asp}.
\begin{align*}
\bE[\Rev] = & \sum_{t=1}^T \sum_{S \in \mathcal{S}} A_t(S) \sum_{ (i,j)\in S }p_j \bE[R_t(i,j) \left| S_t = S, \bm{B}_t = \bm{B}_{t-1}\right.] \\
\geq & \sum_{t=1}^T \sum_{S \in \mathcal{S}} A_t(S) \sum_{ (i,j)\in S }p_j \bE_{Q \sim F_{t,(i,j,S)}}[\min\{B_{t-1}(i), Q\}]
\end{align*}
Under static substitution, this inequality takes equality, and requires no assumption;
under dynamic substitution, this inequality is true due to Assumption~\ref{ass::subs}.

Taking expectation we have the following:
\begin{align}
\bE[\Rev] & = \sum_{t=1}^T \sum_{S \in \mathcal{S}} \bE[A_t(S)] \sum_{ (i,j)\in S }p_j \bE_{Q \sim F_{t,(i,j,S)}}[\min\{B_{t-1}(i), Q\}] \nonumber \\
& \geq \sum_{t=1}^T \sum_{S \in \mathcal{S}} \bE[A_t(S)] \sum_{ (i,j)\in S }p_j \bE_{Q \sim F_{t,(i,j,S)}}[\min\{B_{t-1}(i), Q\} \left| B_{t-1}(i)>1 \right.] \cdot \Pr\{B_{t-1}(i)>1\} \nonumber \\
& = \sum_{t=1}^T \sum_{S \in \mathcal{S}} \bE[A_t(S)] \sum_{ (i,j)\in S }p_j \bE_{Q \sim F_{t,(i,j,S)}}[Q \left| B_{t-1}(i)>1 \right.] \cdot \Pr\{B_{t-1}(i)>1\} \label{eqn:non:asy:1}
 \\
& \geq \sum_{t=1}^T \sum_{S \in \mathcal{S}} \bE[A_t(S)] \sum_{ (i,j)\in S }p_j \bE_{Q \sim F_{t,(i,j,S)}}[Q] \cdot \Pr\{B_T(i)>1\} \nonumber \\
& = \sum_{i=1}^n \Pr\{B_T(i)>1\} \sum_{t=1}^T \sum_{S \in \mathcal{S}} (1-\delta) x^*_t(S) \sum_{j:(i,j)\in S}p_j q_t(i,j,S) \nonumber
\end{align}
where the first equality is because the random variables $A_t(S), B_{t-1}(i)$ and $Q \sim F_{t,(i,j,S)}$ are independent;
the first inequality is re-writing the expectation by a conditional expectation, while ignoring the happening of other events ``$B_{t-1}(i)\leq1$'';
the second equality is because conditioning on $B_{t-1}(i)>1$, there must not be truncated demands during periods $t\in[T]$;
the second inequality is because random variables $B_{t-1}(i)$ and $Q \sim F_{t,(i,j,S)}$ are independent, and that $B_{t}(i)$ is non-increasing in $t, \forall t\in[T]$.

We first lower bound $\Pr\{B_T(i) > 1\} = \Pr\{\sum_{t=1}^T \sum_{S \in \mathcal{S}} A_t(S) \sum_{j:(i,j) \in S} Q_{t}(i,j,S) < b_i-1\}$, the probability that inventory $i$ never runs out.
Conditioning on the event that inventory never runs out, we know that dynamic substitution will never happen.
In all the remaining proof, we will only use notations like $Q(i,j,S)$, for the random quantity that customers attempt to purchase product $(i,j)$, should assortment $S$ be offered, no matter if any of the items from the assortment is stocked out.
Note that the expected amount of inventory sold is strictly less than $b_i-1$:
\begin{equation}
\label{eqn:strictlySmaller}
\bE[\sum_{t=1}^T \sum_{S \in \mathcal{S}} A_t(S) \sum_{j:(i,j) \in S} Q_{t}(i,j,S)] = (1-\delta) \sum_{t=1}^T \sum_{S \in \mathcal{S}} x^*_t(S) \sum_{j:(i,j) \in S} q_{t}(i,j,S) \leq (1-\delta) b_i < b_i -1,
\end{equation}
where the first inequality is due to constraint \eqref{eqn::constr::assortmentInventory};
the second inequality due to $\underline{b} \geq 6$.
Since strict inequality holds, we can lower bound $\Pr\{B_T(i) > 0\}$ as follows.
\begin{align}
\Pr\{B_T(i) > 0\} = & 1 - \Pr \left\{\sum_{t=1}^T \sum_{S \in \mathcal{S}} A_t(S) \sum_{j:(i,j) \in S} Q_{t}(i,j,S) \geq b_i \right\} \nonumber \\
\geq & 1 - \Pr\left\{ \sum_{t=1}^T \sum_{S \in \mathcal{S}} A_t(S) \sum_{j:(i,j) \in S} Q_{t}(i,j,S) - \bE [\sum_{t=1}^T \sum_{S \in \mathcal{S}} A_t(S) \sum_{j:(i,j) \in S} Q_{t}(i,j,S) ] \geq \delta b_i \right\} \nonumber \\
\geq & 1 - \exp\left(- \frac{(\hat{\delta} b_i)^2}{2 \text{Var}(\sum_{t=1}^T \sum_{S \in \mathcal{S}} A_t(S) \sum_{j:(i,j) \in S} Q_{t}(i,j,S)) + 2/3 \hat{\delta} b_i}\right) \nonumber \\
\geq & 1 - \exp\left(- \frac{\hat{\delta}^2 b_i}{2}\right) \nonumber \\
\geq & 1 - \frac{1}{\underline{b}}, \label{eqn:non:asy:2}
\end{align}
where the first inequality is due to \eqref{eqn:strictlySmaller};
second inequality is Bernstein Inequality, where $\forall t, \left| \sum_{S \in \mathcal{S}} A_t(S) \sum_{j:(i,j) \in S} Q_{t}(i,j,S) - \bE[\sum_{S \in \mathcal{S}} A_t(S) \sum_{j:(i,j) \in S} Q_{t}(i,j,S)] \right| \leq 1$ are zero-mean random variables and almost surely bounded by 1, and because $\delta b_i - 1 \geq \hat{\delta} b_i, \forall i \in [n]$;
third inequality is because
\begin{align*}
& \text{Var}(\sum_{t=1}^T \sum_{S \in \mathcal{S}} A_t(S) \sum_{j:(i,j) \in S} Q_{t}(i,j,S)) \\
\leq & \sum_{t=1}^T \left(\sum_{S \in \mathcal{S}} (1-\delta) x_t^*(S) \sum_{j:(i,j)\in S} q_t(i,j,S) \cdot \left(1-\sum_{S \in \mathcal{S}} (1-\delta) x_t^*(S) \sum_{j:(i,j)\in S} q_t(i,j,S) \right) \right) \\
\leq & \sum_{t=1}^T \sum_{S \in \mathcal{S}} (1-\delta) x_t^*(S) \sum_{j:(i,j)\in S} q_t(i,j,S) \\
\leq & (1-\delta) b_i \leq (1-\hat{\delta}) b_i.
\end{align*}
This is because if one random variable with bounded support over $[0,1]$ has the same mean as a Bernoulli random variable, its variance should be smaller than that of the Bernoulli random variable.

Finally putting \eqref{eqn:non:asy:2} into \eqref{eqn:non:asy:1} we have
\begin{align*}
\bE[\Rev] \geq & \sum_{i=1}^n \Pr\{B_T(i)>0\} \sum_{t=1}^T \sum_{S \in \mathcal{S}} (1-\delta) x^*_t(S) \sum_{j:(i,j)\in S}p_j q_t(i,j,S) \\
\geq & (1-\frac{1}{\underline{b}}) (1-\delta) \sum_{i=1}^n \sum_{t=1}^T \sum_{S \in \mathcal{S}} x^*_t(S) \sum_{j:(i,j)\in S}p_j q_t(i,j,S) \\
= & (1-\frac{1}{\underline{b}}) (1 - \sqrt{\frac{3 \log{(\underline{b})}}{\underline{b}}}) \OPT = (1 - \sqrt{\frac{3 \log{(\underline{b})}}{\underline{b}}} + o(\sqrt{\frac{\log{(\underline{b})}}{\underline{b}}})) \OPT
\end{align*}
which finishes the proof. By taking $b \to \infty$ we see the calendar is asymptotically optimal.
\Halmos\endproof

\section{Proof of Theorem~\ref{thm:derandomize}} \label{sec:proof:thm:derandomize}

\proof{Proof of Theorem~\ref{thm:derandomize}.}
We introduce the following notations.
Denote $$\hat{\mu}_K(\bm{z} \left| S_\tau = S_\tau(\bm{z}), \forall \tau \leq t, S_t = S \right.)$$
Denote ``$\tilde{\bm{z}} \left| S_\tau = \tilde{S}_\tau, \forall \tau \in \Gamma \subseteq \mathcal{S} \right.$'' to be a tweaked vector from $\bm{z}$, such that $\tilde{z}_\tau(\mathsf{X}) = \bI_{\mathsf{X}=\tilde{S}_\tau}, \forall \tau \in \Gamma, \forall \mathsf{X} \in \mathcal{S}; \tilde{z}_\varrho(X) = z_\varrho(X), \forall \varrho \notin \Gamma, \forall X \in \mathcal{S}$.
For example, $\tilde{\bm{z}} \left| S_1 = S \right.$ is defined by $\tilde{z}_1(\mathsf{X}) = \bI_{\mathsf{X}=S}, \forall \mathsf{X} \in \mathcal{S}; \tilde{z}_\varrho(X) = z_\varrho(X), \forall \varrho \geq 2, X \in \mathcal{S}$.
Similarly, denote $\bE[\Rev \left| S_\tau = \tilde{S}_\tau, \forall \tau \in \Gamma \right.]$ to be the expected revenue earned from a probabilistic offering of $\tilde{\bm{z}} \left| S_\tau = \tilde{S}_\tau, \forall \tau \in \Gamma \right.$.

Notice that, rigorously, $S_t(\bm{z})$ should be a random assortment based on the simulation results that are random.
So rigorously, we should use $\bE_{\bm{\xi}}[\bE[\Rev \left| S_\tau = S_\tau(\bm{z}), \forall \tau \leq t \right.]]$, because the conditioned event is a random event based on the simulation results.
We take the outer expectation over the simulation randomness $\bm{\xi}$.
In the inner expectation where we do not epecifically designate the source of randomness, the expectation is taken to find the expected revenue.

In each iteration of Algorithm~\ref{def::derandomize}, denote also $S^*_t(\bm{z})$ to be any element from $$S^*_t(\bm{z}) \in \argmax_{S \in \{S \in \mathcal{S} \left| z_t(S)>0 \right.\}} \bE[ \Rev \left| S_\tau = S_\tau(\bm{z}), \forall \tau \leq t, S_t = S \right.],$$ which is the true best assortment to select in this iteration, if we were given a perfect oracle to query the expected revenue of a policy (instead of a simulator).

In each iteration, denote $O_t = \{S \in \mathcal{S} \left| z_t(S)>0 \right.\}$ to be the set of candidate assortments to choose from.
Let $o_t = \left| O_t \right|$.
In each iteration, we can upper bound the sampling error incurred due to selecting the best empirical assortment, instead of (possibly) the true best assortment. $\forall t \in [T]$
\begin{align*}
 & \bE[\Rev \left| S_\tau = S_\tau(\bm{z}), \forall \tau \leq t-1, S_t = S^*_t(\bm{z}) \right.] - \bE_{\bm{\xi}}[ \bE[\Rev \left| S_\tau = S_\tau(\bm{z}), \forall \tau \leq t \right.] ]\\
= & \bE[\Rev \left| S_\tau = S_\tau(\bm{z}), \forall \tau \leq t-1, S_t = S^*_t(\bm{z}) \right.] - \bE[\hat{\mu}_K(\bm{z} \left| S_\tau = S_\tau(\bm{z}), \forall \tau \leq t-1, S_t = S^*_t(\bm{z}) \right.)] + \\
& \bE[\hat{\mu}_K(\bm{z} \left| S_\tau = S_\tau(\bm{z}), \forall \tau \leq t-1, S_t = S^*_t(\bm{z}) \right.)] - \bE_{\bm{\xi}}[\hat{\mu}_K(\bm{z} \left| S_\tau = S_\tau(\bm{z}), \forall \tau \leq t \right.)] + \\
& \bE_{\bm{\xi}}[\hat{\mu}_K(\bm{z} \left| S_\tau = S_\tau(\bm{z}), \forall \tau \leq t \right.)] - \bE_{\bm{\xi}}[ \bE[\Rev \left| S_\tau = S_\tau(\bm{z}), \forall \tau \leq t \right.] ] \\
\leq & \bE_{\bm{\xi}}[\hat{\mu}_K(\bm{z} \left| S_\tau = S_\tau(\bm{z}), \forall \tau \leq t \right.)] - \bE_{\bm{\xi}}[ \bE[\Rev \left| S_\tau = S_\tau(\bm{z}), \forall \tau \leq t \right.] ] + 0 + 0\\
\leq & \bE_{\bm{\xi}}[ \max_{S \in O_t} \left\{ \hat{\mu}_K(\bm{z} \left| S_\tau = S_\tau(\bm{z}), \forall \tau \leq t-1, S_t=S \right.) - \bE[\Rev \left| S_\tau = S_\tau(\bm{z}), \forall \tau \leq t-1, S_t=S \right.] \right\}]
\end{align*}
where the equality is how we decompose the difference by three differences;
the first inequality is because the second difference is non-negative, since we are selecting the optimizer for the empirical performance, and because the first difference is zero, since given any fixed assortment, the empirical estimation is an unbiased estimation.

Now we further bound the sampling error. $\forall h > 0,$
\begin{align*}
& \bE_{\bm{\xi}}[ \max_{S \in O_t} \left\{ \hat{\mu}_K(\bm{z} \left| S_\tau = S_\tau(\bm{z}), \forall \tau \leq t-1, S_t=S \right.) - \bE[\Rev \left| S_\tau = S_\tau(\bm{z}), \forall \tau \leq t-1, S_t=S \right.] \right\}] \\
= & \frac{1}{h} \log{ \exp{\left( h \cdot \bE_{\bm{\xi}}[ \max_{S \in O_t} \left\{ \hat{\mu}_K(\bm{z} \left| S_\tau = S_\tau(\bm{z}), \forall \tau \leq t-1, S_t=S \right.) - \bE[\Rev \left| S_\tau = S_\tau(\bm{z}), \forall \tau \leq t-1, S_t=S \right.] \right\}] \right) }} \\
\leq & \frac{1}{h} \log{ \bE_{\bm{\xi}}\left[ \exp{\left( h \cdot \max_{S \in O_t} \left\{ \hat{\mu}_K(\bm{z} \left| S_\tau = S_\tau(\bm{z}), \forall \tau \leq t-1, S_t=S \right.) - \bE[\Rev \left| S_\tau = S_\tau(\bm{z}), \forall \tau \leq t-1, S_t=S \right.] \right\} \right) } \right] } \\
= & \frac{1}{h} \log{ \bE_{\bm{\xi}}\left[ \max_{S \in O_t} \left\{ \exp{\left( \frac{h}{K} \cdot ( \sum_{k=1}^K \nu(\bm{z} \left| S_\tau = S_\tau(\bm{z}), \forall \tau \leq t-1, S_t=S \right., \xi_k) - \bE[\Rev \left| S_\tau = S_\tau(\bm{z}), \forall \tau \leq t-1, S_t=S \right.] ) \right) } \right\} \right] } \\
\leq & \frac{1}{h} \log{ \bE_{\bm{\xi}}\left[ \sum_{S \in O_t} \exp{\left( \frac{h}{K} \cdot ( \sum_{k=1}^K \nu(\bm{z} \left| S_\tau = S_\tau(\bm{z}), \forall \tau \leq t-1, S_t=S \right., \xi_k) - \bE[\Rev \left| S_\tau = S_\tau(\bm{z}), \forall \tau \leq t-1, S_t=S \right.] ) \right) } \right] } \\
= & \frac{1}{h} \log{ \sum_{S \in O_t} \prod_{k=1}^K \bE_{\bm{\xi}}\left[ \exp{\left( \frac{h}{K} \cdot ( \nu(\bm{z} \left| S_\tau = S_\tau(\bm{z}), \forall \tau \leq t-1, S_t=S \right., \xi_k) - \bE[\Rev \left| S_\tau = S_\tau(\bm{z}), \forall \tau \leq t-1, S_t=S \right.] )  \right) } \right] } \\
\leq & \frac{1}{h} \log{ \sum_{S \in O_t} \prod_{k=1}^K \exp{ \left( \frac{h^2}{K^2} \cdot \frac{(2 (b_1+\cdots+b_n) p_{\mathrm{max}})^2}{8} \right)} } \\
= & \frac{\log{o_t}}{h} + \frac{h (b_1+\cdots+b_n)^2 p_{\mathrm{max}}^2}{2K}
\end{align*}
where the first equality is re-writing the same expression;
the first inequality is due to Jensen's inequality, because $\forall h > 0, f(x)=\exp{(h \cdot x)}$ is a convex function in $x$;
the second equality is from the definition of $\hat{\mu}_K(\cdot)$;
the second inequality is because all the exponentials are positive, and we are replacing the maximum with their sum;
the third equality is due to linearity of expectations and due to the independence of different simulations under randomness $\xi_k$;
the third inequality is due to Hoeffding's Lemma, where each random simulation yields a number bounded by $(b_1+\cdots+b_n) p_{\mathrm{max}}$.

The above bound holds for any $h>0$. If we pick $h=\sqrt{\frac{2K\log{o_t}}{(b_1+\cdots+b_n)^2 p_{\mathrm{max}}^2}}$, then the above term can be simplified as
\begin{align}
\bE[\Rev \left| S_\tau = S_\tau(\bm{z}), \forall \tau \leq t-1, S_t = S^*_t(\bm{z}) \right.] - & \bE_{\bm{\xi}}[ \bE[\Rev \left| S_\tau = S_\tau(\bm{z}), \forall \tau \leq t \right.] ] \leq \sqrt{\frac{\log{o_t}}{2K}} (b_1+\cdots+b_n) p_{\mathrm{max}} \nonumber \\
= & \sqrt{\frac{\log{o_t}}{2(\log{n} + \log{T})}} \cdot \epsilon \cdot \OPT. \label{eqn:samplingerrorbound}
\end{align}

Finally we conclude the proof by induction on $t$.
In the first iteration of Algorithm~\ref{def::derandomize}, we can re-write $\bE[\Rev]$ as follows,
\begin{align}
\label{eqn:derandomizeExpectation}
\bE[\Rev] = \sum_{S\in \{ S \in \mathcal{S} \left| z_1(S)>0 \right. \}} \bE[\Rev \left| S_1=S \right.] \cdot z_1(S) \geq \alpha \cdot \OPT,
\end{align}
where the inequality holds due to Theorems~\ref{thm::assortmentStationary} -- \ref{thm:astmt:non:asy}.
Since we have selected the $S$'s such that $z_1(S) > 0$, and that $\sum_{S\in \{ S \in \mathcal{S} \left| z_1(S)>0 \right. \}} z_1(S) = 1$. So $$\bE[\Rev \left| S_1=S_1^*(\bm{z}) \right.] = \max_{S\in \{ S \in \mathcal{S} \left| z_1(S)>0 \right. \}} \bE[\Rev \left| S_1=S \right.] \geq \alpha \cdot \OPT,$$ because otherwise the summation in \eqref{eqn:derandomizeExpectation} is strictly smaller than $\alpha \cdot \OPT$.
Then we can plug in inequality~\eqref{eqn:samplingerrorbound}, so that $\bE_{\bm{\xi}}[ \bE[\Rev \left| S_1 = S_1(\bm{z}) \right.] ] \geq (\alpha - \sqrt{\frac{\log{o_1}}{2(\log{n} + \log{T})}} \epsilon) \cdot \OPT$

Suppose we have shown that $\bE_{\bm{\xi}}[ \bE[\Rev \left| S_\tau = S_\tau(\bm{z}), \forall \tau \leq t \right.] ] \geq (\alpha - \sum_{\tau=1}^{t}\sqrt{\frac{\log{o_\tau}}{2(\log{n} + \log{T})}} \epsilon) \cdot \OPT$.
In the $(t+1)^{\text{th}}$ iteration of Algorithm~\ref{def::derandomize}, we can re-write the expected revenue as follows,
\begin{align*}
\bE_{\bm{\xi}}[\bE[\Rev \left| S_\tau = S_\tau(\bm{z}), \forall \tau \leq t \right.]] = & \sum_{S\in \{ S \in \mathcal{S} \left| z_{t+1}(S)>0 \right. \}} \bE[\Rev \left| S_\tau = S_\tau(\bm{z}), \forall \tau \leq t, S_{t+1}=S \right.] \cdot z_{t+1}(S) \\
\geq & (\alpha - \sum_{\tau=1}^{t}\sqrt{\frac{\log{o_\tau}}{2(\log{n} + \log{T})}} \epsilon) \cdot \OPT,
\end{align*}
Similarly, we have $\bE_{\bm{\xi}}[\bE[\Rev \left| S_\tau = S_\tau(\bm{z}), \forall \tau \leq t, S_{t+1}=S_{t+1}^*(\bm{z}) \right.]] \geq (\alpha - \sum_{\tau=1}^{t}\sqrt{\frac{\log{o_\tau}}{2(\log{n} + \log{T})}} \epsilon) \cdot \OPT,$ and then using inequality~\eqref{eqn:samplingerrorbound} we have $\bE_{\bm{\xi}}[\bE[\Rev \left| S_\tau = S_\tau(\bm{z}), \forall \tau \leq t+1 \right.]] \geq (\alpha - \sum_{\tau=1}^{t+1}\sqrt{\frac{\log{o_\tau}}{2(\log{n} + \log{T})}} \epsilon) \cdot \OPT.$

By induction, we have
\begin{align*}
\bE_{\bm{\xi}}[\bE[\Rev \left| S_\tau = S_\tau(\bm{z}), \forall \tau \leq T\right.]] \geq & (\alpha - \sum_{\tau=1}^{T}\sqrt{\frac{\log{o_\tau}}{2(\log{n} + \log{T})}} \epsilon) \cdot \OPT \\
\geq & (\alpha - \sqrt{\frac{\log{\left( \sum_{\tau=1}^{T} o_\tau \right) }}{2(\log{n} + \log{T})}} \epsilon) \cdot \OPT \\
\geq & (\alpha - \epsilon) \cdot \OPT
\end{align*}
where the first inequality is due to concavity of $\sqrt{\log{x}}$;
the second inequality is because $\sum_{\tau=1}^T o_\tau \leq 2(\log{n} + \log{T})$.
In the context of non-stationary arrivals, $\sum_{\tau=1}^T o_\tau$ is the total number of non-zero variables from CDLP, can be upper bounded by the total number of constraints, which is $n+T$.
In the context of stationary arrivals, $\sum_{\tau=1}^T o_\tau$ has bounded supports of $n+1$ in each of the $T$ periods -- so it could be bounded by $(n+1)\cdot T$.

Finally, we analyze the time complexity.
In each iteration, we enumerate over all the assortments $\{S \in \mathcal{S} \left| z_t(S)>0 \right.\}$ to find the empirical maximizer $\hS_t$, which involves $K$ queries to the simulator $\hat{\mu}(\cdot)$.
There are at most $n+1$ non-zero variables to enumerate.
So the total number of queries are no more than $K \cdot (n+1) \cdot T = O(KnT)$.
In each query of the simulator $\hat{\mu}(\cdot)$, it takes no more than $(n+1)T$ operations to generate a sequence of $T$ assortments.
On the other hand, since we know the CDF of the demand in each period, it takes only $O(1)$ operations to generate a random demand.
So the total number of arithmetic operations for our de-randomization method is $O(Kn^2T^2)$.
As suggested by Algorithm~\ref{def::assortmentNonstationary::asp}, $K$ is also polynomial in $n$, $T$, and $1/\epsilon$.
\Halmos\endproof

\section{Lemmas for the Proof of Theorem~\ref{thm::pricingStationary}}
\label{sec::proof:thm::gen:pricingStationary}

In this section we prove Lemmas~\ref{lem:structure1} and \ref{lem::convexity}.

\subsection{Proof of Lemma~\ref{lem:structure1}}
\proof{Proof of Lemma~\ref{lem:structure1}.}
Let $(x)^+ = \max\{x, 0\}$ denote the maximum of $x$ and $0$. Let $Q_{v_t}$ denote the random demand if we offer price $v_t$ on day $t$, which follows a distribution of $F_{v_t}(\cdot)$.

Similar to the proof to Lemma~\ref{lem:structure1}, the idea is to exchange a pair of two consecutive prices. Given any calendar $\bm{v} = (v_1, v_2, ..., v_T)$, if there exists $t \in [T-1]$, such that $p_{v_t} < p_{v_{t+1}}$, we compare to another calendar: $$\bm{v}^* = (v_1, v_2, ..., v_{t-1}, v_{t+1}, v_{t}, v_{t+2}, ..., v_T).$$ Since we only exchange this pair of two prices, the expected revenue change comes only from these two periods: Before $t$ the expected revenue is trivially not changed. Since under both calendars, the distribution of total units of inventory consumed during $t$ and $t+1$ are exactly the same, we know that the distribution of initial inventory at the end of $t+1$ will also be the same. So after period $t+1$ the revenue will not be changed, neither.

Let $A(c, t)$ denote the event ``$c$ units of inventory left at the beginning of period $t$'', where $c \in [0,b], t \in [T]$.
Let $\Rev(\bm{v})$ be a random variable for the expected revenue from all time periods under $\bm{v}$.
Let $\Rev_{t,t+1}(\bm{v})$ be a random variable for the expected revenue from $t$ and $t+1$ under $\bm{v}$.
Then for any $c \in [0,b]$, conditioning on $A(c,t)$ we have:
\begin{equation*}
\begin{split}
\bE[\Rev_{t,t+1}(\bm{v}) \left| A(c,t) \right.] = & p_{v_t} \bE\left[\min\{c, Q_{v_t}\}\right] + p_{v_{t+1}} \bE \left[ \min\{(c - Q_{v_t})^+, Q_{v_{t+1}}\} \right]\\
= & (p_{v_t} - p_{v_{t+1}}) \bE\left[\min\{c, Q_{v_t}\}\right] + p_{v_{t+1}} \bE \left[ \min\{c, Q_{v_t} + Q_{v_{t+1}}\} \right]\\
\end{split}
\end{equation*}
where the second equation holds because $\min\{(c - Q_{v_t})^+, Q_{v_{t+1}}\} = \min\{c, Q_{v_t} + Q_{v_{t+1}}\} - \min\{c, Q_{v_t}\}$, and linearity of expectations.

Now let us compare the expected revenue from two calendars, still conditioning on $A(c,t)$:
\begin{equation*}
\begin{split}
& \bE[\Rev(\bm{v}^*) \left| A(c,t) \right.] - \bE[\Rev(\bm{v}) \left| A(c,t) \right.] \\
= & (p_{v_{t+1}} - p_{v_{t}}) \bE\left[\min\{c, Q_{v_{t+1}}\}\right] + p_{v_{t}} \bE \left[ \min\{c, Q_{v_{t+1}} + Q_{v_{t}}\} \right] \\
& - (p_{v_t} - p_{v_{t+1}}) \bE\left[\min\{c, Q_{v_t}\}\right] - p_{v_{t+1}} \bE \left[ \min\{c, Q_{v_t} + Q_{v_{t+1}}\} \right]\\
= & (p_{v_{t+1}} - p_{v_{t}}) (\bE\left[\min\{c, Q_{v_t}\}\right] + \bE\left[\min\{c, Q_{v_{t+1}}\}\right] - \bE \left[ \min\{c, Q_{v_t} + Q_{v_{t+1}}\} \right]) \\
\geq & 0
\end{split}
\end{equation*}
where the last inequality is due to the fact that $\min\{c,x\} + \min\{c,y\} \geq \min\{c,x+y\}, \forall c,x,y \geq 0$, which is proved in the appendix as Lemma~\ref{lem:trc:ineq}. Notice that this is true for all $c \in [0,b]$.

By integrating over $c \in [0,b]$ we have $\bE[\Rev(\bm{v}^*)] \geq \bE[\Rev(\bm{v})]$, which finishes the proof.
\Halmos\endproof

\begin{corollary}
\label{thm:monotone}
There exists an optimal static calendar whose prices are non-increasing over time, i.e. $p_{v^*_t} \geq p_{v^*_{t+1}}, \forall t \in [T-1]$.
\end{corollary}

\proof{Proof of Corollary~\ref{thm:monotone}.}
Directly follows from Lemma~\ref{lem:structure1}. We can start from any calendar and use a finite number (no more than $T!$) of exchange operations to achieve the optimal non-decreasing structure.
\Halmos\endproof

\subsection{Bridge from Lemma~\ref{lem:structure1} to Lemma~\ref{lem::convexity}}

\begin{lemma}
\label{lem:gen:structure2}
Under [0,1]-demand, in any two-price randomized policy $\bm{v}$, if two consecutive probabilities $v_t$, $v_{t+1}$ are such that $v_t<v_{t+1}$, then probabilities $v_t$, $v_{t+1}$ can be exchanged in the calendar without decreasing its expected revenue.
\end{lemma}

\proof{Proof of Lemma~\ref{lem:gen:structure2}.}
Let $(x)^+ = \max\{x, 0\}$ denote the maximum of $x$ and $0$. Let $Q_{\pH,t}$ and $Q_{\pL,t}$ denote the random demand if we offer the higher price and the lower price on day $t$, which follows a distribution of $F_{H}(\cdot)$ and $F_{L}(\cdot)$, respectively.

The idea is to exchange a pair of two consecutive prices. Given any calendar $\bm{v} = (v_1, v_2, ..., v_T)$, if there exists $t \in [T-1]$, such that $v_t < v_{t+1}$, we compare to another calendar: $$\bm{v}^* = (v_1, v_2, ..., v_{t-1}, v_{t+1}, v_{t}, v_{t+2}, ..., v_T).$$ Since we only exchange this pair of two prices, the expected revenue before $t$ is trivially not changed.

Let $J_{\bm{v}, t}(c)$ denote the expected revenue we would earn under calendar $\bm{v}$ if we were endowed with $c$ units of inventory at the beginning of period $t$. Its expectation is taken over future demand randomness.
Let $A(c, t)$ denote the event ``$c$ units of inventory at the beginning of period $t$'', where $c \in [0,b], t \in [T]$.
Then for any $c \in [0,b]$, conditioning on $A(c,t)$ we have:
\begin{align*}
J_{\bm{v}, t}(c) = & v_t \bE\left[ \min\{c,Q_{\pH,t}\} \right] p_\pH + v_t v_{t+1} \bE\left[ \min\{(c-Q_{\pH,t})^+, Q_{\pH,t+1}\} \right] p_\pH \\
& \qquad \qquad \qquad \qquad + v_t (1-v_{t+1}) \bE\left[ \min\{(c-Q_{\pH,t})^+, Q_{\pL,t+1}\} \right] p_\pL\\
& + v_t v_{t+1} \bE\left[J_{\bm{v}, t+2}((c - Q_{\pH,t} - Q_{\pH,t+1})^+)\right]\\
& + v_t (1-v_{t+1}) \bE\left[J_{\bm{v}, t+2}((c - Q_{\pH,t} - Q_{\pL,t+1})^+)\right]\\
& + (1-v_t) \bE\left[ \min\{c,Q_{\pL,t}\} \right] p_\pL + (1-v_t) v_{t+1} \bE\left[ \min\{(c-Q_{\pL,t})^+, Q_{\pH,t+1}\} \right] p_\pH \\
& \qquad \qquad \qquad \qquad + (1-v_t) (1-v_{t+1}) \bE\left[ \min\{(c-Q_{\pL,t})^+, Q_{\pL,t+1}\} \right] p_\pL\\
& + (1-v_t) v_{t+1} \bE\left[J_{\bm{v}, t+2}((c - Q_{\pL,t} - Q_{\pH,t+1})^+)\right]\\
& + (1-v_t) (1-v_{t+1}) \bE\left[J_{\bm{v}, t+2}((c - Q_{\pL,t} - Q_{\pL,t+1})^+)\right]
\end{align*}

Plugging in $\min\{(c - x)^+, y\} = \min\{c, x+y\} - \min\{c, x\}, \forall c,x,y \geq 0$ we have:
\begin{align*}
J_{\bm{v}, t}(c) = & v_t \bE\left[ \min\{c,Q_{\pH,t}\} \right] p_\pH + v_t v_{t+1} \bE\left[ \min\{c, Q_{\pH,t} + Q_{\pH,t+1}\} \right] p_\pH \\
& \qquad \qquad \qquad \qquad + v_t (1-v_{t+1}) \bE\left[ \min\{c, Q_{\pH,t} + Q_{\pL,t+1}\} \right] p_\pL\\
& - v_t v_{t+1} \bE\left[ \min\{c,Q_{\pH,t}\} \right] p_\pH - v_t (1 - v_{t+1}) \bE\left[ \min\{c,Q_{\pH,t}\} \right] p_\pL \\
& + v_t v_{t+1} \bE\left[J_{\bm{v}, t+2}((c - Q_{\pH,t} - Q_{\pH,t+1})^+)\right]\\
& + v_t (1-v_{t+1}) \bE\left[J_{\bm{v}, t+2}((c - Q_{\pH,t} - Q_{\pL,t+1})^+)\right]\\
& + (1-v_t) \bE\left[ \min\{c,Q_{\pL,t}\} \right] p_\pL + (1-v_t) v_{t+1} \bE\left[ \min\{c, Q_{\pL,t} + Q_{\pH,t+1}\} \right] p_\pH \\
& \qquad \qquad \qquad \qquad + (1-v_t) (1-v_{t+1}) \bE\left[ \min\{c, Q_{\pL,t} + Q_{\pL,t+1}\} \right] p_\pL\\
& - (1-v_t) v_{t+1} \bE\left[ \min\{c,Q_{\pL,t}\} \right] p_\pH - (1-v_t) (1 - v_{t+1}) \bE\left[ \min\{c,Q_{\pL,t}\} \right] p_\pL \\
& + (1-v_t) v_{t+1} \bE\left[J_{\bm{v}, t+2}((c - Q_{\pL,t} - Q_{\pH,t+1})^+)\right]\\
& + (1-v_t) (1-v_{t+1}) \bE\left[J_{\bm{v}, t+2}((c - Q_{\pL,t} - Q_{\pL,t+1})^+)\right]
\end{align*}

Merging similar expressions we have the following:
\begin{equation}
\label{eqa:gen:exp1}
\begin{split}
J_{\bm{v}, t}(c) = & v_t (1 - v_{t+1}) (p_\pH - p_\pL) \bE\left[ \min\{c,Q_{\pH,t}\} \right] \\
& - (1 - v_t) v_{t+1} (p_\pH - p_\pL) \bE\left[ \min\{c,Q_{\pL,t}\} \right] \\
& + v_t v_{t+1} p_\pH \bE\left[ \min\{c, Q_{\pH,t} + Q_{\pH,t+1}\} \right]\\
& + v_t (1-v_{t+1}) p_\pL \bE\left[ \min\{c, Q_{\pH,t} + Q_{\pL,t+1}\} \right]\\
& + (1-v_t) v_{t+1} p_\pH \bE\left[ \min\{c, Q_{\pL,t} + Q_{\pH,t+1}\} \right]\\
& + (1-v_t) (1-v_{t+1}) p_\pL \bE\left[ \min\{c, Q_{\pL,t} + Q_{\pL,t+1}\} \right]\\
& + v_t v_{t+1} \bE\left[J_{\bm{v}, t+2}((c - Q_{\pH,t} - Q_{\pH,t+1})^+)\right]\\
& + v_t (1-v_{t+1}) \bE\left[J_{\bm{v}, t+2}((c - Q_{\pH,t} - Q_{\pL,t+1})^+)\right]\\
& + (1-v_t) v_{t+1} \bE\left[J_{\bm{v}, t+2}((c - Q_{\pL,t} - Q_{\pH,t+1})^+)\right]\\
& + (1-v_t) (1-v_{t+1}) \bE\left[J_{\bm{v}, t+2}((c - Q_{\pL,t} - Q_{\pL,t+1})^+)\right]
\end{split}
\end{equation}

Similarly we have $J_{\bm{v}^*, t}(c)$, the expected revenue under calendar $\bm{v}^*$, as following:
\begin{equation}
\begin{split}
J_{\bm{v}^*, t}(c) = & v_{t+1} (1 - v_{t}) (p_\pH - p_\pL) \bE\left[ \min\{c,Q_{\pH,t}\} \right] \\
& - (1 - v_{t+1}) v_{t} (p_\pH - p_\pL) \bE\left[ \min\{c,Q_{\pL,t}\} \right] \\
& + v_{t+1} v_{t} p_\pH \bE\left[ \min\{c, Q_{\pH,t} + Q_{\pH,t+1}\} \right]\\
& + v_{t+1} (1-v_{t}) p_\pL \bE\left[ \min\{c, Q_{\pH,t} + Q_{\pL,t+1}\} \right]\\
& + (1-v_{t+1}) v_{t} p_\pH \bE\left[ \min\{c, Q_{\pL,t} + Q_{\pH,t+1}\} \right]\\
& + (1-v_{t+1}) (1-v_{t}) p_\pL \bE\left[ \min\{c, Q_{\pL,t} + Q_{\pL,t+1}\} \right]\\
& + v_{t+1} v_{t} \bE\left[J_{\bm{v}, t+2}((c - Q_{\pH,t} - Q_{\pH,t+1})^+)\right]\\
& + v_{t+1} (1-v_{t}) \bE\left[J_{\bm{v}, t+2}((c - Q_{\pH,t} - Q_{\pL,t+1})^+)\right]\\
& + (1-v_{t+1}) v_{t} \bE\left[J_{\bm{v}, t+2}((c - Q_{\pL,t} - Q_{\pH,t+1})^+)\right]\\
& + (1-v_{t+1}) (1-v_{t}) \bE\left[J_{\bm{v}, t+2}((c - Q_{\pL,t} - Q_{\pL,t+1})^+)\right]
\end{split}
\end{equation}
The equality holds because calendars $\bm{v}$ and $\bm{v}^*$ are the same from period $t+2$.

On the other hand, we know that $\forall c,x,y \geq 0, \min\{c, x+y\} \geq 0$ and $J_{\bm{v}, t+2}((c - x - y)^+) \geq 0$. Due to Fubini's theorem, we can exchange the double integration:
\begin{equation}
\label{eqa:gen:exp2}
\begin{split}
\bE\left[ \min\{c, Q_{\pH,t} + Q_{\pL,t+1}\} \right] & = \bE\left[ \min\{c, Q_{\pL,t} + Q_{\pH,t+1}\} \right]\\
\bE\left[J_{\bm{v}, t+2}((c - Q_{\pH,t} - Q_{\pL,t+1})^+)\right] & = \bE\left[J_{\bm{v}, t+2}((c - Q_{\pL,t} - Q_{\pH,t+1})^+)\right]
\end{split}
\end{equation}

Then we calculate their difference, conditioning on $A(c,t)$:
\begin{align*}
& J_{\bm{v}^*, t}(c) - J_{\bm{v}, t}(c)\\
= & (v_{t+1} - v_t) (p_\pH - p_\pL) \left( \bE\left[ \min\{c,Q_{\pH,t}\} \right] + \bE\left[ \min\{c,Q_{\pL,t}\} \right] - \bE\left[ \min\{c,Q_{\pH,t}+Q_{\pL,t+1}\} \right] \right)\\
= & (v_{t+1} - v_t) (p_\pH - p_\pL) \left( \bE\left[ \min\{c,Q_{\pH,t}\} \right] + \bE\left[ \min\{c,Q_{\pL,t+1}\} \right] - \bE\left[ \min\{c,Q_{\pH,t}+Q_{\pL,t+1}\} \right] \right)\\
\geq & 0
\end{align*}
where the inequality holds because three components are all greater or equal to zero, the third component due to Lemma~\ref{lem:trc:ineq}. Notice that this is true for all $c \in [0,b]$.

By integrating over $c \in [0,b]$ we have $\bE[\Rev(\bm{v}^*)] \geq \bE[\Rev(\bm{v})]$, which finishes the proof.
\Halmos\endproof

\begin{corollary}
\label{thm:gen:monotone2}
There exists an optimal two-price randomized policy whose probabilities are non-increasing over time, i.e. $v_t \geq v_{t+1}, \forall t \in [T-1]$.
\end{corollary}

\proof{Proof of Corollary~\ref{thm:gen:monotone2}.}
Directly follows from Lemma~\ref{lem:structure1}. We can start from any calendar and use a finite number (no more than $T!$) of exchange operations to achieve the optimal non-increasing structure.
\Halmos\endproof

Notice that Lemma~\ref{lem:gen:structure2} and Corollary~\ref{thm:gen:monotone2} are assumption-free.

\subsection{Proof of Lemma~\ref{lem::convexity}}

\begin{lemma}
\label{lem:gen:structure3}
Under Assumption \ref{asp:sta1}, in any two-price randomized policy $\bm{v}$, if two consecutive probabilities $v_t$, $v_{t+1}$ are such that $v_t \geq v_{t+1}, v_t < 1, v_{t+1} > 0$, then the last pair of probabilities indexed by $t = \sup_{\tau \in [T-1]} \left\{ v_{\tau} < 1, v_{\tau+1} > 0, v_{\tau} \geq v_{\tau+1} \right\}$ can be changed from $(v_t, v_{t+1})$ to $(v_t+\delta, v_{t+1}-\delta)$ where $0 \leq \delta \leq \max \{1-v_t, v_{t+1}\}$, without decreasing its expected revenue.
\end{lemma}

\proof{Proof.}
Let $(x)^+ = \max\{x, 0\}$ denote the maximum of $x$ and $0$. Let $Q_{\pH,t}$ and $Q_{\pL,t}$ denote the random demand if we offer the higher price and the lower price on day $t$, which follows a distribution of $F_{H}(\cdot)$ and $F_{L}(\cdot)$, respectively.

Again our idea is to modify a pair of consecutive probabilities, to achieve a greater revenue. Given any calendar $\bm{v} = (v_1, v_2, ..., v_n)$, where $v_1 \geq v_2 \geq ... \geq v_T$. If $\exists \tau \in [T-1]$, such that $v_{\tau} < 1, v_{\tau+1} > 0, v_{\tau} \geq v_{\tau+1}$, we compare to another calendar: $$\bm{v}^* = (v_1, v_2, ..., v_{t-1}, v_t+\delta, v_{t+1}-\delta, v_{t+2}, ..., v_T)$$ where $t = \sup_{\tau \in [T-1]} \left\{ v_{\tau} < 1, v_{\tau+1} > 0, v_{\tau} \geq v_{\tau+1} \right\}$ and $0 \leq \delta \leq \min \{1-v_t, v_{t+1}\}$. Notice that $t$ is the largest element in this set $\left\{ v_{\tau} < 1, v_{\tau+1} > 0, v_{\tau} \geq v_{\tau+1} \right\}$, which indicates that $\forall \tau \geq t+2, v_{\tau}=0$. Since we are changing probabilities from period $t$, the expected revenue does not change before period $t$.

Let $J_{\bm{v}, t}(c)$ denote the expected revenue we would earn under calendar $\bm{v}$ if we were endowed with $c$ units of inventory at the beginning of period $t$. Its expectation is taken over future demand randomness.

Let $A(c, t)$ denote the event ``$c$ units of inventory at the beginning of period $t$'', where $c \in [0,b], t \in [T]$. Then for any $c \in [0,b]$, conditioning on $A(c,t)$ we can expand the expression of the expected revenue. We proceed from equation \eqref{eqa:gen:exp1}:
\begin{equation}
\label{eqa:gen:exp3}
\begin{split}
J_{\bm{v}^*, t}(c) = & (v_t + \delta) (1 - v_{t+1} + \delta) (p_\pH - p_\pL) \bE\left[ \min\{c,Q_{\pH,t}\} \right] \\
& - (1 - v_t - \delta) (v_{t+1} - \delta) (p_\pH - p_\pL) \bE\left[ \min\{c,Q_{\pL,t}\} \right] \\
& + (v_t + \delta) (v_{t+1} - \delta) p_\pH \bE\left[ \min\{c, Q_{\pH,t} + Q_{\pH,t+1}\} \right]\\
& + (v_t + \delta) (1-v_{t+1} + \delta) p_\pL \bE\left[ \min\{c, Q_{\pH,t} + Q_{\pL,t+1}\} \right]\\
& + (1 - v_t - \delta) (v_{t+1} - \delta) p_\pH \bE\left[ \min\{c, Q_{\pL,t} + Q_{\pH,t+1}\} \right]\\
& + (1 - v_t - \delta) (1-v_{t+1} + \delta) p_\pL \bE\left[ \min\{c, Q_{\pL,t} + Q_{\pL,t+1}\} \right]\\
& + (v_t + \delta) (v_{t+1} - \delta) \bE\left[J_{\bm{v}, t+2}((c - Q_{\pH,t} - Q_{\pH,t+1})^+)\right]\\
& + (v_t + \delta) (1-v_{t+1} + \delta) \bE\left[J_{\bm{v}, t+2}((c - Q_{\pH,t} - Q_{\pL,t+1})^+)\right]\\
& + (1 - v_t - \delta) (v_{t+1} - \delta) \bE\left[J_{\bm{v}, t+2}((c - Q_{\pL,t} - Q_{\pH,t+1})^+)\right]\\
& + (1 - v_t - \delta) (1-v_{t+1} + \delta) \bE\left[J_{\bm{v}, t+2}((c - Q_{\pL,t} - Q_{\pL,t+1})^+)\right]
\end{split}
\end{equation}
where equality holds because calendars $\bm{v}$ and $\bm{v}^*$ are the same from period $t+2$.

Using \eqref{eqa:gen:exp2}, we subtract \eqref{eqa:gen:exp3} and \eqref{eqa:gen:exp1} to calculate their difference, conditioning on $A(c,t)$:
\begin{align*}
J_{\bm{v}^*, t}(c) - J_{\bm{v}, t}(c) = & \delta (1 - v_{t+1} + v_t + \delta) (p_\pH - p_\pL) \bE\left[ \min\{c,Q_{\pH,t}\} \right] \\
& + \delta (1 - v_t + v_{t+1} - \delta) (p_\pH - p_\pL) \bE\left[ \min\{c,Q_{\pL,t}\} \right] \\
& + \delta (v_{t+1} - v_t - \delta) p_\pH \bE\left[ \min\{c, Q_{\pH,t} + Q_{\pH,t+1}\} \right]\\
& + \delta (1 - v_{t+1} + v_t + \delta) p_\pL \bE\left[ \min\{c, Q_{\pH,t} + Q_{\pL,t+1}\} \right]\\
& + \delta (- 1 + v_t - v_{t+1} + \delta) p_\pH \bE\left[ \min\{c, Q_{\pL,t} + Q_{\pH,t+1}\} \right]\\
& + \delta (- v_t + v_{t+1} - \delta) p_\pL \bE\left[ \min\{c, Q_{\pL,t} + Q_{\pL,t+1}\} \right]\\
& + \delta (v_{t+1} - v_t - \delta) \bE\left[J_{\bm{v}, t+2}((c - Q_{\pH,t} - Q_{\pH,t+1})^+)\right]\\
& + \delta (- v_{t+1} + v_t + \delta) \bE\left[J_{\bm{v}, t+2}((c - Q_{\pH,t} - Q_{\pL,t+1})^+)\right]\\
& + \delta (v_t - v_{t+1} + \delta) \bE\left[J_{\bm{v}, t+2}((c - Q_{\pL,t} - Q_{\pH,t+1})^+)\right]\\
& + \delta (- v_t + v_{t+1} - \delta) \bE\left[J_{\bm{v}, t+2}((c - Q_{\pL,t} - Q_{\pL,t+1})^+)\right]
\end{align*}

Merging similar expressions we have the following:
\begin{align*}
& \frac{J_{\bm{v}^*, t}(c) - J_{\bm{v}, t}(c)}{\delta} \\
= & (p_\pH - p_\pL) \left\{ \bE\left[ \min\{c,Q_{\pH,t}\} \right] + \bE\left[ \min\{c,Q_{\pL,t}\} \right] - \bE\left[ \min\{c, Q_{\pH,t} + Q_{\pL,t+1}\} \right] \right\} \\
& + (v_t - v_{t+1} + \delta) (p_\pH - p_\pL) \bE\left[ \min\{c,Q_{\pH,t}\} \right] \\
& - (v_t - v_{t+1} + \delta) (p_\pH - p_\pL) \bE\left[ \min\{c,Q_{\pL,t}\} \right] \\
& - (v_t - v_{t+1} + \delta) p_\pH \bE\left[ \min\{c, Q_{\pH,t} + Q_{\pH,t+1}\} \right]\\
& + (v_t - v_{t+1} + \delta) p_\pL \bE\left[ \min\{c, Q_{\pH,t} + Q_{\pL,t+1}\} \right]\\
& + (v_t - v_{t+1} + \delta) p_\pH \bE\left[ \min\{c, Q_{\pL,t} + Q_{\pH,t+1}\} \right]\\
& - (v_t - v_{t+1} + \delta) p_\pL \bE\left[ \min\{c, Q_{\pL,t} + Q_{\pL,t+1}\} \right]\\
& - (v_t - v_{t+1} + \delta) \bE\left[J_{\bm{v}, t+2}((c - Q_{\pH,t} - Q_{\pH,t+1})^+)\right]\\
& + (v_t - v_{t+1} + \delta) \bE\left[J_{\bm{v}, t+2}((c - Q_{\pH,t} - Q_{\pL,t+1})^+)\right]\\
& + (v_t - v_{t+1} + \delta) \bE\left[J_{\bm{v}, t+2}((c - Q_{\pL,t} - Q_{\pH,t+1})^+)\right]\\
& - (v_t - v_{t+1} + \delta) \bE\left[J_{\bm{v}, t+2}((c - Q_{\pL,t} - Q_{\pL,t+1})^+)\right]\\
\geq & (v_t - v_{t+1} + \delta) (p_\pH - p_\pL) \left\{ \bE\left[ \min\{c,Q_{\pH,t}\} \right] + \bE\left[ \min\{c,Q_{\pL,t}\} \right] \right. \\
& \qquad \qquad \qquad \left. -\bE\left[ \min\{c, Q_{\pH,t} + Q_{\pL,t+1}\} \right] \right\} \\
& + (v_t - v_{t+1} + \delta) (p_\pH - p_\pL) \bE\left[ \min\{c,Q_{\pH,t}\} \right] \\
& - (v_t - v_{t+1} + \delta) (p_\pH - p_\pL) \bE\left[ \min\{c,Q_{\pL,t}\} \right] \\
& - (v_t - v_{t+1} + \delta) p_\pH \bE\left[ \min\{c, Q_{\pH,t} + Q_{\pH,t+1}\} \right]\\
& + (v_t - v_{t+1} + \delta) p_\pL \bE\left[ \min\{c, Q_{\pH,t} + Q_{\pL,t+1}\} \right]\\
& + (v_t - v_{t+1} + \delta) p_\pH \bE\left[ \min\{c, Q_{\pL,t} + Q_{\pH,t+1}\} \right]\\
& - (v_t - v_{t+1} + \delta) p_\pL \bE\left[ \min\{c, Q_{\pL,t} + Q_{\pL,t+1}\} \right]\\
& - (v_t - v_{t+1} + \delta) \bE\left[J_{\bm{v}, t+2}((c - Q_{\pH,t} - Q_{\pH,t+1})^+)\right]\\
& + (v_t - v_{t+1} + \delta) \bE\left[J_{\bm{v}, t+2}((c - Q_{\pH,t} - Q_{\pL,t+1})^+)\right]\\
& + (v_t - v_{t+1} + \delta) \bE\left[J_{\bm{v}, t+2}((c - Q_{\pL,t} - Q_{\pH,t+1})^+)\right]\\
& - (v_t - v_{t+1} + \delta) \bE\left[J_{\bm{v}, t+2}((c - Q_{\pL,t} - Q_{\pL,t+1})^+)\right]
\end{align*}
where the inequality holds because $\delta \leq \min\{1 - v_t, v_{t+1}\}$ so $v_t - v_{t+1} + \delta \leq 1$, and because of Lemma~\ref{lem:trc:ineq}.

Further merging similar expressions, while using \eqref{eqa:gen:exp2} we have the following:
\begin{align*}
\frac{J_{\bm{v}^*, t}(c) - J_{\bm{v}, t}(c)}{\delta (v_t - v_{t+1} + \delta)} \geq & (p_\pH - p_\pL) \left\{ 2 \bE\left[ \min\{c,Q_{\pH,t}\} \right] - \bE\left[ \min\{c, Q_{\pH,t} + Q_{\pH,t+1}\} \right] \right\} \\
& - p_\pL \bE\left[ \min\{c, Q_{\pH,t} + Q_{\pH,t+1}\} \right]\\
& + p_\pL \bE\left[ \min\{c, Q_{\pH,t} + Q_{\pL,t+1}\} \right]\\
& + p_\pL \bE\left[ \min\{c, Q_{\pL,t} + Q_{\pH,t+1}\} \right]\\
& - p_\pL \bE\left[ \min\{c, Q_{\pL,t} + Q_{\pL,t+1}\} \right]\\
& - \bE\left[J_{\bm{v}, t+2}((c - Q_{\pH,t} - Q_{\pH,t+1})^+)\right]\\
& + \bE\left[J_{\bm{v}, t+2}((c - Q_{\pH,t} - Q_{\pL,t+1})^+)\right]\\
& + \bE\left[J_{\bm{v}, t+2}((c - Q_{\pL,t} - Q_{\pH,t+1})^+)\right]\\
& - \bE\left[J_{\bm{v}, t+2}((c - Q_{\pL,t} - Q_{\pL,t+1})^+)\right]\\
\geq & - p_\pL \bE\left[ \min\{c, Q_{\pH,t} + Q_{\pH,t+1}\} \right]\\
& + p_\pL \bE\left[ \min\{c, Q_{\pH,t} + Q_{\pL,t+1}\} \right]\\
& + p_\pL \bE\left[ \min\{c, Q_{\pL,t} + Q_{\pH,t+1}\} \right]\\
& - p_\pL \bE\left[ \min\{c, Q_{\pL,t} + Q_{\pL,t+1}\} \right]\\
& - \bE\left[J_{\bm{v}, t+2}((c - Q_{\pH,t} - Q_{\pH,t+1})^+)\right]\\
& + \bE\left[J_{\bm{v}, t+2}((c - Q_{\pH,t} - Q_{\pL,t+1})^+)\right]\\
& + \bE\left[J_{\bm{v}, t+2}((c - Q_{\pL,t} - Q_{\pH,t+1})^+)\right]\\
& - \bE\left[J_{\bm{v}, t+2}((c - Q_{\pL,t} - Q_{\pL,t+1})^+)\right]
\end{align*}
where the second inequality is due to Lemma~\ref{lem:trc:ineq}.

Now let us introduce a coupling argument: suppose $U_{\tau}, \tau = t, t+1 $ are two uniform distributions over $[0,1]$, and let $Q_{\pH,\tau} = F_\pH^{-1}(U_{\tau}), Q_{\pL,\tau} = F_\pL^{-1}(U_{\tau}), \tau = t, t+1$. Since $F_\pH$ and $F_\pL$ are CDF's of distributions, they are monotone increasing, thus the inverse function exists. And for any specific realization of $U_{\tau}, \tau = t, t+1$, we know that $Q_{\pH,\tau} \leq Q_{\pL,\tau}$ or $Q_{\pH,\tau} \geq Q_{\pL,\tau}$, due to Assumption~\ref{asp:sta1}.

Now we only need to understand $J_{\bm{v}, t+2}(c)$ as a function of $c \in [0,b]$. Apparently $J_{\bm{v}, t+2}(0) = 0$. Let us denote the distribution of random demands in the last $T-t+1$ periods as $X$, whose CDF denoted as $\hat{F}$. $X$ is a non-negative random variable. We sell at the lower price from period $t+2$, and so the expected revenue in the last $T-t+1$ periods is calculated as $J_{\bm{v}, t+2}(c) = p_\pL \bE[\min\{c, X\}] = p_\pL \int_{[0,b]} \min\{c,u\} \mathrm{d} \hat{F}(u)$. Due to Fubini's theorem, we can perform integration by part, and have: 
\begin{align*}
J_{\bm{v}, t+2}(c) = & p_\pL \int_{[0,b]} \min\{c,u\} \mathrm{d} \hat{F}(u) \\
= & p_\pL \int_{[0,c]} u \mathrm{d} \hat{F}(u) + p_\pL \int_{[c,b]} c \mathrm{d} \hat{F}(u)\\
= & p_\pL ( c \hat{F}(c)- \int_{[0,c]} \hat{F}(u) \mathrm{d}u) + p_\pL (c - c\hat{F}(c)) \\
= & p_\pL ( c - \int_{[0,c]} \hat{F}(u) \mathrm{d}u)
\end{align*}
This is a differentiable function with respect to $c$. Taking derivative we have: $J'_{\bm{v}, t+2}(c) = p_\pL (1 - \hat{F}(c)) \in [0, p_\pL]$. Since $\hat{F}(c)$ is a CDF, thus non-decreasing, we know that $J'_{\bm{v}, t+2}(c)$ is non-increasing. So $J_{\bm{v}, t+2}(c)$ is concave.
Due to Lemma~\ref{lem:trc:ineq2} we finish our proof.
\Halmos\endproof

Now let us prove Lemma~\ref{lem::convexity}.

\proof{Proof of Lemma~\ref{lem::convexity}.}
Let us start from the second calendar described in Lemma~\ref{lem:structure1}. We will show that by finitely applying Corollary \ref{thm:gen:monotone2} and Lemma~\ref{lem:gen:structure3} we can obtain the first calendar described in Lemma~\ref{lem:structure1}, which earns a greater revenue.

First deduct from Corollary \ref{thm:gen:monotone2} that the randomized policy is non-increasing. Then by Lemma~\ref{lem:gen:structure3} we can change the last pair of probabilities with non-increasing order, i.e. $t = \sup_{\tau \in [T-1]} \left\{ v_{\tau} < 1, v_{\tau+1} > 0, v_{\tau} \geq v_{\tau+1} \right\}$, from $(v_t, v_{t+1})$ to $(v_t+\delta, v_{t+1}-\delta)$ with $\delta = \max \{1-v_t, v_{t+1}\} \geq 0$ to achieve a larger revenue. After such change, either $v_t = 1$ or $v_{t+1} = 0$ so we know that the total number of fractional (strictly greater than $0$ and smaller than $1$) periods have decreased by $1$. By repeatedly applying Corollary~\ref{thm:gen:monotone2} and Lemma~\ref{lem:gen:structure3} we will obtain the optimal randomized policy as described.
\Halmos\endproof

\section{Necessity of Two Prices in Single-Item Pricing Problem: Proof of Proposition~\ref{prop:singlePriceInfeasible}}
\label{sec:proof:singlePriceInfeasible}

\proof{Proof of Proposition~\ref{prop:singlePriceInfeasible}.}
Construct the following instance. Let there be $T$ periods and $b=1$ unit of initial inventory. Let $T$ be some large number. There are two prices $p_1=3, p_2=2$. Let random demand be Bernoulli random variables. The purchase probability of offering the higher price $p_1$ is $q_1 = 1/2T$; of offering the lower price $p_2$ is $q_2=1$.

The LP upper bound suggests a total of $\frac{4T-3}{2T-1} \approx 2$ units of revenue. 
Only offering the higher price $p_1$ suggests a total of $3 \cdot \bE[\min\{\Bin(T,1/2T),1\}] = 3 \cdot (1-1/2T)^T \approx 1 \cdot (1-e^{-1}) \approx 1.180$ units of revenue.
Only offering the lower price $p_2$ suggests a total of $1$ unit of revenue.
$\max\{1.180, 1\} < 2 \cdot (1-1/e) \approx 1.264$.
So the expected revenue of either single-price policy is strictly smaller than expression~(\ref{eqn::binomialBound}), which finishes the tightness proof.
\Halmos\endproof

\section{Proof of Theorem~\ref{thm::pricingNonstationary}} \label{sec:proof:thm::pricingNonstationary}

We prove Theorem~\ref{thm::pricingNonstationary} in the general [0,1]-demand setting; the \{0,1\}-demand setting is a special case of it.

\proof{Proof of Theorem~\ref{thm::pricingNonstationary}.}
Let $j_t, \forall t \in [T]$ denote the prices selected from expression~\eqref{eqn::policyPricingNonstationary}. Denote $B_t$ to be the remaining inventory at the \textit{end} of time $t$, with $B_0=0$. Denote $Q_t$ to be the inventory at time $t$ that customer \textit{would have} demanded if price $p_{j_t}$ is offered in time period $t$, which can take any value in $[0,1]$. Then $\min\{Q_t,B_{t-1}\}$ is the actual sales at time $t$.

We let $\bE[\Rev]$ denote the expected revenue earned by the deterministic calendar suggested in Algorithm~\ref{def::pricingNonstationary}, which can be written as:
\begin{align*}
\bE[\Rev] &=\sum_{t=1}^Tp_{j_t} \bE[\min\{Q_t,B_{t-1}\}] \\
&=\sum_{t=1}^T(p_{j_t}-\frac{\OPT}{2b}) \bE[\min\{Q_t,B_{t-1}\}] +\frac{\OPT}{2b}\sum_{t=1}^T \bE[\min\{Q_t,B_{t-1}\}] \\
&=\sum_{t=1}^T(p_{j_t}-\frac{\OPT}{2b})\bE[\min\{Q_t,B_{t-1}\}]+\frac{\OPT}{2b}(b-\bE[B_T])
\end{align*}

Now, note that the inventory level $B_t$ is decreasing in $t$ and that each $(p_{a_t}-\frac{\OPT}{2b})$ term is non-negative, so we can bound each $B_{t-1}$ from below by $B_T$. The following can then be derived:
\begin{align*}
\bE[\Rev] &=\sum_{t=1}^T(p_{j_t}-\frac{\OPT}{2b})\bE[\min\{Q_t,B_{t-1}\}\cdot\frac{Q_tB_{t-1}}{\min\{Q_t,B_{t-1}\}\max\{Q_t,B_{t-1}\}}]+\frac{\OPT}{2}(1-\frac{\bE[B_T]}{b}) \\
&\ge\sum_{t=1}^T(p_{j_t}-\frac{\OPT}{2b})\bE[\frac{Q_tB_{t-1}}{b}]+\frac{\OPT}{2}(1-\frac{\bE[B_T]}{b}) \\
&\ge\frac{\bE[B_T]}{b}\sum_{t=1}^T(p_{j_t}-\frac{\OPT}{2b})q_{t j_t}+\frac{\OPT}{2}(1-\frac{\bE[B_T]}{b}) \\
&\ge\frac{\bE[B_T]}{b}\sum_{t=1}^T\sum_{j=1}^J(p_j-\frac{\OPT}{2b})q_{tj}\ssx_{tj}+\frac{\OPT}{2}(1-\frac{\bE[B_T]}{b}) \\
&\ge\frac{\bE[B_T]}{b}\Big(\OPT-\frac{\OPT}{2b}\cdot b\Big)+\frac{\OPT}{2}(1-\frac{\bE[B_T]}{b})\\
&=\frac{\OPT}{2}
\end{align*}
where the first inequality is because $\forall t \in [T], Q_t \leq b$ and $B_{t}\le b$; second inequality is because $B_t$ is decreasing in $t$; third inequality follows from the optimality of $j_t$ in expression~\eqref{eqn::policyPricingNonstationary}; and the fourth inequality follows from constraint~(\ref{eqn::constr::pricingInventory}) in DLP-N.
\Halmos\endproof


\end{document}